\documentclass{article}

\usepackage[hidelinks]{hyperref}
\usepackage{a4wide}

\usepackage{subcaption}
\newcommand{\RNum}[1]{\uppercase\expandafter{\romannumeral #1\relax}}

\usepackage[numbers,sort&compress]{natbib}
\usepackage{booktabs}
\usepackage{tabularx}
\usepackage{adjustbox}
\usepackage{multirow}
\usepackage{xcolor}
\usepackage{empheq}
\usepackage{amssymb}
\usepackage{amsmath}
\usepackage{amsthm}
\usepackage{mathrsfs}  
\usepackage{siunitx}
\usepackage{cleveref}

\newtheorem{remark}{Remark}

\crefname{equation}{equation}{equations}
\Crefname{equation}{Equation}{Equations}

\usepackage{authblk} 

\title{Bayesian Model Selection for Complex Flows \\ of Yield Stress Fluids}

\author[1]{Aricia Rinkens}
\author[1]{Clemens V.\ Verhoosel}
\author[1]{Alexandra Alicke}
\author[1]{Patrick D.\ Anderson}
\author[1]{Nick O.\ Jaensson}

\affil[1]{\small Eindhoven University of Technology, Department of Mechanical Engineering, The Netherlands}

\begin{document}

\maketitle

\begin{abstract}
Modeling yield stress fluids in complex flow scenarios presents significant challenges, particularly because conventional rheological characterization methods often yield material parameters that are not fully representative of the intricate constitutive behavior observed in complex geometries and flow conditions. In this work, we propose a Bayesian uncertainty quantification framework for the calibration and selection of constitutive models for yield stress fluids, explicitly accounting for uncertainties in both modeling accuracy and experimental observations. The framework addresses the challenge of complex flow modeling by making discrepancies that emanate from rheological measurements explicit and quantifiable. We apply the Bayesian framework to rheological measurements and squeeze flow experiments on Carbopol 980, comparing a range of constitutive models. Our analysis demonstrates that Bayesian model selection yields robust probabilistic predictions and provides an objective assessment of model suitability through evaluated plausibilities, particularly when model bias is incorporated into the inference process. The framework naturally penalizes unnecessary complexity, in line with Occam’s razor, and shows that the optimal model choice depends on the incorporated physics, the specification of prior information, and the availability of calibration data. In rheological settings, the Herschel-Bulkley and biviscous power law models perform well, with observational noise dominating the uncertainty. However, when these rheological outcomes are used as prior information for a rheo-informed squeeze flow analysis, a significant mismatch with the experimental data is observed. This is due to the yield stress inferred from rheological measurements not being representative of the complex squeeze flow case. In contrast, an expert-informed squeeze flow analysis, based on broader priors, yields accurate predictions. These findings highlight the limitations of translating rheological measurements to complex flows and underscore the value of Bayesian approaches in quantifying model bias and guiding model selection under uncertainty.
\end{abstract}



\section{Introduction}
\label{sec:intro}

Soft matter refers to a class of materials that are typically described as liquids with a complex internal microstructure, such as emulsions, suspensions, and polymer solutions \cite{de1992soft, bonn2017yield}. The microstructure leads to an intricate relationship between deformation and stress, known as its constitutive behavior \cite{larson1999structure}. Soft matter plays a vital role in a wide range of industries, including food processing, pharmaceuticals, and energy. In such sectors, accurate flow modeling is crucial for process optimization, control, and scalability, among other aspects. We herein focus on yield stress fluids -- \emph{i.e.}, materials that behave like a solid until a certain stress threshold is exceeded, after which they flow like a liquid -- which form a sub-class of soft matter with many industrial applications (\emph{e.g.}, toothpaste, cement, ketchup).

Two prominent modeling objectives in the study of complex fluid flows are to characterize the material itself, which we refer to as \emph{descriptive modeling}, and to employ models that make useful predictions about the system under consideration, here referred to as \emph{predictive modeling}. A good descriptive model is capable of capturing all essential physical phenomena related to the behavior of the material. To also be a good predictive model, it is crucial that the model parameters can be uniquely and accurately characterized in application scenarios. For yield stress fluids it is well known that material characterization is challenging, and the success of predictive modeling efforts is sensitive to the specific flow conditions encountered in complex applications \cite{Coussot2014YieldReview,de2021yield}. This challenge was highlighted during discussions at the VPF workshop, which forms the basis of this special issue.

The conventional approach to improving predictive modeling capabilities for yield stress fluids is to refine both the experimental procedures and the mathematical-physical models, tailoring them to specific flow scenarios of interest. Such an approach is warranted for cases that require very high levels of certainty. However, in engineering systems, such a deterministic approach may not always be justified due to all other inherent uncertainties. Refinements of experiments and models to practically eliminate uncertainties is often impractical or infeasible. From an engineering perspective, uncertainties are not necessarily prohibitive, provided that they can be quantified with confidence and remain within well-defined margins. Through incorporation of uncertainties into the modeling process, the \emph{uncertainty quantification (UQ)} modeling paradigm provides an alternative to the deterministic approach.

In UQ the discrepancy between a model prediction and an experimental observation, referred to as the error, can be due to two sources: the model bias, \emph{i.e.}, the epistemic error, and the observational noise, \emph{i.e.}, the aleatoric error \cite{oden2017predictive}. In principle, classical mechanics treats all errors as model biases. However, in practice, the underlying model is often intractable and non-essential for achieving modeling objectives. UQ then seeks a practical balance between these errors through probabilistic modeling. A key aspect of this probabilistic modeling process is to identify which elements of the model and observations require a detailed physics-based description and which can be treated as random variables. In practice, balancing these sources of error involves selecting an appropriate model and representing both modeling and observational uncertainties in a stochastic manner. UQ and model selection provide a natural way to achieve this balance.

Two main paradigms are commonly used to address stochasticity: the \emph{frequentist} and the \emph{Bayesian} approaches. The key distinction lies in their treatment of uncertainty. In the frequentist framework, the model is considered deterministic, while the data are viewed as stochastic; it seeks to determine which data are most probable given a fixed model. In contrast, the Bayesian approach treats the model parameters as random variables and answers the question of which models are most probable given the observed data, thereby providing a natural framework for model selection and uncertainty quantification. For this reason we adopt the Bayesian approach to UQ in this work.

Bayesian UQ is based on Bayes' rule, which essentially provides a model calibration framework through the updating of probabilistic model parameters using observational data. From the point of view of model selection, Bayesian UQ is a probabilistic interpretation of Occam's razor\footnote{Occam's Razor is the principle that, among competing models that equally explain the data, the simplest one should be preferred.} \cite{mackay2003information}. From a UQ perspective, more complex models are not inherently better, as a price is paid for additionally introduced parametric uncertainty, particularly when sufficient data for calibration are lacking. As a result, a model that performs well under controlled laboratory conditions may not be suitable in industrial environments where opportunities for parameter calibration are limited.

Yield stress fluids in complex flow scenarios exemplify the relevance of probabilistic model selection. The literature reports the difficulty of characterizing and modeling yield stress materials, with Carbopol being a prominent example \cite{Dinkgreve2016YieldStress, dinkgreve2018carbopol, r2019rheological, gutowski2012scaling, jaworski2022carbomer}. Even when the same material is prepared using the same protocol, the measured yield stress values and other rheological properties often vary significantly due to minor variations in preparation, physical state, and experimental environment. These variations become even more pronounced when considering real-world applications involving intricate flows, which often correlate poorly with the simple shear flow conditions under which rheological characterizations typically takes place. This highlights the importance of understanding the behavior of yield stress fluids beyond simple shear, including squeeze flow applications \cite{Rabideau2009SqueezeFlow} and scenarios emanating from emerging techniques such as protorheology \cite{Hossain2024Protorheology, hossain2024protorheologymis, marsh2025egg}. In such cases, it is essential to select and calibrate models \cite{Coussot2025FiftyShades}.

Yield stress fluids research has focused on refining measurements and models to achieve a near-perfect fit between theory and observation. Traditional constitutive models such as Bingham \cite{bingham1917investigation,bingham1922fluidity} and Herschel–Bulkley \cite{herschel1926konsistenzmessungen} have long been used, assuming a sharp transition from solid-like to fluid-like behavior. While the Herschel–Bulkley model extends Bingham’s formulation by incorporating shear-thinning effects, both rely on a fixed yield stress threshold and do not incorporate phenomena such as thixotropy, aging, and viscoelasticity \cite{moller2006yield, deSouza2012, deSouza2013, joshi2018yield, larson2019review}. To address these limitations, more sophisticated constitutive models have been developed. For example, \citet{saramito2007new} introduced a thermodynamically consistent elastoviscoplastic model that combines viscoelastic and viscoplastic behavior, allowing for more accurate simulations of complex materials such as Carbopol. \citet{pagani2024no} challenged the concept of a fixed yield stress by proposing a stress-activated relaxation time model, which treats yielding as a gradual transition rather than a discrete threshold. \citet{kamani2024brittle} further emphasized the diversity of yielding responses, such as brittle versus ductile, based on microstructural dynamics, emphasizing the need for rheological models tailored to the specific behavior of soft matter systems.

The research on yield stress fluids can predominantly be characterized as a conventional, deterministic, approach to improving predictive modeling capabilities.  Although this approach has been successful in both controlled (laboratory) and engineering settings, obtaining accurate predictions remains challenging when considering intricate flow scenarios, where the complexity of the system and the limitations of available data complicate  model calibration. While the UQ modeling paradigm is better suited to such uncertain conditions, it is still relatively unexplored in non-Newtonian fluid mechanics and rheology. Uncertainty quantification in the context of rheological modeling was systematically introduced by the pioneering work of \citet{freund2015quantitative}, covering Bayesian inference and model selection. Subsequent studies have demonstrated that Bayesian calibration of constitutive model parameters, combined with uncertainty propagation, enables robust and reliable predictions for rheological flows, \emph{e.g.}, Refs.~\cite{kim2019uncertainty,Singh2019OnRepresentation,ranftl2023bayesian}. Bayesian inference has also been extended to more complex rheological models (\emph{e.g.}, Refs.~\cite{Chakraborty2021, paul2021bayesian, Ran2023UnderstandingInference,miranda2025bayesian}) and increasingly complex flow scenarios (\emph{e.g.}, Refs.~\cite{rinkens2023uncertainty,kontogiannis2025learning}). Recent model selection research in the context of rheology also considers data-driven approaches \cite{saadat2022data,lardy2025}.

In this contribution we propose and study a UQ framework for the calibration and selection of yield stress fluid models in complex flow scenarios. Our framework explicitly accounts for the inherent uncertainties in both the modeling process and experimental observations. We study how model complexity and data availability interact in the model selection process, and demonstrate that Bayesian model selection naturally penalizes unnecessary complexity \emph{cf.} the principle of Occam’s razor. We show that, as a result, the optimal model choice may differ between rheometric experiments, where data are abundant and controlled, and more complex flow scenarios, where data are sparse and noisy.

Given the critical importance of controlling all aspects of uncertainty, we adopt a combined experimental and modeling approach. This integrated approach enables us to systematically quantify both model bias and experimental errors. For the development of our UQ framework for complex yield stress fluid flow scenarios, we follow a two-step approach. In the first step, we consider rheological measurements, where we have a well-defined and controlled experimental setting. These measurements benefit from sophisticated instrumentation and allow for precise quantification of uncertainties. Moreover, the associated models are typically analytical, offering robustness and computational efficiency, thereby significantly reducing the challenges commonly encountered in uncertainty quantification. In the second step, we extend our analysis to a squeeze flow, which serves as a typical case of a less controlled complex flow experiment capable of characterizing yield stress fluids \cite{covey1981use}. We consider a tailor-made experimental setup and semi-analytical models for which sampling remains tractable. We consider two distinct approaches for Bayesian inference. In the first, which we refer to as \emph{rheo-informed inference}, the rheological measurements are incorporated as prior information, simulating a situation where complex flow experiments are informed by experimental data. In the second, which we refer to as \emph{expert-informed inference}, we exclude these measurements from the priors, representing a scenario where such supporting data are unavailable. This approach allows us to assess how prior knowledge influences model choice and performance.

We structure our work as follows. We introduce Bayesian inference and model selection in \autoref{sec:uq}. The constitutive models to which we apply Bayesian model selection are described in \autoref{sec:rheologymmodels}. We discuss the results on Bayesian model selection in the rheological setting in \autoref{sec:results_rheometry}, where we make a distinction between two cases: one that assumes negligible model bias and another that treats model bias as an inferred parameter. In \autoref{sec:squeezeflow}, we present the results of constitutive model selection for the squeeze flow, considering both the rheo-informed and the expert-informed approaches. Finally, we conclude with a summary of our findings and recommendations for future work in \autoref{sec:conclusion}.


\section{Bayesian model selection}
\label{sec:uq}

\subsection{Probabilistic modeling}
A central objective in science and engineering is to make reliable predictions about physical phenomena. As systems grow in complexity, so do the sources of uncertainty. These uncertainties arise from multiple origins: limitations in experimental measurements (used for model validation and calibration), incomplete knowledge of system behavior, and simplifications inherent in the models themselves. Moreover, in many engineering applications, models are calibrated using simplified experiments that may not fully capture the behavior of the real system. Uncertainty Quantification (UQ) provides a framework for making predictions while explicitly accounting for these uncertainties.

In UQ, we model uncertain quantities as random variables that are described by a probability distribution. When we make predictions of a quantity of interest, the result in itself is also a probability distribution, reflecting the uncertainty introduced by the model parameters. To formalize the concept of UQ, we define the ground truth of the quantity of interest as $\boldsymbol{g}$ \cite{oden2017predictive}. The observations of the ground truth are described by a set $\boldsymbol{y} = \{ \boldsymbol{y}_1, \boldsymbol{y}_2, ..., \boldsymbol{y}_n \}$, where $n$ represents the number of measurements. The observational (aleatoric) error between the ground truth and the experiments is defined as
\begin{equation}
    \boldsymbol{\varepsilon}_{{\rm noise},i} = \boldsymbol{y}_i - \boldsymbol{g}_i, 
\end{equation}
where $\boldsymbol{g}_i$ is the realization of $\boldsymbol{g}$ corresponding to the observation $\boldsymbol{y}_{i}$. Since the ground truth is unknown in practice, we typically model the observational error probabilistically through a noise model. 

Next to an observational error, we have a model error (or model bias) $\boldsymbol{\varepsilon}_{{\rm bias},i}(\boldsymbol{\theta})$ that quantifies the error between the model and the ground truth as
\begin{equation}
    \boldsymbol{\varepsilon}_{{\rm bias},i}(\boldsymbol{\theta}) = \boldsymbol{g}_i - \boldsymbol{d}_i(\boldsymbol{\theta}),
\end{equation}
where $\boldsymbol{d}_i(\boldsymbol{\theta})$ represents a model evaluation. By combining the observational error and model bias, we can eliminate the unknown ground truth as
\begin{equation}
    \boldsymbol{y}_{i} - \boldsymbol{d}_i(\boldsymbol{\theta}) = \boldsymbol{\varepsilon}_{{\rm noise},i} + \boldsymbol{\varepsilon}_{{\rm bias},i}(\boldsymbol{\theta}).
\end{equation}
This equation formalizes that the mismatch between model predictions and observed data arises from two sources: measurement noise and model inadequacy.

A key task in UQ is to construct appropriate probabilistic models for these error terms, as they directly influence the reliability of predictions. Commonly, these two contributions are combined into a single effective noise model that encapsulates both observational and model uncertainties. However, in this work we separate these two terms from one another and use Bayesian inference to investigate both sources of uncertainty. 

\subsection{Bayesian model selection}
In Bayesian model selection, we consider a set of $m$ model classes $\boldsymbol{M} = [M_1, M_2, ... M_m]$. Each model class $M_j$ is a function of $p_j$ model parameters $\boldsymbol{\theta}_j=[\theta_1, \theta_2, ..., \theta_{p_j}] \in \boldsymbol{\Theta}_j$, where $\boldsymbol{\Theta}_j$ is the parameter domain corresponding to the model class $M_j$\footnote{We formally define a \emph{model class} as a mathematical formulation, and a \emph{model} as a model class together with its parameters. However, for improved readability, we use these terms interchangeably throughout the remainder of this work.}.

Bayes' rule expresses the probability density of the parameters $\boldsymbol{\theta}_j$ of model class $M_j$ given the observed data $\boldsymbol{y}$, referred to as the posterior, as
\begin{equation}
\begin{aligned}
    \pi(\boldsymbol{\theta}_j \mid \boldsymbol{y}, M_j) &= \frac{\pi(\boldsymbol{y} \mid \boldsymbol{\theta}_j, M_j) \, \pi(\boldsymbol{\theta}_j \mid M_j)}{\pi(\boldsymbol{y} \mid  M_j)} \\
    &= \frac{\mathcal{L}(\boldsymbol{\theta}_j \mid \boldsymbol{y}, M_j) \,\pi(\boldsymbol{\theta}_j \mid M_j)}{z_j},
    \label{eq:bayesianposterior}
\end{aligned}
\end{equation}
where $\pi(\boldsymbol{y} \mid \boldsymbol{\theta}_j, M_j) \equiv \mathcal{L}(\boldsymbol{\theta}_j \mid \boldsymbol{y}, M_j)$ is the likelihood, $\pi(\boldsymbol{\theta}_j \mid M_j$) the prior, and $\pi(\boldsymbol{y} \mid  M_j) \equiv z_j$ the evidence \cite{oden2017predictive,Lambert2017AStatistics}.

In model selection, the evidence $z_j$ is interpreted as the probability of observing the data $\boldsymbol{y}$ given the model class $M_j$, which corresponds to the marginal likelihood
\begin{equation}
z_j = \int_{\boldsymbol{\Theta}_j} \mathcal{L}(\boldsymbol{\theta}_j \mid \boldsymbol{y}, M_j) \, \pi(\boldsymbol{\theta}_j \mid M_j) \, d\boldsymbol{\theta}_j.
\label{eq:marginalevidence}
\end{equation}
Through the Bayesian model plausibility
\begin{equation}
    \rho_j \equiv \pi(M_j \mid \boldsymbol{y}) = \frac{\pi(\boldsymbol{y} \mid M_j)\,  \pi(M_j)}{\pi(\boldsymbol{y})} = \frac{z_j \, \pi(M_j)}{\pi(\boldsymbol{y})},
    \label{eq:bayesianplausibility}
\end{equation}
the model evidence allows us to evaluate which model class best explains the data. In \cref{eq:bayesianplausibility}, $\pi(M_j)$ is the prior probability of model class $M_j$, which is for example based on expert opinion, and
\begin{equation}
    \pi(\boldsymbol{y}) = \sum_{j=1}^m z_j \,\pi( M_j)
\end{equation}
is the normalization constant which guarantees that $\sum_{j=1}^m \rho_j = 1$. Therefore, the plausibilities are a discrete probability distribution over the set of model classes, $\boldsymbol{M}$. The higher the plausibility of a model class, the more likely it is that it explains the data compared to the other model classes.

\subsection{Model selection based on Markov chain Monte Carlo sampling}\label{sec:MCMC}
To evaluate the posterior distribution \eqref{eq:bayesianposterior}, we employ a Markov Chain Monte Carlo (MCMC) method, which is the standard for problems involving a significant number of parameters \cite{kaipio2005statistical, gamerman2006markov}. For each model class, MCMC generates a sequence $\boldsymbol{\mathcal{C}}_j = \{ \boldsymbol{\theta}_{j,n} \}_{n=1}^{N}$ of length $N$, where each $\boldsymbol{\theta}_{j,n}$ contains the $p_j$ parameter values at iteration $n$, effectively exploring the posterior landscape.

We use the Affine Invariant Stretch Move (AISM) sampler, which is advantageous in settings where the gradient of the log-posterior is not analytically available \cite{goodman2010ensemble}. This sampler operates with an ensemble of walkers, reducing the risk of convergence to local maxima and improving the exploration of complex posterior distributions. See \autoref{rem:samplerimplementation} for a note on important implementation aspects, and \citet{rinkens2025comparison} for a further discussion of the AISM algorithm, including aspects such as burn-in removal and convergence assessment.

To evaluate the plausibilities \eqref{eq:bayesianplausibility}, the marginal likelihood \eqref{eq:marginalevidence} must be evaluated for each model class. For problems with a significant number of parameters, directly computing the marginal likelihood becomes intractable due to the need to integrate over the entire parameter domain. To avoid this, we estimate the model evidence using the samples drawn from the posterior distribution, \emph{i.e.}, $\boldsymbol{\theta}_{j,n} \in \boldsymbol{\mathcal{C}}_j$, using the harmonic mean estimator \cite{newton1994approximate}
\begin{align}
    z_j^{-1} &= \text{E}_{\pi(\boldsymbol{\theta}_j \mid M_j)} \left[ \frac{\pi(\boldsymbol{\theta}_j \mid \boldsymbol{y}, M_j)}{\mathcal{L}(\boldsymbol{\theta}_{j}  \mid \boldsymbol{y}, M_j)\, \pi(\boldsymbol{\theta}_j \mid M_j )} \right]  \nonumber \\
    &= \text{E}_{\pi(\boldsymbol{\theta}_j \mid \boldsymbol{y}, M_j)} \left[ \frac{1}{\mathcal{L}(\boldsymbol{\theta}_{j}  \mid \boldsymbol{y}, M_j)} \right] \nonumber \\
    &\approx  \frac{1}{N} \sum_{n=1}^N \frac{1}{\mathcal{L}(\boldsymbol{\theta}_{j,n}  \mid \boldsymbol{y}, M_j)}.
\end{align}
This estimator is attractive because it only requires posterior samples and likelihood evaluations. However, it suffers from high variance, as it heavily weighs samples with very low likelihood, which are typically found in the tails of the posterior. These low-likelihood samples dominate the harmonic mean, leading to unreliable estimates or even divergence \cite{neal1994contribution}.

To stabilize the estimator, an arbitrary probability density $\varphi(\boldsymbol{\theta})$ can be defined to ameliorate the impact of very small likelihood values. An importance sampling estimator for the model evidence is then obtained as
\begin{align}
    z_j^{-1}  &= \text{E}_{\varphi(\boldsymbol{\theta}_j )} \left[ \frac{\pi(\boldsymbol{\theta}_j \mid \boldsymbol{y}, M_j)}{\mathcal{L}(\boldsymbol{\theta}_{j}  \mid \boldsymbol{y}, M_j)\, \pi(\boldsymbol{\theta}_j \mid M_j )} \right]  \nonumber \\
    &= \text{E}_{\pi(\boldsymbol{\theta}_j \mid \boldsymbol{y}, M_j)} \left[ \frac{\varphi(\boldsymbol{\theta}_j)}{\mathcal{L}(\boldsymbol{\theta}_{j}  \mid \boldsymbol{y}, M_j) \pi(\boldsymbol{\theta}_{j} \mid M_j)} \right] \nonumber \\
       &\approx  \frac{1}{N} \sum_{n=1}^N \frac{\varphi(\boldsymbol{\theta}_{j,n})}{\mathcal{L}(\boldsymbol{\theta}_{j,n}  \mid \boldsymbol{y}, M_j) \pi(\boldsymbol{\theta}_{j,n} \mid M_j)} \equiv \hat{z}_j^{-1},
    \label{eq:importancesamplingestimator}
\end{align}
where $\pi(\boldsymbol{\theta}_{j,n} \mid M_j)$ are the prior probability densities of the samples \cite{gelfand1994bayesian}.

To exclude low-likelihood regions from the estimator \eqref{eq:importancesamplingestimator}, we consider the uniform density
\begin{equation}
    \varphi(\boldsymbol{\theta}_{j,n}) = \frac{I_s(\boldsymbol{\theta}_{j,n})}{V_{s,j}} ,
    \label{eq:phidistribution}
\end{equation}
where $I_s(\boldsymbol{\theta}_{j,n})$ is the indicator function for the mean-centered ellipsoid defined by
\begin{equation}
    I_s(\boldsymbol{\theta}_{j,n}) =
    \begin{cases}
        1, & (\boldsymbol{\theta}_{j,n} - \bar{\boldsymbol{\theta}}_j)^{\mathrm{T}} \boldsymbol{\Sigma}_j^{-1} (\boldsymbol{\theta}_{j,n} - \bar{\boldsymbol{\theta}}_j) < R_s^2 \\
        0, & \text{otherwise}
    \end{cases},
    \label{eq:indicator}
\end{equation}
with posterior mean $\bar{\boldsymbol{\theta}}_j$ and posterior covariance $\boldsymbol{\Sigma}_{j}$. The volume $V_{s,j}$ of this ellipsoid is given by
\begin{equation}
    V_{s,j} = \frac{\pi^{p_j/2}}{\Gamma(p_j/2+1)} R_s^{p_j} |\Sigma_j |^{1/2},
\end{equation}
where $p_j$ is the dimension of the parameter space and $\Gamma(\cdot)$ is the Gamma function. The estimator based on this particular choice for the density $\varphi$ is known as the truncated harmonic mean estimator \cite{vanhaasteren2013marginal}. We note that alternative, more versatile, stabilization strategies exist \cite{mcewen2021machine}, but have found the truncated estimator to be adequate for the purposes of the current work.

For the truncated harmonic mean estimator \eqref{eq:importancesamplingestimator} to be accurate, the truncation radius $R_s$ in \cref{eq:indicator} must be selected appropriately. The optimal choice for the truncation radius minimizes the variance of the estimator, ensuring that the distribution \eqref{eq:phidistribution} is not too broad, which would still incorporate low-likelihood samples, and not too narrow, which would exclude informative regions of the posterior. To approximate this optimum we estimate the variance of the inverse of the truncated mean harmonic estimator \eqref{eq:importancesamplingestimator} as
\begin{align}
    \text{Var}\left[ \hat{z}_j^{-1} \right] &\approx \frac{1}{N^{\rm eff}} \text{Var}_{\pi(\boldsymbol{\theta}_j \mid \boldsymbol{y}, M_j)} \left[ \frac{\varphi(\boldsymbol{\theta}_j)}{\mathcal{L}(\boldsymbol{\theta}_{j}  \mid \boldsymbol{y}, M_j) \pi(\boldsymbol{\theta}_{j} \mid M_j)} \right] \nonumber \\
       &\approx \frac{1}{N^{\rm eff}} \cdot \frac{1}{N-1} \sum_{n=1}^N \left( \frac{\varphi(\boldsymbol{\theta}_{j,n})}{\mathcal{L}(\boldsymbol{\theta}_{j,n}  \mid \boldsymbol{y}, M_j) \pi(\boldsymbol{\theta}_{j,n} \mid M_j)} - \hat{z}_j^{-1}\right)^2,
    \label{eq:varianceestimator}
\end{align}
where $N^{\rm eff}$ is the effective sample size of the posterior sample \cite{geyer1992practical}, and determine the radius for which this variance is minimal. In Appendix~\ref{sec:APPvariance} we study the accuracy of this variance estimator.

\begin{figure}
    \centering
    \includegraphics[width=0.5\linewidth]{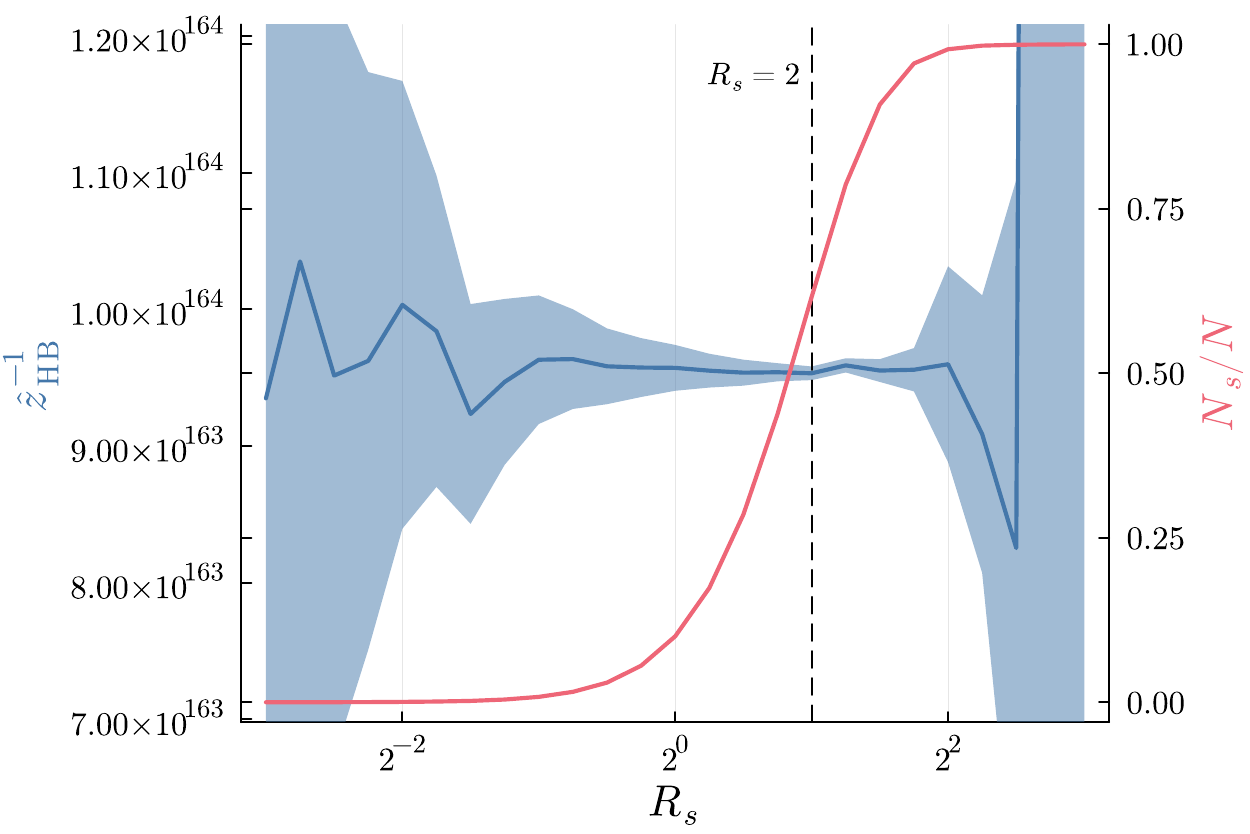}
    \caption{Example of the dependence of the evidence estimator on the truncation radius $R_s$. The uncertainty band shows the 95\% confidence interval of the estimator. The fraction of samples included in the truncation is represented by $N_s/N$.}
    \label{fig:LHE_rheo_R}
\end{figure}

\autoref{fig:LHE_rheo_R} shows an example of how we select the truncation radius. We plot the evidence estimator \eqref{eq:importancesamplingestimator} versus the truncation radius $R_s$, with the uncertainty band based on \cref{eq:varianceestimator}. On the right axis we display the posterior samples included in the ellipsoid ($N_s$) normalized by the total number of samples ($N$). It is observed that the variance of the evidence estimator minimizes around $R_s \approx 2$. Although the optimum value of $R_s$ is generally problem-specific, for all considered cases we have found that $R_s=2$ is close to the optimum. Due to this constant $R_s$ value, the proportion of posterior samples included in the truncation is virtually constant, which facilitates a fair comparison between models.

We use the estimator \eqref{eq:importancesamplingestimator} and its variance \eqref{eq:varianceestimator} to estimate the plausibilities \eqref{eq:bayesianplausibility} as 
\begin{equation}
    \hat{\rho}_j = \frac{1}{N} \sum_{n=1}^N \frac{\hat{z}_{j,n} \pi(M_j) }{\sum_{k=1}^m \hat{z}_{k,n} \pi(M_k) },
    \label{eq:sampledplausibilities}
\end{equation}
where
\begin{equation}
    \hat{z}_{j,n}^{-1} \sim \text{log}\mathcal{N}\left(\hat{z}_{j}^{-1}, \sqrt{\text{Var}\left[ \hat{z}_{j}^{-1} \right]} \right).
\end{equation}
From the sampled plausibilities \eqref{eq:sampledplausibilities} we also derive uncertainty bounds for the plausibilities. While these bounds are based on the estimator \eqref{eq:varianceestimator}, they are indicative of the sampler-related uncertainty of the plausibilities (see \autoref{sec:APPvariance}).

\begin{remark}[Sampler implementation aspects]\label{rem:samplerimplementation}
With respect to the implementation of the sampling procedures considered in this work, there are two important aspects to note:
\begin{itemize}
    \item To mitigate floating-point underflow and overflow, all sampling procedures operate on the logarithms of the probabilities and likelihoods.
    \item To enforce constraints on (a selection of) the parameters, it is not sufficient to select adequate prior distributions (\emph{e.g.}, uniform, log-normal or exponential), as the proposal step in an MCMC sampler may then still violate the constraint. To robustly enforce constraints, in our implementation we map constrained parameters to the real axis.  For the log-normal distribution this, \emph{e.g.}, means that the transformed parameter is normally distributed.
\end{itemize}

\end{remark}


\section{Generalized Newtonian models for yield stress fluids}
\label{sec:rheologymmodels}
In fluid mechanics, the stress tensor is split in the hydrostatic pressure tensor $p\boldsymbol{I}$  and the extra stress tensor $\boldsymbol{\tau}$ as 
\begin{equation}
    \boldsymbol{\sigma} = \boldsymbol{\tau} - p\boldsymbol{I}.
\end{equation}
The extra stress tensor characterizes the relationship between the deformation of a material and the stresses acting on it, referred to as the constitutive behavior. In the case of generalized Newtonian fluids, the stress response is governed entirely by the instantaneous rate of deformation and not by its history, such that the constitutive relation is of the form 
\begin{equation}
    \boldsymbol{\tau} = \boldsymbol{\tau} (  \boldsymbol{D} ),
\end{equation}
where the rate-of-deformation tensor is defined as $\boldsymbol{D}=\frac{1}{2}(\nabla \boldsymbol{v} + (\nabla \boldsymbol{v})^T)$, \emph{i.e.}, the symmetric gradient of the velocity field $\boldsymbol{v}$.

For the simple shear flow and lubrication flow in the thin-gap limit considered in this work we limit ourselves to the scalar setting of the constitutive relation, \emph{i.e.}, $\tau=\tau( \dot{\gamma} )$, where $\tau$ and $\dot{\gamma}$ are the shear components of the extra stress tensor and rate-of-deformation tensor, respectively. For notational brevity, in the presentation of the constitutive models we consider $\dot{\gamma} \geq 0$.

\begin{figure}
    \centering
    \includegraphics[width=0.5\linewidth]{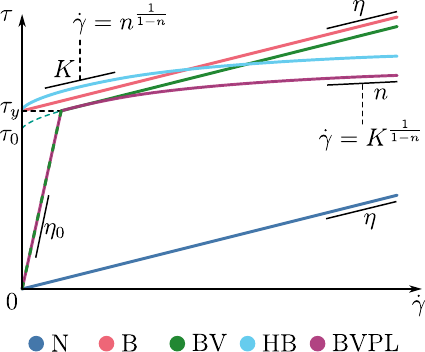}
    \caption{Overview of the considered rheological models: (N) Newtonian; (B) Bingham; (BV) biviscous; (HB) Herschel-Bulkley; and (BVPL) biviscous power law. Note that parameters may have model-dependent interpretations, \emph{e.g.}, $\eta$ in the biviscous power law should not be interpreted as an effective Newtonian viscosity.}
    \label{fig:rheo_taugd_all}
\end{figure}

We consider a set of model classes for yield stress fluids -- \emph{i.e.}, materials that behave as a solid below a certain stress threshold and flow as a liquid above it -- each offering a different representation of how materials transition from solid-like to fluid-like states. The critical stress at which this transition occurs is known as the shear yield stress $\tau_y$. For brevity, $\tau_y$ will hereafter be denoted simply as the yield stress, with the implicit assumption that the shear yield stress is intended. We compare five parametrized constitutive model classes, with non-decreasing level of complexity: the Newtonian model (N); the Bingham model (B); a biviscous model (BV); the Herschel-Bulkley model (HB); and a biviscous power law model (BVPL). The complexity of these model classes, which are illustrated in \autoref{fig:rheo_taugd_all}, is defined as the number of parameters. In the remainder of this section, these model classes will be elaborated. 

\subsection{The Newtonian model}
The Newtonian model assumes a liner relationship between the shear stress and shear rate as
\begin{equation}
    \tau = \eta \dot{\gamma},
\end{equation}
where the constant viscosity $\eta$ is the only parameter, \emph{i.e.}, $\boldsymbol{\theta}_{N} = [\eta]$ and $p_{N}=1$. As the Newtonian model does not incorporate the yield stress as a parameter, it cannot be expected to provide accurate representations of the constitutive behavior of a yield stress fluid. However, we include it in our hierarchy of model classes as a baseline model, with the goal of demonstrating how our framework addresses model biases.

\subsection{The Bingham model}
In the Bingham model \cite{bingham1922fluidity}, the fluid is assumed to behave as an infinitely rigid solid in the pre-yield regime. Once the stress exceeds the yield stress, the fluid enters the flow regime, where it deforms with a constant viscosity, $\eta$, similar to a Newtonian fluid. The Bingham model is defined by
\begin{equation}
\begin{aligned}
\dot{\gamma} &= 0 && \tau < \tau_y \\
\tau &= \tau_y + \eta \dot{\gamma} && \tau \ge \tau_y .
\end{aligned}
\end{equation}
The Bingham model contains two parameters, \emph{i.e.}, $\boldsymbol{\theta}_{B} = [\eta, \tau_y]$ and $p_{B}=2$. In the case that the yield stress is set to zero, the Bingham model reduces to the Newtonian model.

\subsection{The biviscous model}
The biviscous model \cite{beverly1992numerical} generalizes the Bingham model by considering a very high constant viscosity $\eta_0$ in the pre-yield regime. Once the stress exceeds the yield threshold, the fluid transitions into the flow regime, where it behaves like a Newtonian fluid with a much lower viscosity $\eta$. This sharp contrast between the pre- and post-yield viscosities allows the model to approximate yield stress behavior while maintaining computational tractability (\autoref{rem:biviscous}). 

The constitutive relation is given by
\begin{equation}
\begin{aligned}
    \dot{\gamma} &= \frac{\tau}{\eta_0}  &  &\tau < \tau_y \\
    \tau &=\tau_0 + \eta\dot{\gamma} & &\tau \geq \tau_y
\end{aligned},
\end{equation}
where $\tau_0=(1-\varepsilon)\tau_y$ with $\varepsilon=\eta/\eta_0$ ensures stress continuity over the yield threshold. This three-parameter model, \emph{i.e.}, $\boldsymbol{\theta}_{BV} = [\eta_0, \eta, \tau_y]$ and $p_{BV}=3$, reduces to the Bingham model when $\eta_0 \rightarrow \infty$, as illustrated in \autoref{fig:rheo_taugd_all}.

\subsection{The Herschel-Bulkley model}
The Herschel–Bulkley model \cite{herschel1926konsistenzmessungen} describes a fluid that behaves like an infinitely rigid solid until the yield stress is exceeded. Once this threshold is surpassed and the fluid enters the flow regime, it exhibits either shear-thinning or shear-thickening behavior, governed by the power-law relation. The Herschel–Bulkley model is defined by 
\begin{equation}
\begin{aligned}
\dot{\gamma} &= 0 && \tau < \tau_y \\
\tau &= \tau_y + K \dot{\gamma}^n && \tau \ge \tau_y,
\end{aligned}
\end{equation}
where $K$ is the consistency index and $n$ the dimensionless flow index. The fluid is shear thinning for $0 < n < 1$ and shear thickening for $n > 1$. For the case that $n=1$ the Bingham model is recovered (with $\eta$ redefined as $K$). Like the biviscous model, the Herschel-Bulkley model has three parameters, \emph{i.e.}, $\boldsymbol{\theta}_{HB} = [K, n, \tau_y]$ and $p_{HB}=3$. However, it generalizes the Bingham model in a different way compared to the biviscous model.

\subsection{The biviscous power law model}
We finally consider a model that incorporates the traits of both the biviscous model and the Herschel-Bulkley model. In the pre-yield regime, the solid-like behavior is characterized by a very high constant viscosity, $\eta_0$, similar to the biviscous model. The model exhibits a power law behavior after yielding similar to the Herschel-Bulkley model, allowing it to capture both shear thinning and shear thickening responses. The constitutive relation of this biviscous power law model reads
\begin{equation}
\begin{aligned}
    \dot{\gamma} &= \frac{\tau}{\eta_0} &  &\tau < \tau_y \\
    \tau &= \tau_0 + K\dot{\gamma}^n & &\tau \geq \tau_y    
\end{aligned},
\end{equation}
where $\tau_0=(1-\varepsilon)\tau_y$ and $\varepsilon= K \tau_y^{n-1} / \eta_0^{n}$ ensure stress continuity over the yield threshold. This four-parameter model, \emph{i.e.}, $\boldsymbol{\theta}_{BVPL} = [\eta_0, K, n, \tau_y]$ and $p_{BVPL}=4$, reduces to the Hershel-Bulkley model when $\eta_0\rightarrow\infty$, as illustrated in \autoref{fig:rheo_taugd_all}, and to the biviscous model for $n=1$.

\begin{remark}[Apparent viscosity in the biviscous model]\label{rem:biviscous}

\leavevmode
In the biviscous model \cite{beverly1992numerical}, the apparent fluid viscosity after yielding is $\tau_0 / \dot{\gamma} + \eta$, which essentially mimics thinning behavior. Note that we refer to the yield stress at zero shear rate as $\tau_0$. In \citet{beverly1992numerical}, where the biviscous model is considered as a numerical regularization technique (\autoref{rem:regularization}), $\tau_y$ is used instead, which in that setting closely resembles $\tau_0$. Since we do not necessarily assume the pre-yield viscosity to be very high, we assign the parameter $\tau_y$ to the stress level at which yielding occurs.
\end{remark}

\begin{remark}[Regularization interpretation]\label{rem:regularization}

\leavevmode

The pre-yield viscosity in the biviscous models (with and without thinning) generalizes the Bingham and Herschel-Bulkley models, making them computationally tractable \cite{Papanastasiou87}. Therefore, these models can be perceived of as regularized versions of the aforementioned models. When considered as regularizations, the pre-yield viscosity should be set as high as possible while retaining computational tractability. However, we consider the pre-yield viscous behavior to be a physical trait of the models, and hence pursue calibration of the pre-yield viscosity based on experimental observations. For this reason, we do not refer to the biviscous models as regularized Bingham or Herschel-Bulkley models.
\end{remark}

\begin{remark}[Model hierarchy]

\leavevmode
The considered models form a hierarchy in the sense that the more complex models can be mathematically reduced to the less complex models, with the side note that the biviscous model and Herschel-Bulkley model are different generalizations of the Bingham model (both with three parameters). It is important to note, however, that the physical interpretation of the parameters can change between model classes, introducing ambiguity in the hierarchy from a physical point of view. For example, when comparing the biviscous model to the Newtonian model, it becomes ambiguous which of the two viscosities is to be interpreted as the Newtonian viscosity. This interpretation ambiguity is important in the context of Bayesian analyses, as it can affect the choice of the priors.
\end{remark}


\section{Constitutive model selection in rheological measurements}
\label{sec:results_rheometry}
In this section, we apply Bayesian model selection in a rheological setting. We begin by outlining the experimental procedures and specifying the material used in \autoref{sec:rheoexperiment}. Next, in \autoref{sec:rheobayesian} we present the essential components of the Bayesian framework, namely the priors, the model bias, and the experimental noise, of which the latter two comprise the likelihood function. In \autoref{sec:rheoresults} we present the results for two distinct cases: one that assumes negligible model bias, and another one in which the model bias is inferred. 

\subsection{Experimental data acquisition}\label{sec:rheoexperiment}
The yield stress material we characterize is Carbopol 980 at a concentration of 0.5 wt\%. 
The material is prepared following the procedures described in Ref.~\cite{r2019rheological}. Briefly, \SI{150}{\milli\liter} of ultrapure water (MilliQ, resistivity \SI{18.2}{\mega\ohm\centi\meter}) is placed in a plastic container under high agitation at 750\,rpm in a mechanical stirrer equipped with a propeller blade. The Carbopol powder (\SI{750}{\milli\gram}) is then slowly added to the stirring liquid. The dispersion is left to stir for 15 minutes and then allowed to rest for 30 minutes. Subsequently, while stirring at 200\,rpm with an anchor blade, \SI{1.44}{\milli\liter} of a 18\,\% w/w NaOH solution is added to the dispersion to neutralize the pH, upon which the viscosity immediately starts to increase. The mixture is then left to stir continuously for 48 hours. This protocol ensures stable Carbopol dispersions and reproducibility across different batches. Two separate batches were prepared following this procedure.

To characterize the material, we performed steady-rate sweep tests on a rotational rheometer (MCR501, Anton Paar) from high to low shear rates to obtain the flow curve over the range $\dot{\gamma} = [10^{-1}, 10^{2}]$ \SI{}{\per\second}. We ensure that the material has sufficient time to reach a steady stress at a given shear rate. For the full shear rate sweep, we measured from $\dot{\gamma} = \SI{100}{\per\second}$ until $\dot{\gamma} = \SI{0.1}{\per\second}$,
for a duration that linearly increased from 1 second at the highest shear rate to 80 seconds at the lowest shear rate. Lower shear rates were not probed due to evaporation effects, which became significant after approximately 20 minutes. The measured shear stress for each shear rate is collected in the observation vector $\boldsymbol{y}_i$, where $i$ is the index corresponding to the experiment.

To assess whether there are significant (apparent) slip effects, as reported in the literature \cite{marchesini2015rheological}, we performed experiments with different rheometer geometries. For the first batch, we conducted three measurements using \SI{50}{\milli\meter} rough (cross-hatched) plates with a gap of \SI{1}{\milli\meter}. For the second batch, we used \SI{50}{\milli\meter} smooth plates at gap heights of \SI{0.7}{\milli\meter} (one experiment), \SI{1}{\milli\meter} (four experiments) and \SI{1.2}{\milli\meter} (one experiment). 

\begin{figure}
     \centering
     \subfloat[Linear scale]{\includegraphics[width=0.5\linewidth]{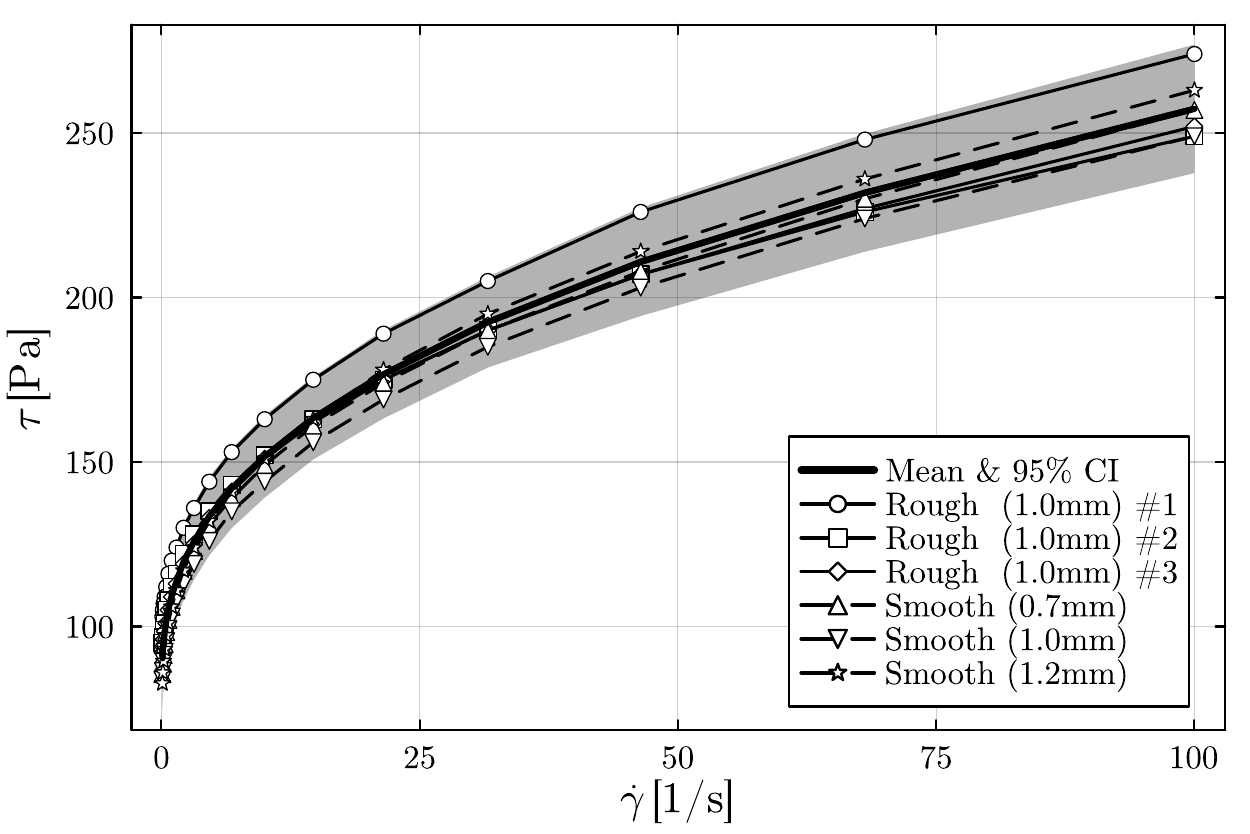}}
     \subfloat[Logarithmic scale]{\includegraphics[width=0.5\linewidth]{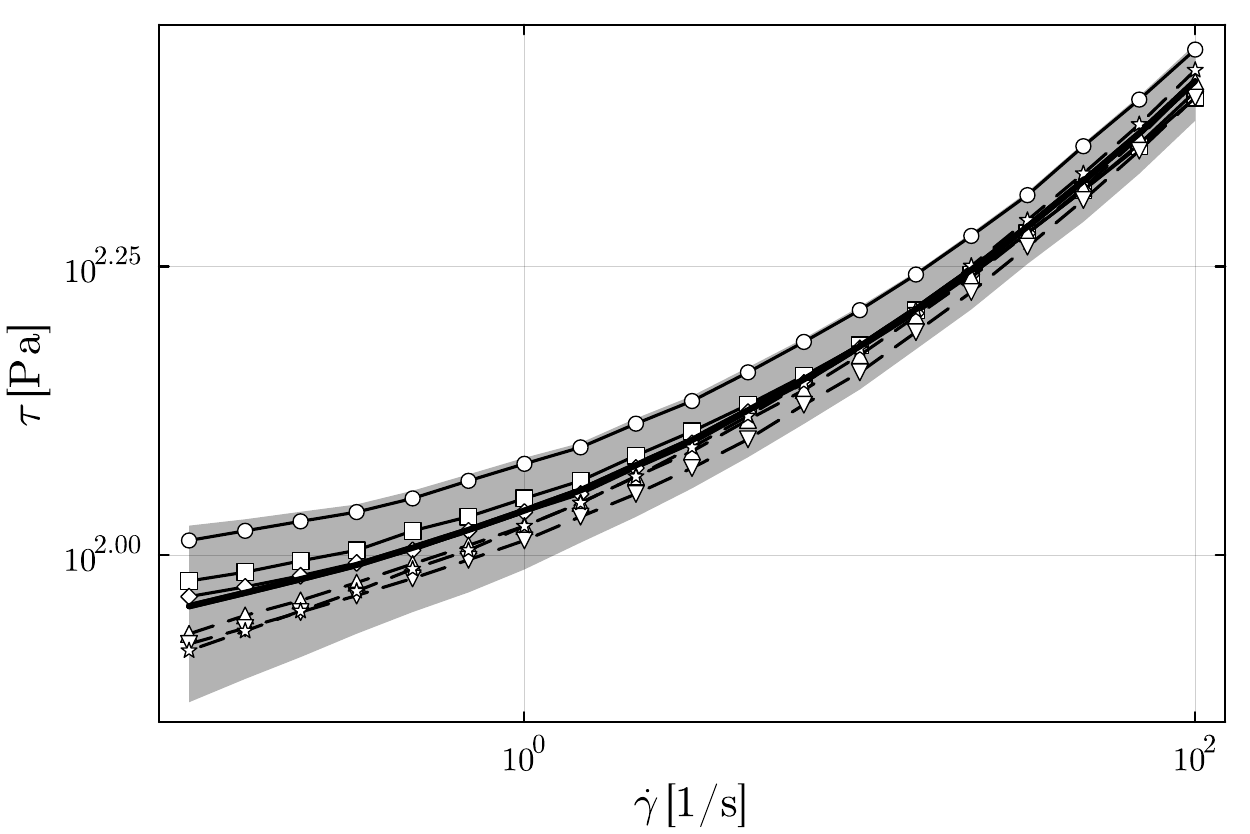}}
    \caption{Rheological data of Carbopol 980 from two batches, where batch 1 is obtained using \SI{50}{\milli\meter} rough plates and batch 2 is obtained using \SI{50}{\milli\meter} smooth plates measured at different gap heights. We combine these batches into a single data set of which the 95\% confidence interval is visualized.}
     \label{fig:rheo_data}
\end{figure} 

The measurement data for all batches are shown in \autoref{fig:rheo_data}. The 95\% confidence interval based on the mean and standard deviation of the combined data of the different measurements for the two batches, \emph{i.e.},
\begin{subequations}
\label{eq:experimentalaverages}
\begin{align}
\boldsymbol{\mu}_{\boldsymbol{y}} &= \frac{1}{n} \sum_{i=1}^n \boldsymbol{y}_i, \label{eq:mean_rheo}\\
\boldsymbol{\sigma}_{\boldsymbol{y}} &= \sqrt{\frac{1}{n-1} \sum_{i=1}^n \left( \boldsymbol{y}_i - \boldsymbol{\mu}_{\boldsymbol{y}} \right)^2},\label{eq:std_rheo}
\end{align}%
\end{subequations}
illustrates the reproducibility of the experiment, both between the batches and between the various measurements. The effects of slip, as exposed by the consideration of different plate geometries, is observed to fall within the uncertainty bounds for the considered range of shear rates. We therefore treat slip-induced variability as part of the stochastic measurement model, \emph{i.e.}, we model it via the likelihood function.

\subsection{Probabilistic modeling}\label{sec:rheobayesian}
To apply Bayesian inference on the aforementioned constitutive models with the experimental data sets gathered through the steady rate sweep tests, the prior information on the model parameters and the likelihood function must be defined.

\subsubsection{Prior information}
For the prior information on the rheological parameters of the models, \emph{i.e.}, $\left\{ \boldsymbol{\theta}_m \right\}_{m=1}^M$, we strive to enforce physically motivated positivity constraints, but otherwise keep them uninformed. We therefore represent each physical parameter by a uniform distribution with wide bounds (\autoref{tab:prior_posterior_rheo}). When a parameter appears in different models and has the same physical interpretation, the same prior is used.
    
In our analyses we consider all models to be equally probable \emph{a priori}. That is, we assign a uniform discrete distribution to the model prior.

{ 

\sisetup{
  detect-all,
  round-mode=figures,
  round-precision=3,
  scientific-notation=true,
  retain-explicit-plus = true
}

\begin{table*}
\centering
\caption{Rheological priors and posteriors for the five rheological models.\label{tab:prior_posterior_rheo}}
\resizebox{\textwidth}{!}{%
\begin{tabular}{
  l@{\hspace{3.em}}  
  l@{\hspace{1.em}}  
  S@{\hspace{1.em}}  
  S@{\hspace{1.em}}  
  l@{\hspace{1.em}}  
  S@{\hspace{1.em}}  
  S@{\hspace{1.em}}  
  l@{\hspace{1.em}}  
}
\toprule
\textbf{Parameter} & \textbf{Prior} & \multicolumn{3}{c}{\textbf{Posterior without model bias}} & \multicolumn{3}{c}{\textbf{Posterior with model bias}} \\
\cmidrule(lr){3-5} \cmidrule(lr){6-8}
 & & \textit{Mean} & \textit{Std.~dev.} & \textit{Coef.~of~var.} & \textit{Mean} & \textit{Std.~dev.} & \textit{Coef.~of~var.} \\
\midrule
\multicolumn{8}{l}{\textit{Newtonian}} \\
\midrule
$\eta$ [\si{Pa \cdot s}] & $\mathcal{U}(1, 100)$ & 4.2317 & 0.026515 & \phantom{0}0.6\% & 3.5862 & 0.22313 & \phantom{0}6.2\% \\
$\sigma_{\rm bias}$ [\si{Pa}] & $\mathcal{E}(50)$ & & & & 74.139 & 4.9914 & \phantom{0}6.7\% \\
\midrule
\multicolumn{8}{l}{\textit{Bingham}} \\
\midrule
$\eta$ [\si{Pa \cdot s}] & $\mathcal{U}(1, 100)$ & 1.8368 & 0.030522 & \phantom{0}1.7\% & 1.5119 & 0.065951 & \phantom{0}4.4\% \\
$\tau_y$ [\si{Pa}] & $\mathcal{U}(1, 200)$ & 114.29 & 0.70408 & \phantom{0}0.6\% & 127.79 & 2.51468 & \phantom{0}2.0\% \\
$\sigma_{\rm bias}$ [\si{Pa}] & $\mathcal{E}(50)$ & & & & 13.656 & 1.14522 & \phantom{0}8.4\% \\
\midrule
\multicolumn{8}{l}{\textit{Biviscous}} \\
\midrule
$\log_{10}{\eta_0}$ [\si{Pa \cdot s}] & $\mathcal{U}(2, 4)$ & 2.8510 & 0.013483 & \phantom{0}0.5\% & 3.4362 & 0.30485 & \phantom{0}8.9\%\\
$\eta_1$ [\si{Pa \cdot s}] & $\mathcal{U}(1, 100)$ & 1.78246 & 0.030791 & \phantom{0}1.7\% & 1.5112 & 0.065672  & \phantom{0}4.3\% \\
$\tau_y$ [\si{Pa}] & $\mathcal{U}(1, 200)$ & 117.23 & 0.74278 & \phantom{0}0.6\% & 127.92 & 2.5014 & \phantom{0}2.0\% \\
$\sigma_{\rm bias}$ [\si{Pa}] &  $\mathcal{E}(50)$ & & & & 13.573 & 1.159 & \phantom{0}8.5\% \\
\midrule
\multicolumn{8}{l}{\textit{Herschel-Bulkley}} \\
\midrule
$K$  [\si{Pa \cdot s^n}] & $\mathcal{U}(1, 100)$ & 30.380 & 2.6699 & \phantom{0}8.8\% & 30.551 & 3.374 & 11.0\%\\
$\tau_y$  [\si{Pa}] & $\mathcal{U}(1, 200)$ & 78.587 & 2.8284 & \phantom{0}3.6\% & 78.376 & 3.856 & \phantom{0}4.9\% \\
$n$  [-] & $\mathcal{U}(0.1, 1.0)$ & 0.38439 & 0.01796 & \phantom{0}4.7\% & 0.38391 & 0.021772 & \phantom{0}5.7\% \\
$\sigma_{\rm bias}$  [\si{Pa}] & $\mathcal{E}(50)$ &  &  &  & 3.3979 & 0.72766 & 21.4\%\\
\midrule
\multicolumn{8}{l}{\textit{Biviscous power law}} \\
\midrule
$\log_{10}{\eta_0}$ [\si{Pa \cdot s}] & $\mathcal{U}(2, 4)$ & 3.4683 & 0.3096 & \phantom{0}8.9\% & 3.47134 & 0.309307 & \phantom{0}8.9\% \\
$K$ [\si{Pa \cdot s^n}] & $\mathcal{U}(1, 100)$ & 30.427 & 2.6993 & \phantom{0}8.9\% & 30.556 & 3.4009 & 11.1\%   \\
$\tau_y$ [\si{Pa}] & $\mathcal{U}(1, 200)$ & 86.792 & 2.9381 & \phantom{0}3.4\% & 86.655 & 3.46881 & \phantom{0}4.0\% \\
$n$ [-] & $\mathcal{U}(0.1, 1.0)$ & 0.3840 & 0.01808 & \phantom{0}4.7\% & 0.38384 & 0.021921 & \phantom{0}5.7\% \\
$\sigma_{\rm bias}$ [\si{Pa}] & $\mathcal{E}(50)$ & & & & 3.38849 &0.72178 & 21.3\% \\
\bottomrule
\end{tabular}
}
\end{table*}
} 

\subsubsection{The likelihood function: experimental noise and model bias}\label{sec:rheolikelihood}
The difference between model predictions and experimental observations,  $\boldsymbol{y}_i - \boldsymbol{d}_i(\boldsymbol{\theta})$, is additively decomposed into the experimental noise $\boldsymbol{\varepsilon}_{{\rm noise},i}$, and the model bias $\boldsymbol{\varepsilon}_{{\rm bias},i}(\boldsymbol{\theta})$, where $i=1,\ldots,n$, with $n$ the number of experiments. That is, for each experiment this difference can be interpreted as a representation of the ground truth, perturbed by experimental noise and modeling bias. In the Bayesian framework, both these terms are modeled probabilistically. 

We assume the conducted experiments $\{ \boldsymbol{y}_i \}_{i=1}^n$ to be independent and identically normally distributed, \emph{i.e.}, $\boldsymbol{y}_i \sim \mathcal{N}(\boldsymbol{\mu}_{\boldsymbol{y}}, \boldsymbol{\Sigma}_{\boldsymbol{y}})$. The diagonal covariance matrix $\boldsymbol{\Sigma}_{\boldsymbol{y}} = {\rm diag}(\boldsymbol{\sigma}_{\boldsymbol{y}})$ corresponds to the standard deviation of the experiments \eqref{eq:std_rheo} and reflects our assumption that the measurement points are uncorrelated between shear rates. The experimental noise for each of the considered experiments can then be postulated as
\begin{equation}
    \boldsymbol{\varepsilon}_{{\rm noise},i} \sim \mathcal{N}\left( \boldsymbol{0}, \boldsymbol{\Sigma}_{\boldsymbol{y}} \right),
    \label{eq:noisemodel}
\end{equation}
which conceptually presumes that the experiment has been conducted a sufficient number of times to make the mean \eqref{eq:std_rheo} representative of the ground truth.

For the model bias it is reasonable to expect correlation between shear rates. We therefore express the model bias over the domain $[0,\dot{\gamma}_{\rm max}]$ by means of the Gaussian process
\begin{equation}
    \varepsilon_{{\rm bias},i} (\dot{\gamma}) \sim \mathcal{GP}( 0, \sigma_{\rm bias}^2 \rho_{\rm bias}(\dot{\gamma}, \dot{\gamma}') ),\label{eq:gaussianprocess}
\end{equation}
where $\sigma_{\rm bias}$ represents the strength of the bias through the point-wise standard deviation of the field. The correlation between model errors for different shear rates is modeled through the autocorrelation kernel 
\begin{equation}
    \rho_{\rm bias}(\dot{\gamma}, \dot{\gamma}') = \exp{\left( -\frac{\left| \dot{\gamma} - \dot{\gamma}' \right|}{l_{\rm bias}} \right)},
    \label{eq:correlationfunction}
\end{equation}
where $l_{\rm bias}$ is the correlation length. When evaluated in the measurement points, the Gaussian process \eqref{eq:gaussianprocess} can be expressed by the multi-variate normal distribution
\begin{equation}
    \boldsymbol{\varepsilon}_{{\rm bias},i} \sim \mathcal{N}\left( \boldsymbol{0}, \sigma_{\rm bias}^2 \boldsymbol{\Lambda}(l_{\rm bias}) \right),
    \label{eq:biasmodel}
\end{equation}
where $\boldsymbol{\Lambda}(l_{\rm bias})$ is the correlation matrix \emph{cf.}~\cref{eq:correlationfunction}.

To acknowledge the uncertainty surrounding the model bias, we consider the bias strength $\sigma_{\rm bias}$ as a hyperparameter to be inferred. The exponential prior distribution for this hyperparameter is included in \autoref{tab:prior_posterior_rheo}. In principle, also the correlation length could be considered as a hyperparameter, but we have found this to over-complicate our analyses. Therefore, in the remainder we assume the bias correlation length fixed at \SI{1}{\per\second}.
 
Since both the experimental noise \eqref{eq:noisemodel} and the model bias \eqref{eq:biasmodel} are Gaussian, the combined error is also Gaussian, \emph{i.e.},
\begin{equation}
    {\boldsymbol{y}_i} - \boldsymbol{d}_i(\boldsymbol{\theta}) \sim \mathcal{N}\left(\boldsymbol{0}, \boldsymbol{\Sigma}_{\boldsymbol{y}} + \sigma_{\rm bias}^2 \boldsymbol{\Lambda}(l_{\rm bias}) \right).
    \label{eq:combinederrordist}
\end{equation}
The likelihood function $\mathcal{L}\left(\boldsymbol{\theta} \mid \boldsymbol{y}, M \right)$ can then be defined as the product of the probability densities of the combined errors over all experiments \cite{oden2017predictive} as
\begin{equation}
\mathcal{L}\left(\boldsymbol{\theta} \mid \boldsymbol{y}, M \right)\propto  \prod \limits_{i=1}^n \exp \left( -\frac{1}{2} \left( \boldsymbol{d} - {\boldsymbol{y}_i} \right)^\mathrm{T} \left[ \boldsymbol{\Sigma} \right]^{-1} \left( \boldsymbol{d} - {\boldsymbol{y}_i} \right) \right),
\label{eq:likelihoodformcombined}
\end{equation}
where $\boldsymbol{\Sigma} = \boldsymbol{\Sigma}_{\boldsymbol{y}} + \sigma_{\rm bias}^2 \boldsymbol{\Lambda}(l_{\rm bias})$ is the covariance matrix of the combined errors.

\subsection{Results}\label{sec:rheoresults}
To demonstrate the effect of including the model bias in the Bayesian framework, we study the model selection with and without the inclusion of the bias. The model bias is omitted from the analysis by setting $\sigma_{\rm bias}=0$. All results are based on MCMC simulations with two walkers per parameter and \num{10000} steps per walker. For all chains the Gelman-Rubin diagnostic converged to a tolerance significantly below 1.1. 

\subsubsection{Omitted model bias}\label{sec:omittedbias}
\begin{figure*}
    \centering
    \subfloat[Newtonian]{\includegraphics[width=0.45\linewidth]{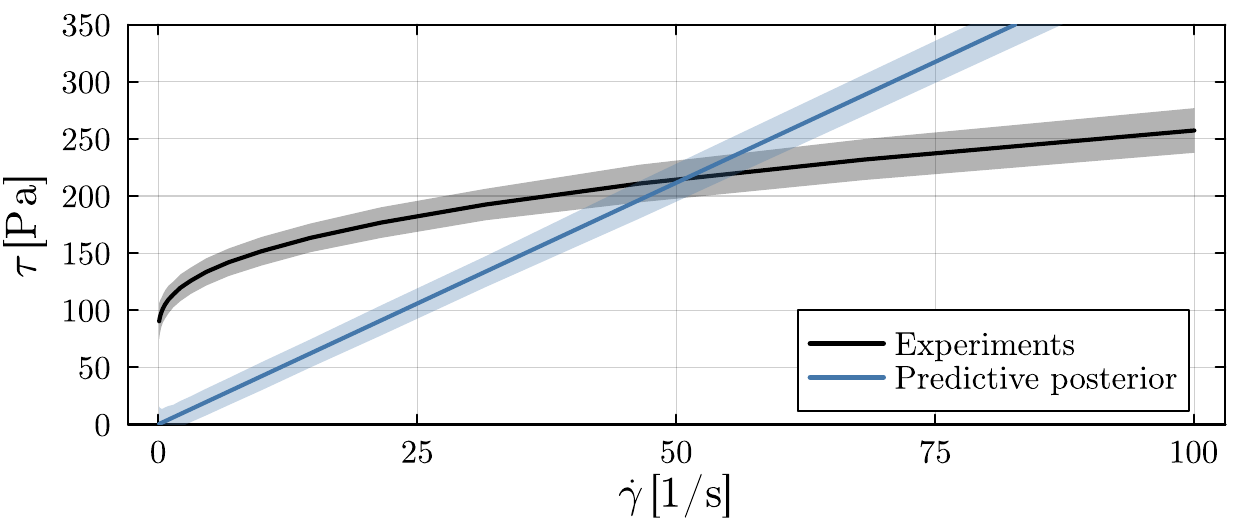}
    \includegraphics[width=0.45\linewidth]{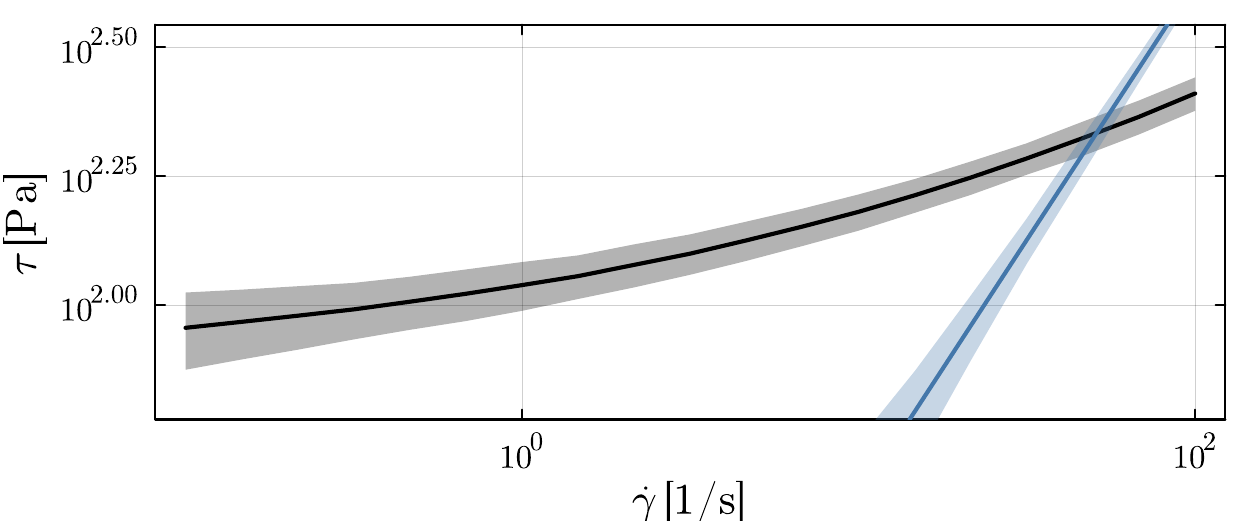}
    \label{fig:ppd_N_nobias}
    }\\[1em]
    \subfloat[Bingham]{\includegraphics[width=0.45\linewidth]{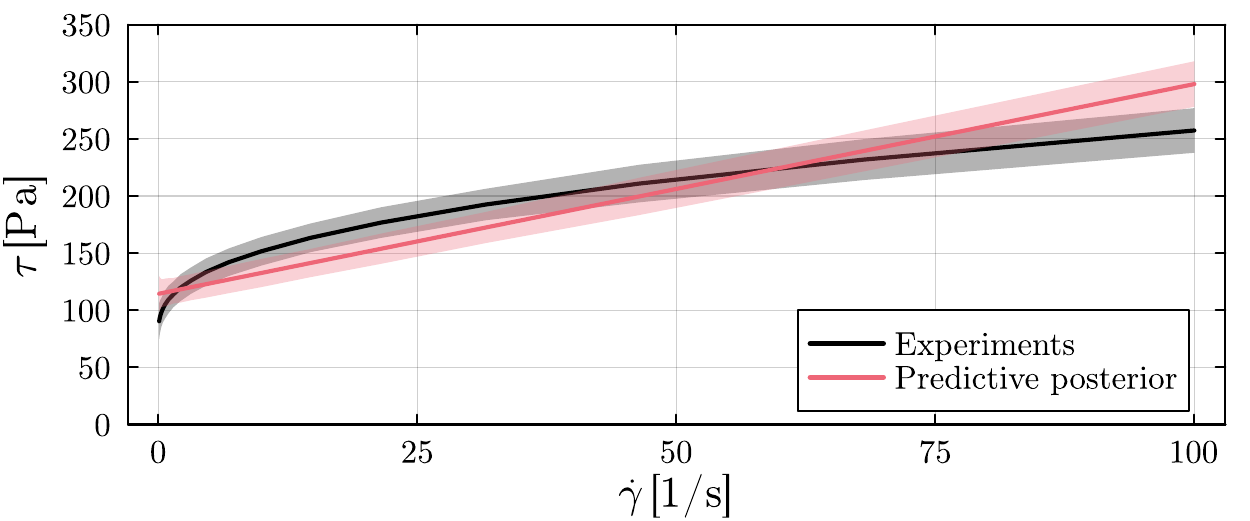}
    \includegraphics[width=0.45\linewidth]{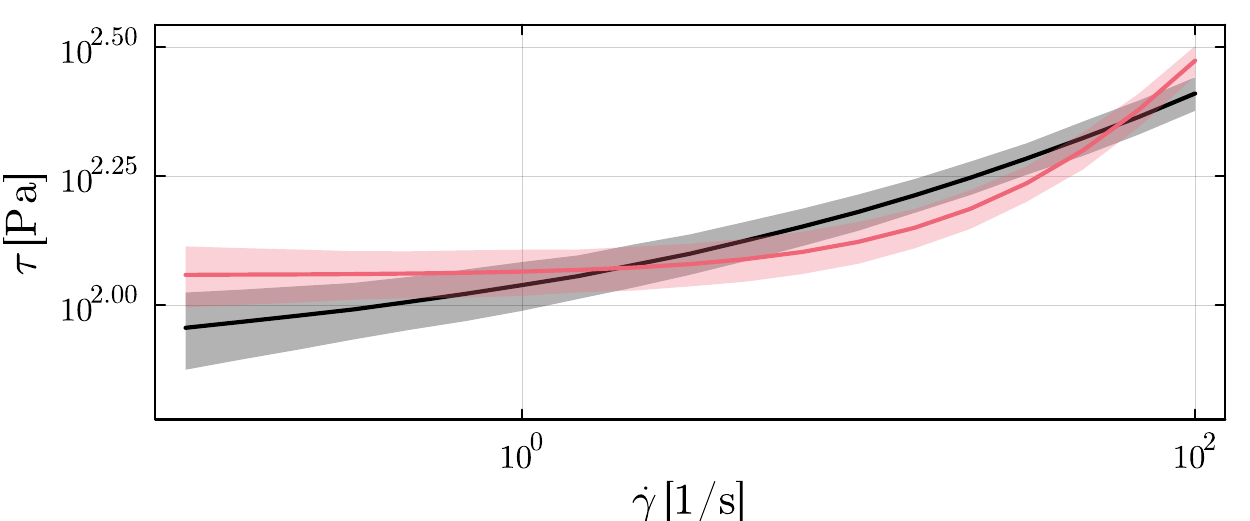}}\\[1em]
    \subfloat[Biviscous]{\includegraphics[width=0.45\linewidth]{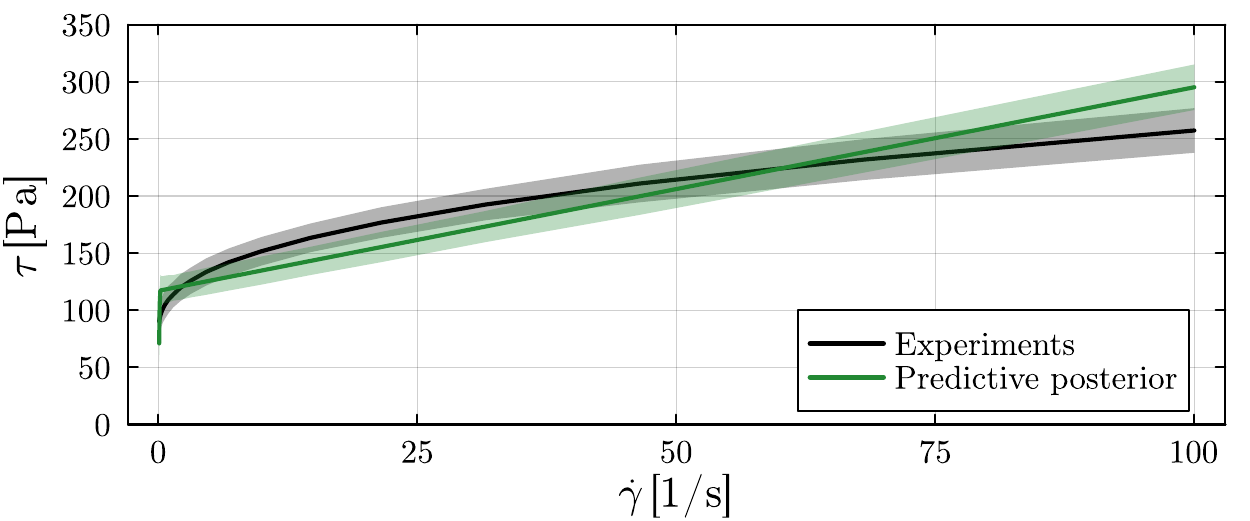}
    \includegraphics[width=0.45\linewidth]{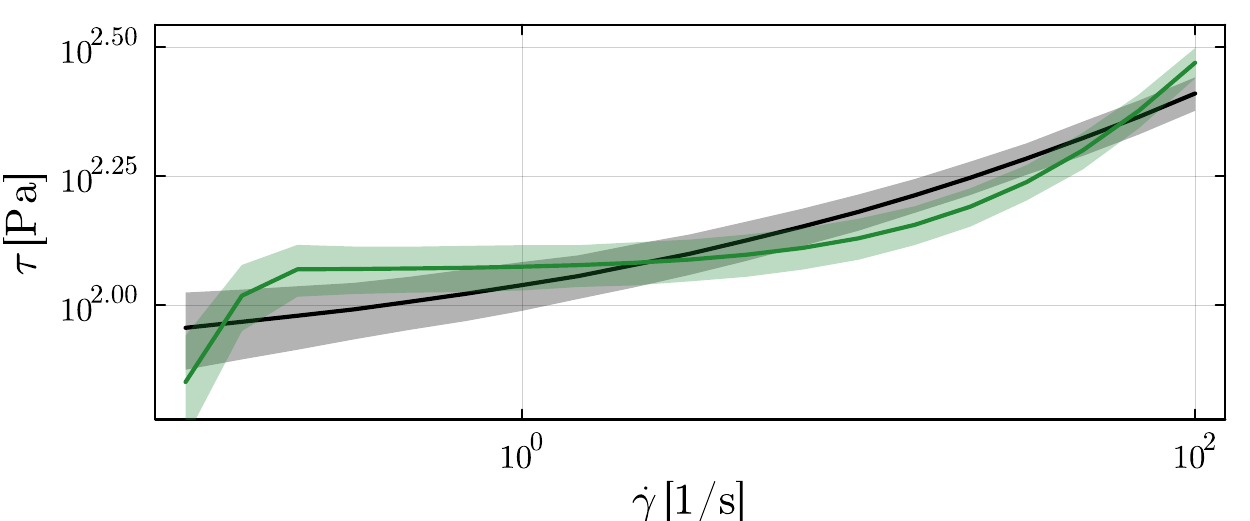}}\\[1em]    
    \subfloat[Herschel-Bulkley]{\includegraphics[width=0.45\linewidth]{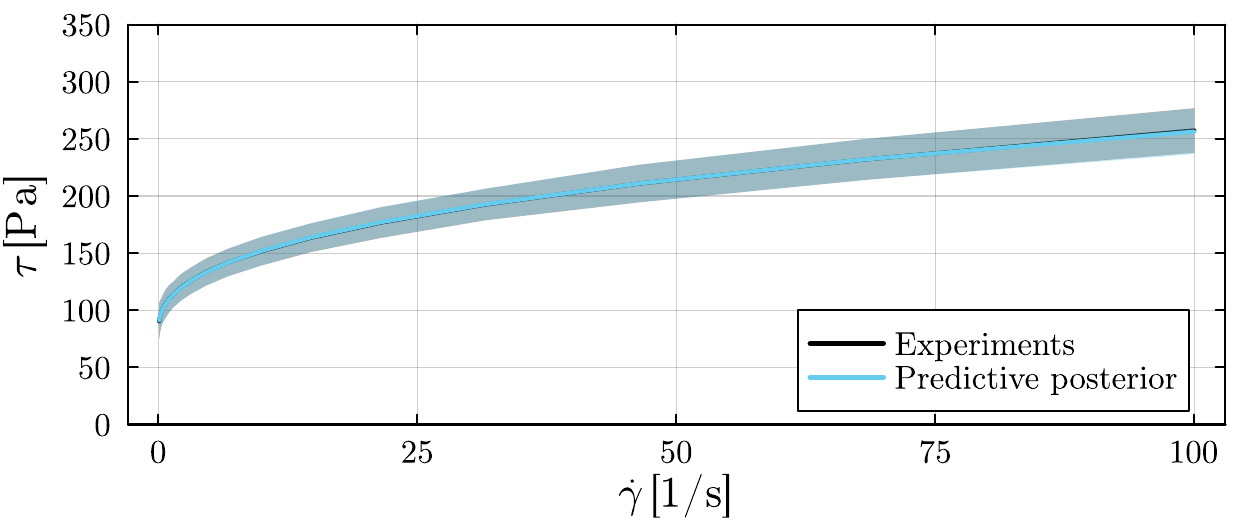}
    \includegraphics[width=0.45\linewidth]{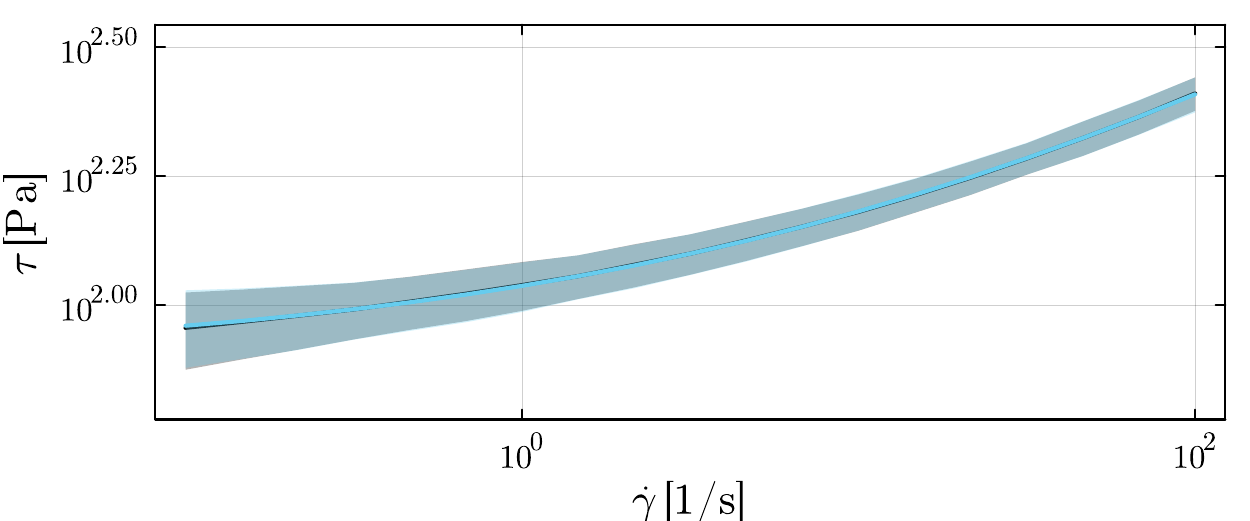}}\\[1em]
    \subfloat[Biviscous power law]{\includegraphics[width=0.45\linewidth]{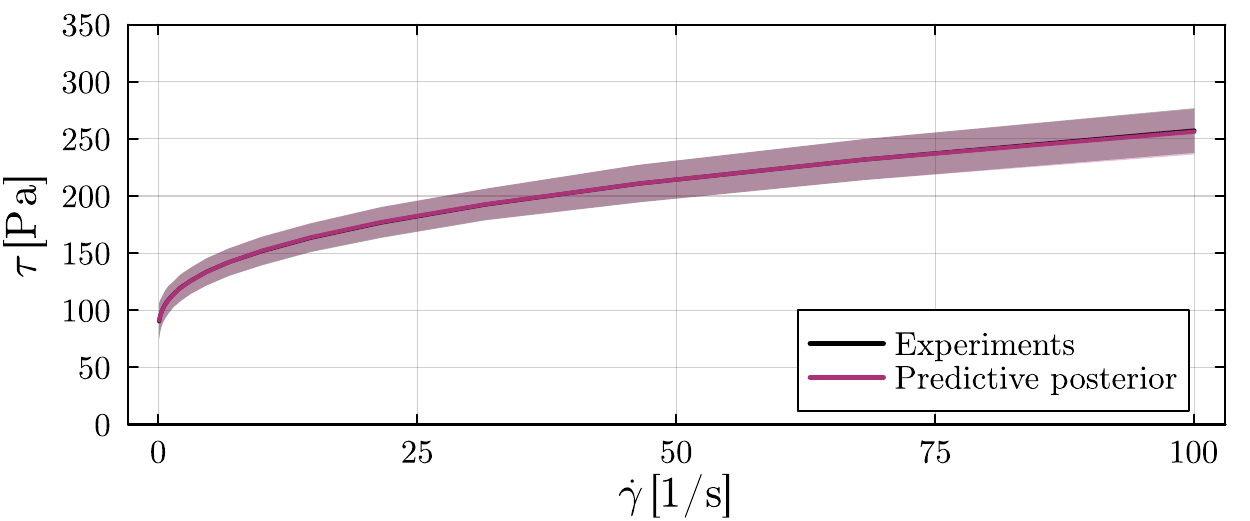}
    \includegraphics[width=0.45\linewidth]{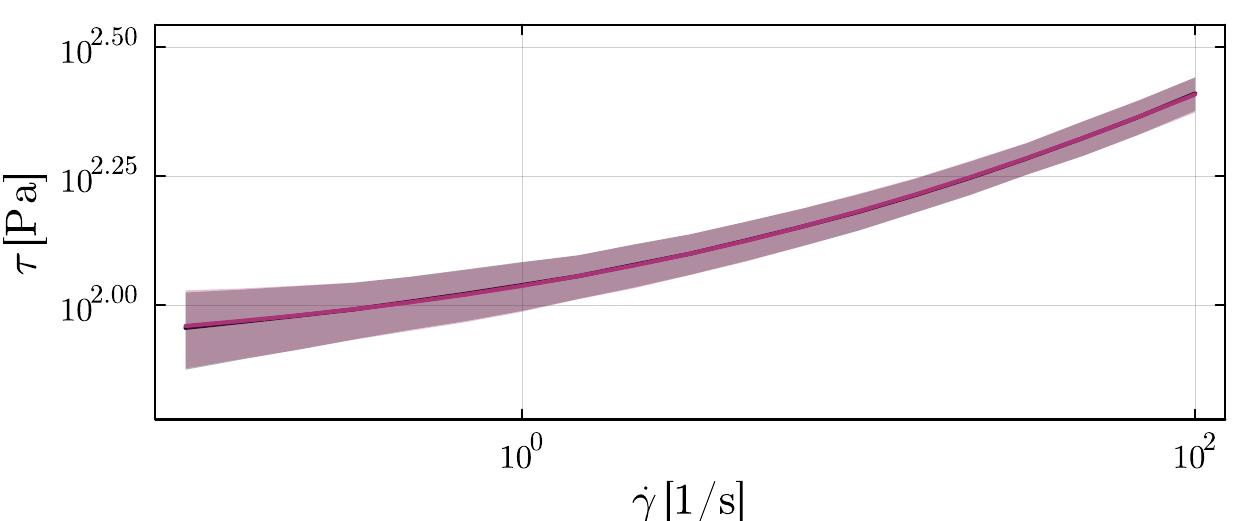}
    \label{fig:ppd_BVPL_nobias}}
    \caption{Predictive posterior distribution per model when omitting model bias. The left column shows all models on a linear scale, whereas the right columns shows the same results on a logarithmic scale.}
    \label{fig:ppdrheo_combined}
\end{figure*}

In \autoref{fig:ppdrheo_combined} we show the posterior predictive distribution of $\tau(\dot{\gamma})$ for all rheological models. These predictive posteriors are obtained from simulated experiments, \emph{i.e.}, with the noise being added. The moments of the (calibrated) posterior parameter distributions are reported in \autoref{tab:prior_posterior_rheo}. The observed substantial reduction in standard deviation of all parameters compared to the prior distribution conveys that the priors were, as intended, uninformative. The uncertainty bands in \autoref{fig:ppdrheo_combined} represent the 95\% credible interval of the predictive posteriors. The 95\% confidence interval of the experimental data is shown for comparison. The Newtonian model exhibits the largest discrepancy with the data, as it neither captures the experimentally observed yield stress, nor the shear thinning. The biviscous model and Bingham model capture the yield stress behavior, but fail to describe the shear thinning effects. The Herschel–Bulkley and biviscous power law model both agree well with the experimental data, as both models incorporate the essential physical phenomena.

For the Newtonian model it is observed that the uncertainty bands permit negative shear stresses at positive shear rates. This non-physical behavior is caused by the additive and Gaussian nature of the probabilistic descriptions of the experimental noise and model bias. In cases that such non-physical attributes are deemed problematic for the modeling purpose, more advanced probabilistic models should be developed.

From \autoref{fig:ppdrheo_combined} it is also observed that the standard deviation of the predictive posterior distribution closely resembles that of the experimental noise for all models, \emph{i.e.}, the bandwidth of the predictive posterior is similar to that of the experimental data. This is a direct consequence of the broad priors and neglected model bias. The broad priors ensure that the posterior uncertainty is dictated by the likelihood function, and since the likelihood now solely contains the experimental noise, the predictive posterior is governed by this experimental noise. Consequently, the predictive posterior uncertainty is essentially independent of the suitability of the selected model. This leads to misleading results, in the sense that for models with a large bias -- in this case the Newtonian model, the Bingham model, and the biviscous model -- the bandwidth is not representative of the discrepancy with the data. The bandwidth suggests a relatively high degree of certainty in comparison to the mismatch with the data. For the models with a small bias -- in this case the Herschel-Bulkley model and the biviscous power law model -- the bandwidth of the predictive posterior matches excellently with the experimental data, indicating that the bias of these models is indeed negligible. 

\begin{figure}
    \centering
    \includegraphics[width=0.5\linewidth]{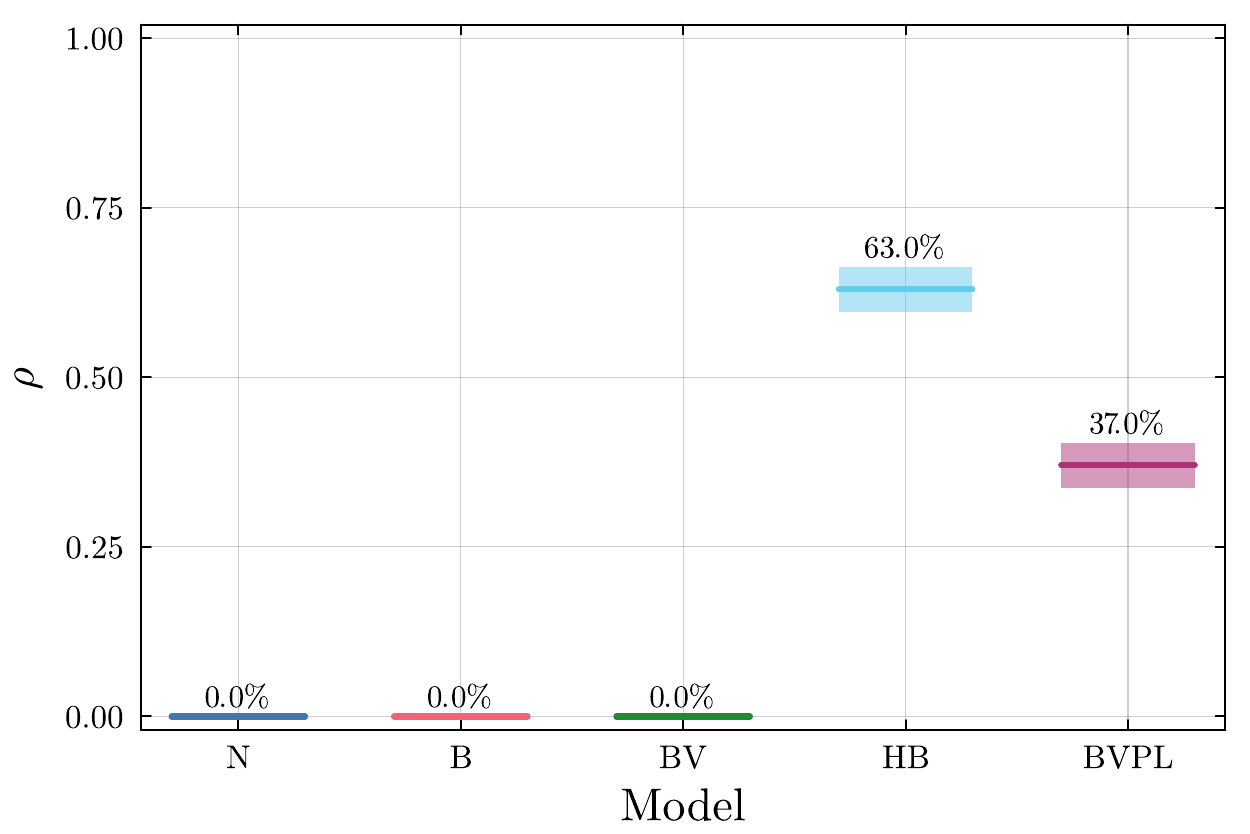}
    \caption{Model plausibilities when omitting model bias for the Newtonian (N), Bingham (B), biviscous (BV), Herschel-Bulkley (HB) and biviscous power law (BVPL) models. For the first three models, the plausibility is zero up to machine precision.}
    \label{fig:rheo_modsel_nobias}
\end{figure}

Although the predictive posterior distributions separate the unsuitable from the suitable models in a qualitative sense, they do not reveal differences in plausibilities emanating from unnecessary model complexity. To assess this aspect of model selection, the plausibilities computed using our Bayesian model selection framework are shown in \autoref{fig:rheo_modsel_nobias}. The uncertainty bands correspond to the 95\% confidence interval of the estimator discussed in \autoref{sec:MCMC}. As expected from the analysis of the predictive posteriors, the plausibility of the Newtonian model, the Bingham model and the biviscous model is negligible, on account of these models being unable of mimicking the yield stress and/or shear thinning behavior. Contrary to the predictive posterior distributions, the plausibilities do reveal that the Herschel-Bulkley model is more plausible than the biviscous power-law model. This is a consequence of the Herschel-Bulkley model being less complex than the biviscous power law model, while being equally capable of explaining the observations. In line with Occam's razor, the simpler model is then preferred.

The additional complexity introduced in the biviscous power law model emanates from the pre-yield viscosity, $\eta_0$. We have used a uniform prior for $\log_{10}{\eta_0}$ with a lower bound of 2 and an upper bound of 4, \emph{i.e.}, the pre-yield viscosity falls between \SI{100}{\pascal\second} and \SI{10000}{\pascal\second}. This encompasses the highest observable pre-yield viscosity in our experimental data, \emph{i.e.}, the viscosity corresponding to a stress of approximately \SI{90}{\pascal} at a shear rate of \SI{0.1}{\per\second}. For pre-yield viscosities larger than \SI{900}{\pascal\second}, all shear stress observations fall in the post-yield regime. The sensitivity of the likelihood on the parameter is then low, with some residual sensitivity resulting from the post-yield curve being influenced by the pre-yield viscosity. For the considered prior, a significant part of the probability mass falls in the region below \SI{900}{\pascal\second}, where the biviscous power law is significantly less likely than the Herschel-Bulkley model. As a consequence, overall, the Herschel-Bulkley model is more plausible.
    
\begin{figure}
    \centering
    \includegraphics[width=0.5\linewidth]{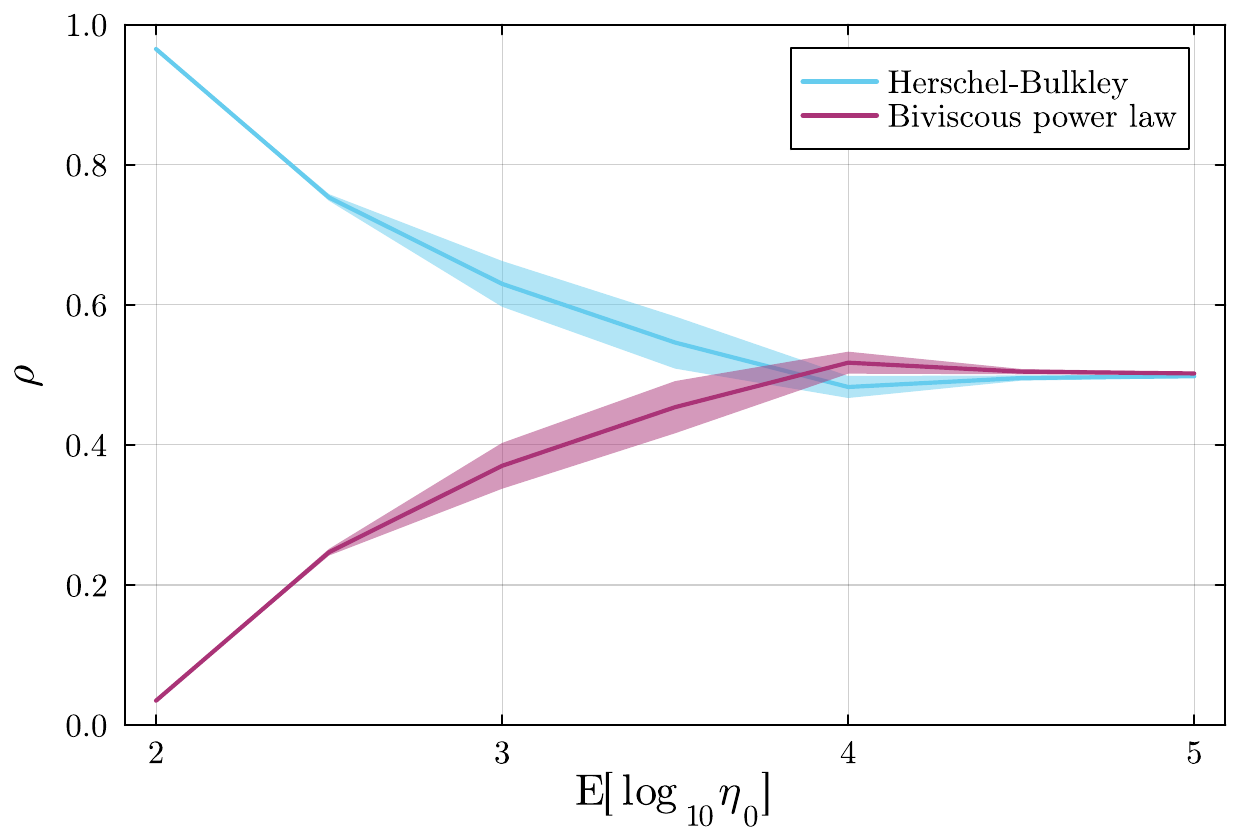}
    \caption{Dependence of the plausibility of the Herschel-Bulkley model and the biviscous power law model on the prior of the pre-yield viscosity $\eta_0$. The prior is defined as $\log_{10}{\eta_0} \sim \mathcal{U}(\ell, \ell+2)$ with mean ${\rm E}[\log_{10}{\eta_0}]=\ell + 1$.}
    \label{fig:rheo_model_study}
\end{figure}

\autoref{fig:rheo_model_study} studies the effect of the prior choice for the pre-yield viscosity in more detail. We keep the difference between the upper and lower bound of the uniform prior for $\log_{10}{\eta_0}$ fixed to 2, but vary its mean. The plausibilities reported in \autoref{fig:rheo_modsel_nobias} correspond to a mean value of ${\rm E}[\log_{10}{\eta_0}]=3$, \emph{i.e.}, $\log_{10}{\eta_0} \sim \mathcal{U}(2,4)$. When the mean value is lowered, the plausibility of the biviscous power law model decreases substantially. The reason for this is that the prior for the pre-yield viscosity then becomes biased, \emph{i.e.}, it assigns a high probability to values that are not in agreement with the data. In this case, the Herschel-Bulkley model is clearly preferred, as it is not affected by this misinformation. When the mean value for $\log_{10}{\eta_0}$ is increased compared to the reference case of 3, the plausibility of the biviscous power law model gradually increases on account of a decrease of the prior bias. When the mean value increases toward 5, \emph{i.e.}, when the pre-yield viscosity falls between \SI{10000}{\pascal\second} and \SI{1000000}{\pascal\second}, the plausibility of the two models is practically equal as a consequence of the biviscous power law model being insensitive to the pre-yield viscosity for all considered data points.

From \autoref{fig:rheo_model_study} it is also observed that when the mean of $\log_{10}{\eta_0}$ is around 4, \emph{i.e.}, for the prior $\log_{10}{\eta_0} \sim \mathcal{U}(3,5)$, the biviscous power law model is favored over the Herschel-Bulkley model, albeit marginal. The explanation for this is that in this case the lower bound of the prior is close to the highest observable pre-yield viscosity of \SI{900}{\pascal\second}. In this regime the biviscous power law is able to match the data slightly better than the Bingham model, while not being penalized by low pre-yield viscosities that do not match the data. An important point to make in this regard is that the plausibility of a model does not only depend on the phenomena it can describe, but, via the parameter calibration, also on the choice of the prior.

\subsubsection{Inferred model bias}\label{sec:includedmodelbias}
\begin{figure*}
    \centering
    \subfloat[Newtonian]{\includegraphics[width=0.45\linewidth]{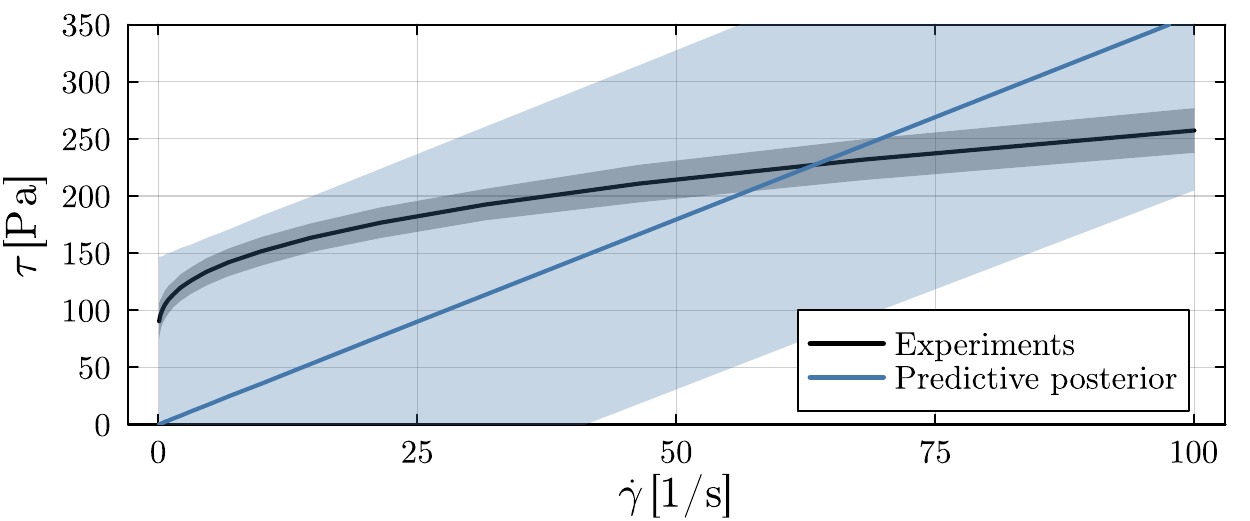}
    \includegraphics[width=0.45\linewidth]{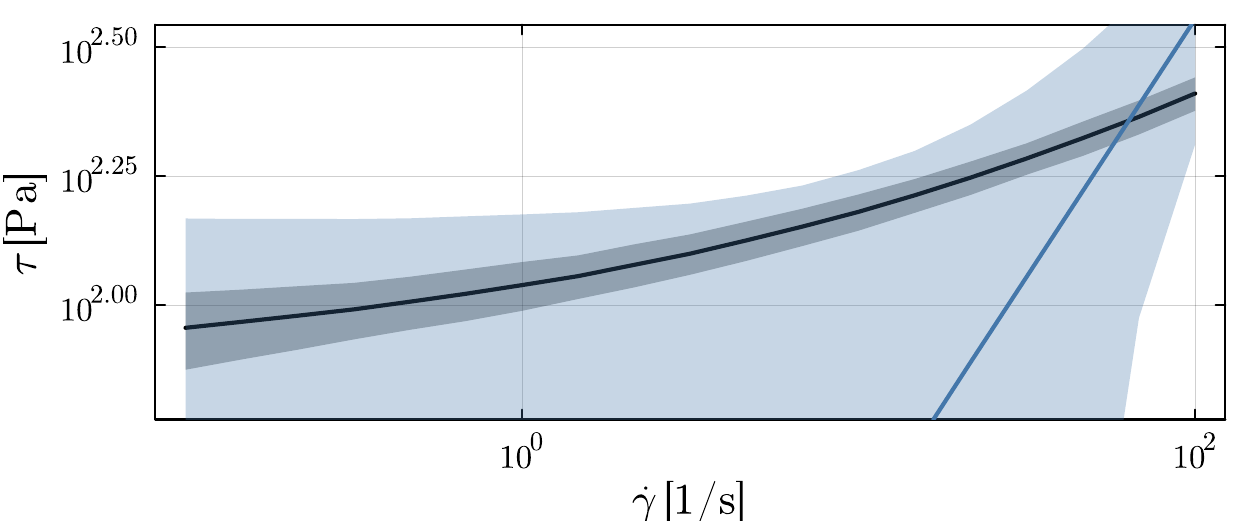}
    \label{fig:ppd_N_bias}}\\[1em]
    \subfloat[Bingham]{\includegraphics[width=0.45\linewidth]{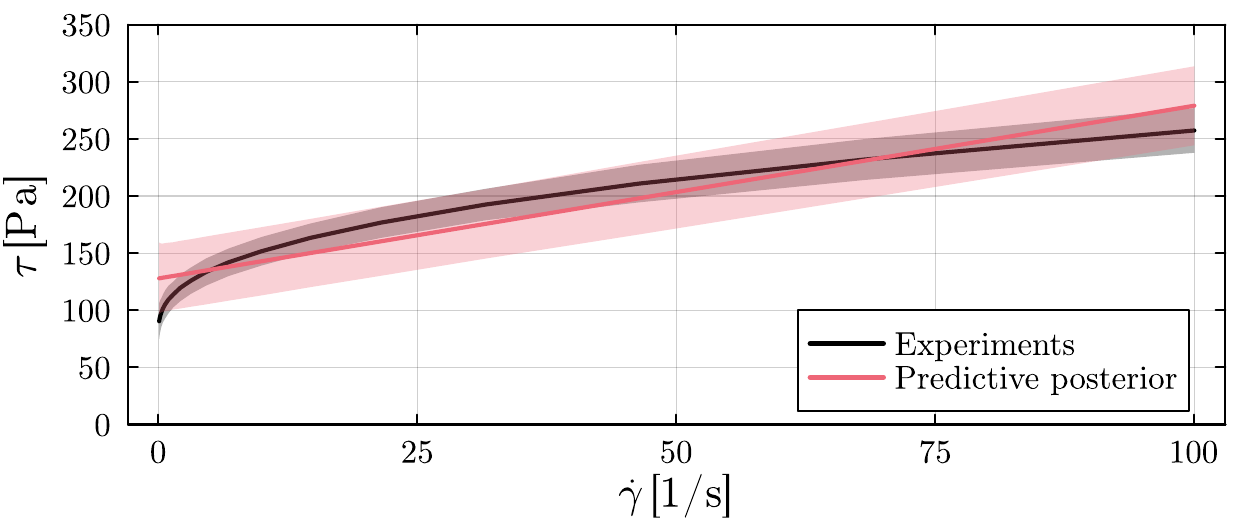}
    \includegraphics[width=0.45\linewidth]{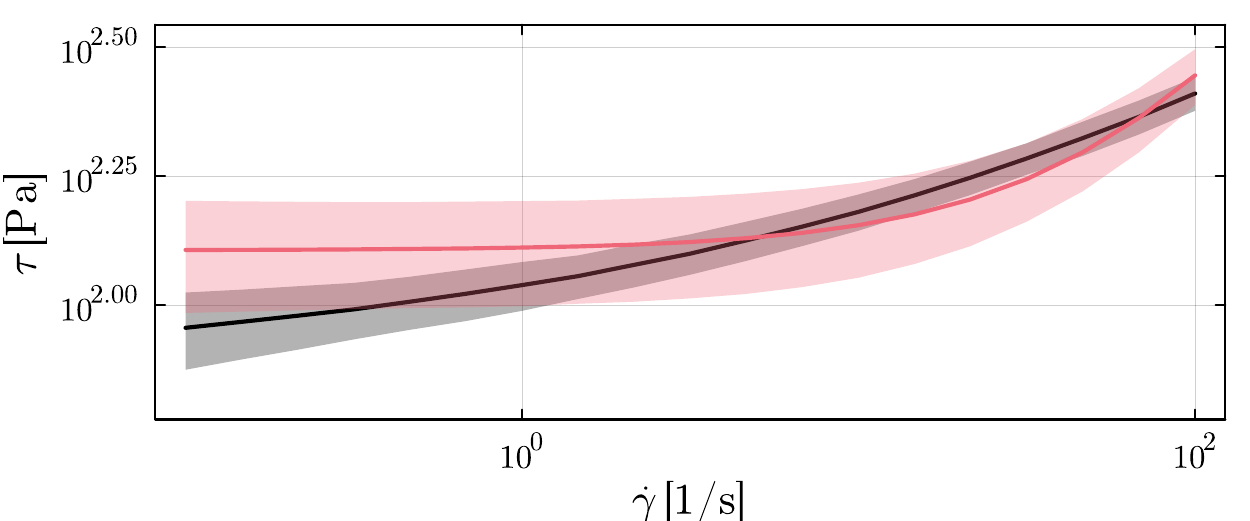}}\\[1em]
    \subfloat[Biviscous]{\includegraphics[width=0.45\linewidth]{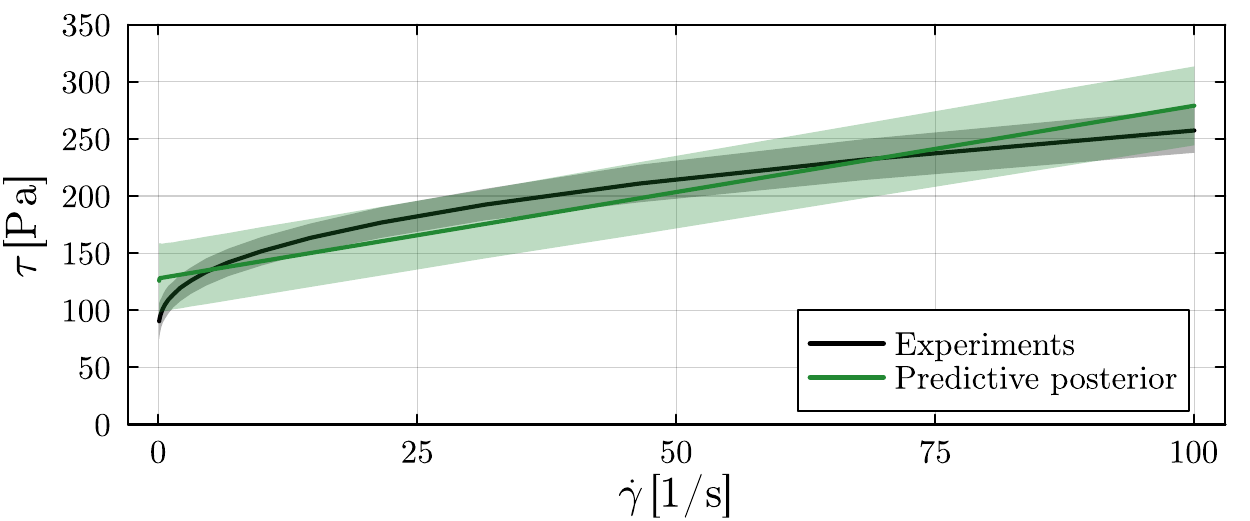}
    \includegraphics[width=0.45\linewidth]{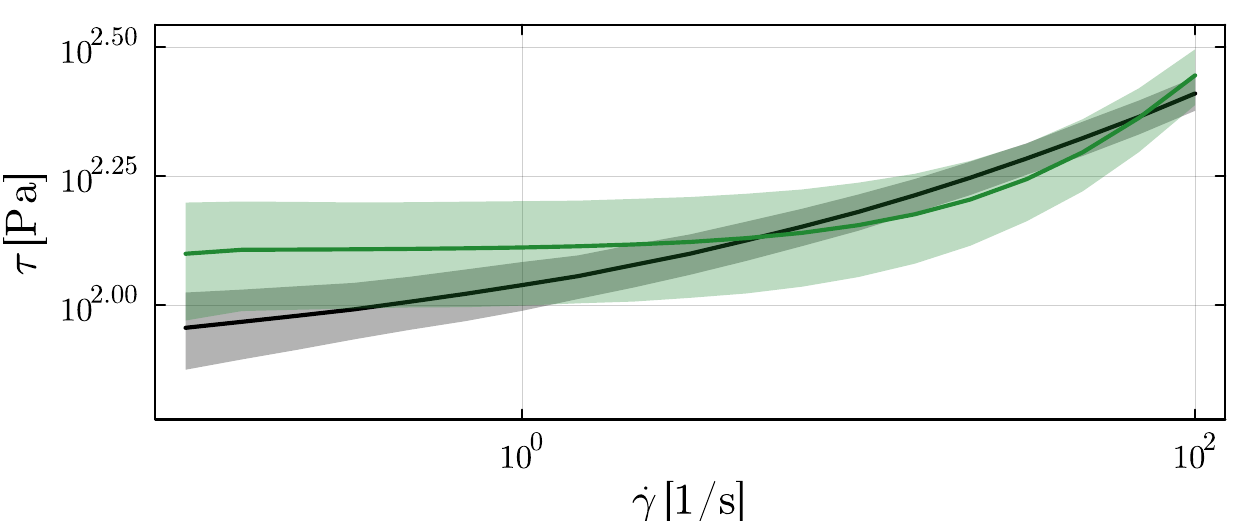}}\\[1em]    
    \subfloat[Herschel-Bulkley]{\includegraphics[width=0.45\linewidth]{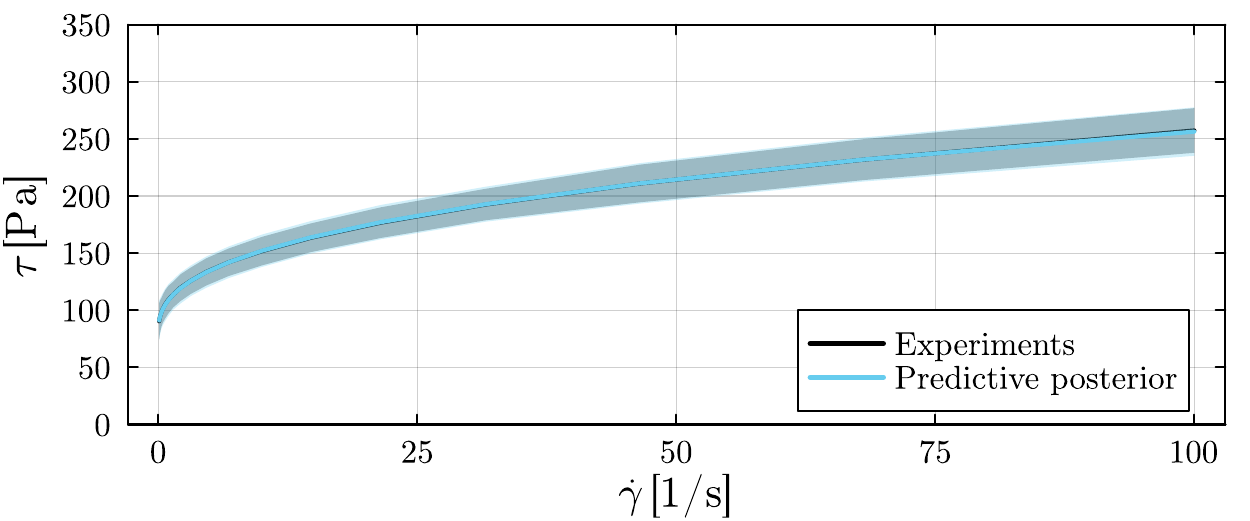}
    \includegraphics[width=0.45\linewidth]{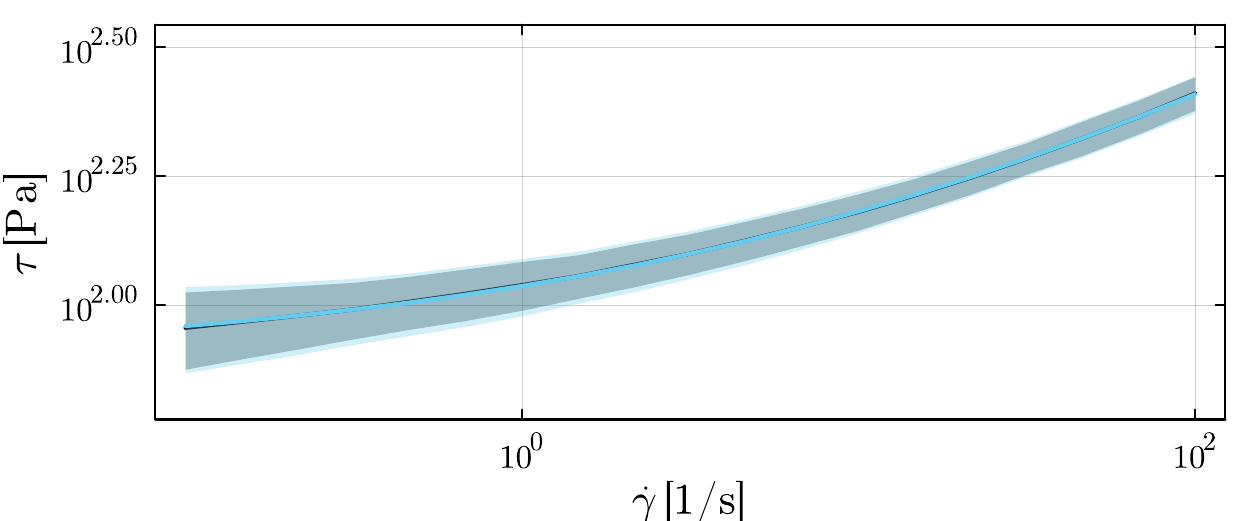}}\\[1em]
    \subfloat[Biviscous power law]{\includegraphics[width=0.45\linewidth]{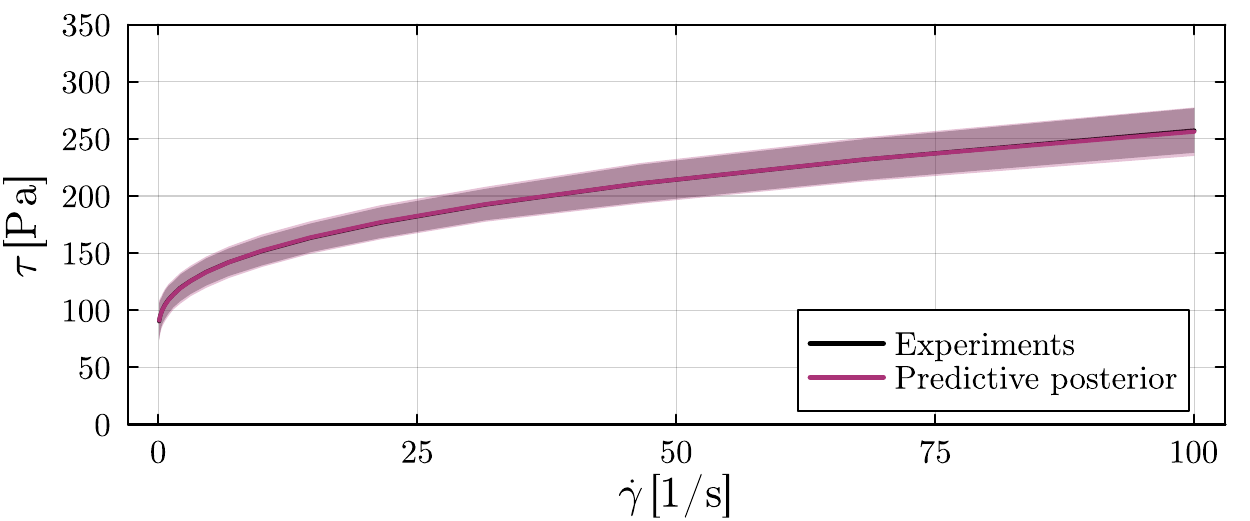}
    \includegraphics[width=0.45\linewidth]{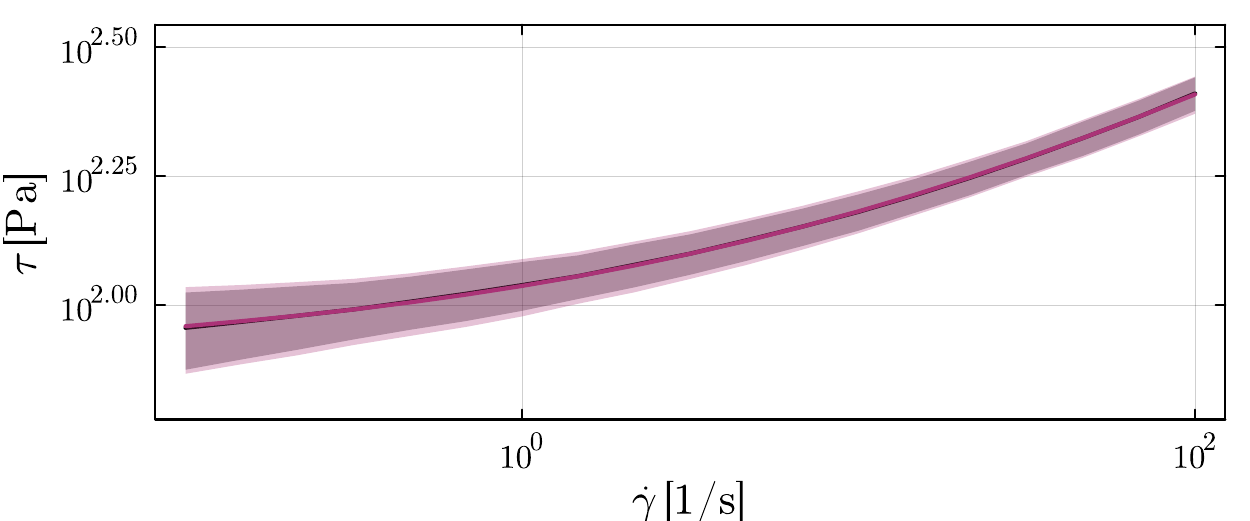}\label{fig:ppd_BVPL_bias}}
    \caption{Predictive posterior distribution per model when inferring the model bias. The left column shows all models on a linear scale, whereas the right columns shows the same results on a logarithmic scale.}
    \label{fig:ppdrheo_noise_combined}
\end{figure*}

\begin{figure*}
    \centering
    \subfloat[Newtonian]{\includegraphics[width=0.33\linewidth]{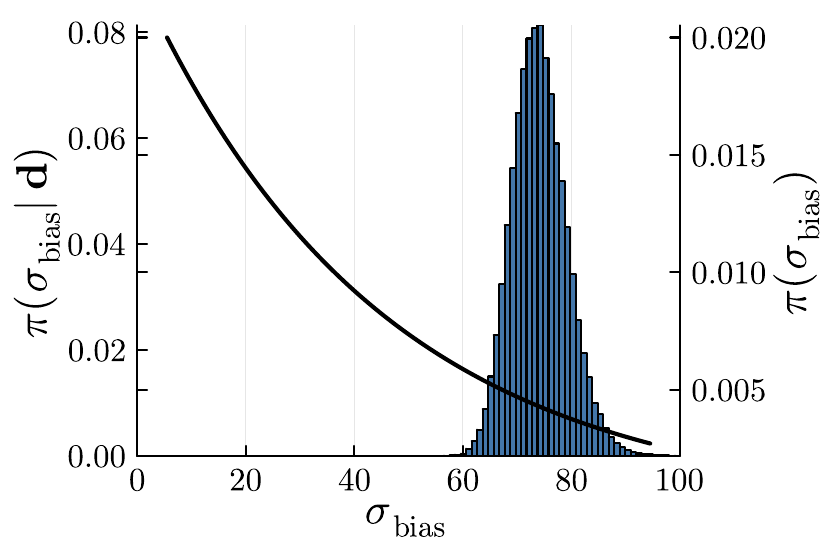}}
    \subfloat[Bingham]{\includegraphics[width=0.33\linewidth]{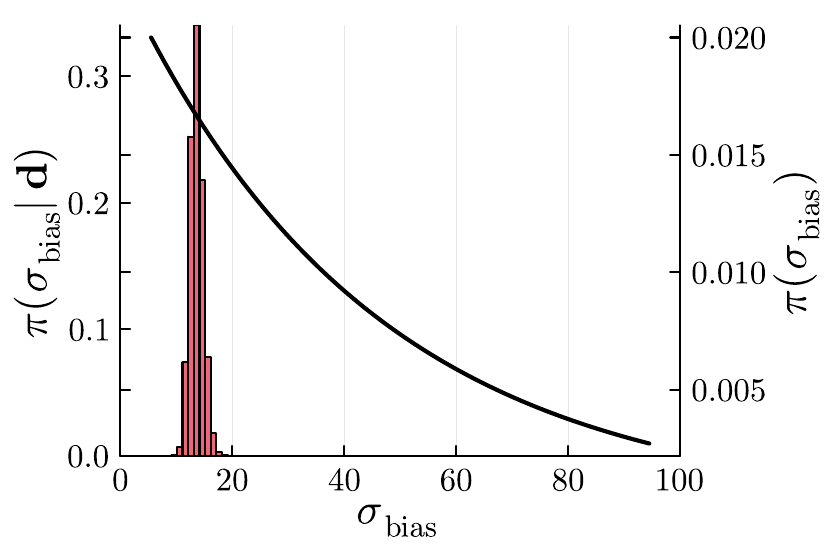}}
    \subfloat[Biviscous]{\includegraphics[width=0.33\linewidth]{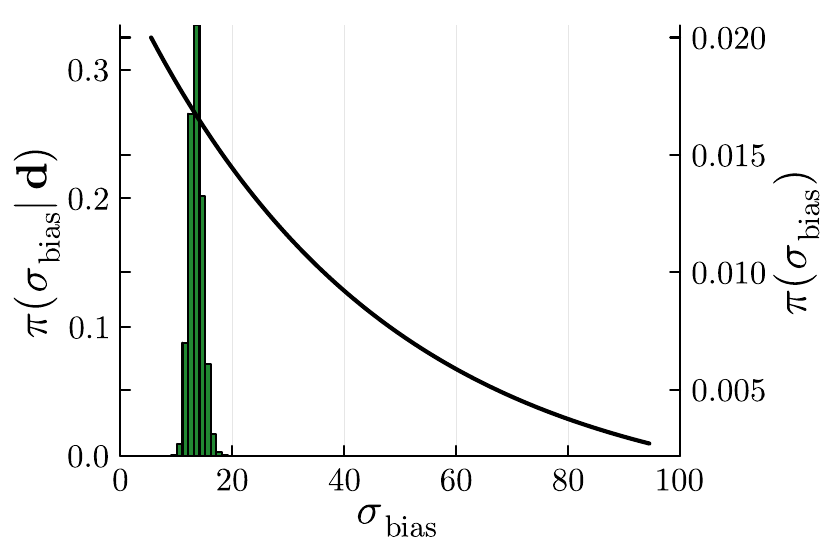}}
    
    \subfloat[Herschel-Bulkley]{\includegraphics[width=0.33\linewidth]{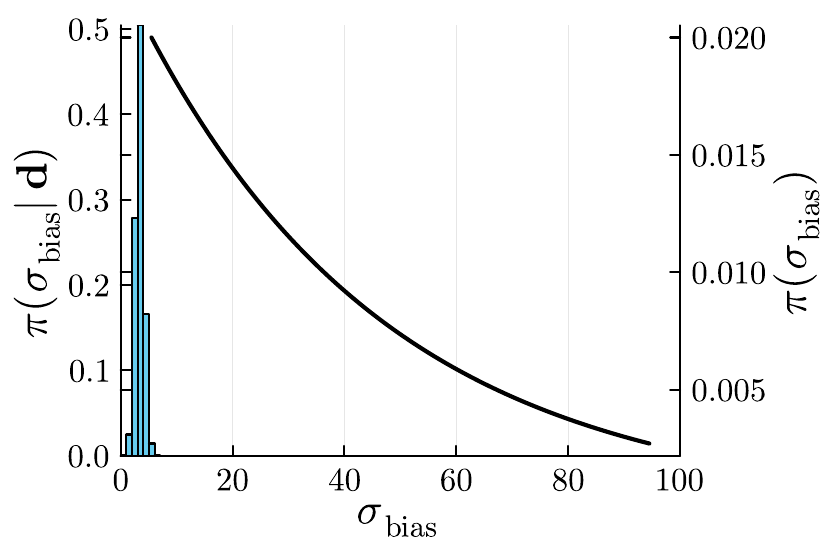}}
    \subfloat[Biviscous power law]{\includegraphics[width=0.33\linewidth]{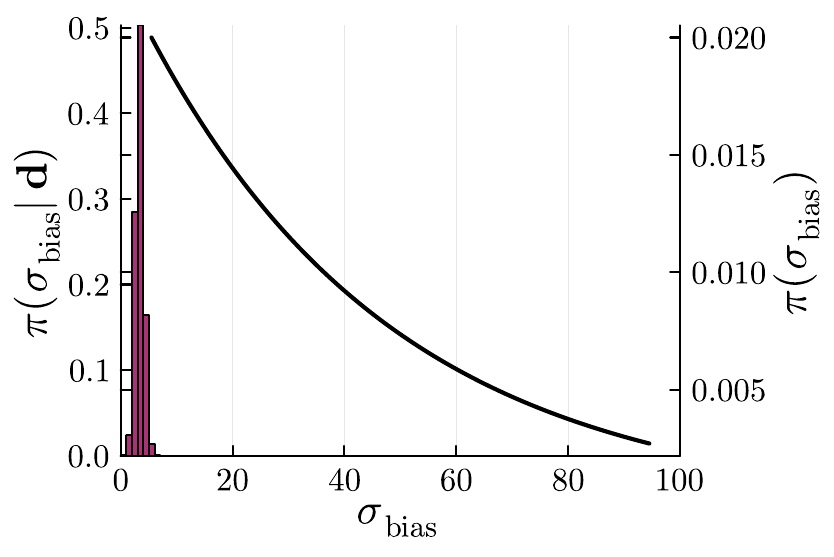}}
    \caption{Posterior (histogram) versus prior (black line) of the inferred model bias for all rheological models.}
    \label{fig:modelbias}
\end{figure*}

\autoref{fig:ppdrheo_noise_combined} shows the posterior predictive distributions for each rheological model when the model bias is inferred. Compared to the results when the model bias in omitted (\autoref{fig:ppdrheo_combined}), the predictive bands of the Newtonian, Bingham, and biviscous models have widened significantly, whereas those of the Herschel–Bulkley and biviscous power law models did not change notably. This is a direct consequence of the inferred model bias, which is relatively large for the first three models compared to the latter two models, as shown in \autoref{tab:prior_posterior_rheo} and \autoref{fig:modelbias}. From \autoref{fig:modelbias} it is observed that the exponential prior for the model bias is relatively uninformed for all models.

Compared to the analyses without model bias, the predictive posterior for the models with a significant bias quantifies the uncertainty more reliably. For example, when comparing the results for the Newtonian model without and with inferred model bias, \autoref{fig:ppd_N_nobias} and \autoref{fig:ppd_N_bias}, respectively, it is observed that the prediction with inferred model bias encompasses the experimental observation. This contrasts the case where the model bias is omitted, which results in an uncertainty band that cannot explain the experimental observation. For a model with a small bias, such as the biviscous power law model shown in \autoref{fig:ppd_BVPL_nobias} and \autoref{fig:ppd_BVPL_bias}, this fundamental difference is insignificant.

\begin{figure}
    \centering
    \includegraphics[width=0.5\linewidth]{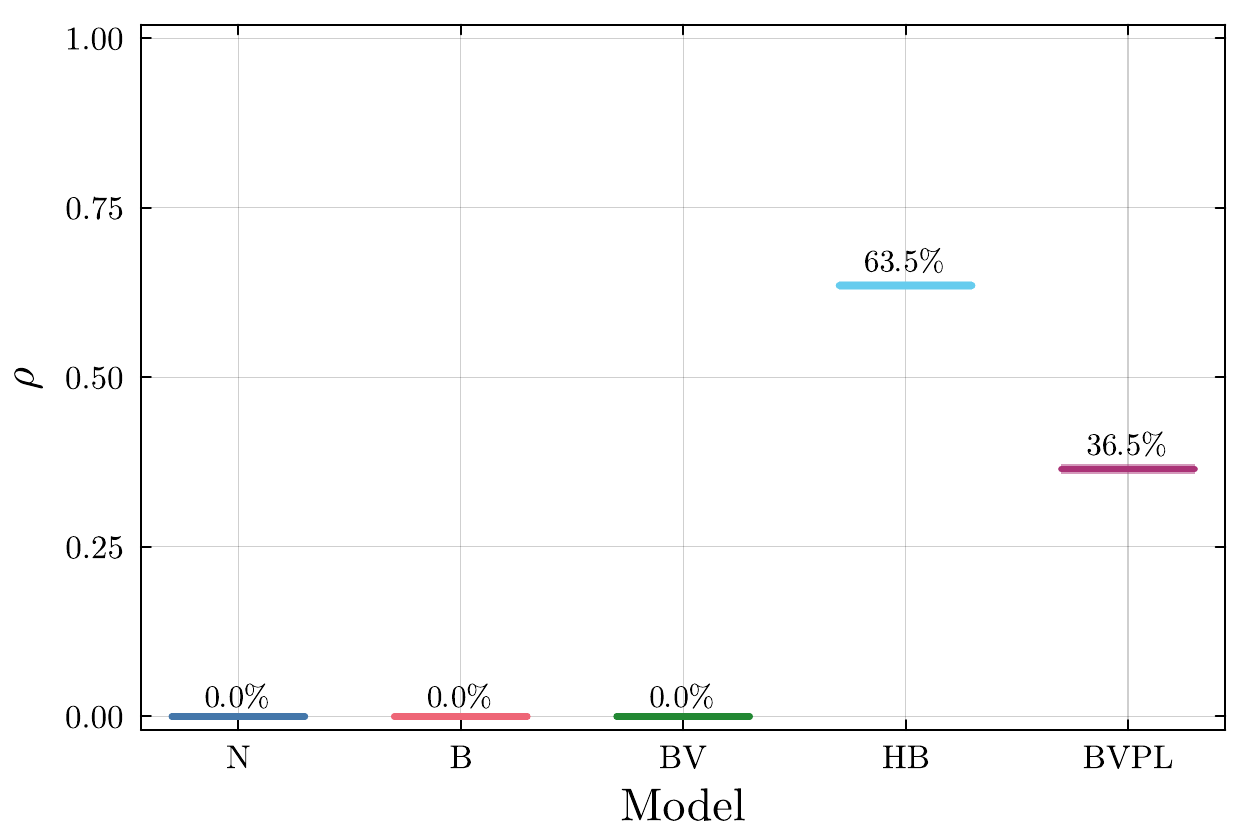}
    \caption{Plausibilities of the rheological models with inferred model bias for the Newtonian (N), Bingham (B), biviscous (BV), Herschel-Bulkley (HB) and biviscous power law (BVPL) models.}
    \label{fig:rheo_modsel_bias}
\end{figure}

\autoref{fig:rheo_modsel_bias} shows the computed model plausibilities, which confirm the qualitative observations in the predictive posterior distributions (\autoref{fig:ppdrheo_noise_combined}). The plausibilities for the Newtonian, Bingham and biviscous models remain negligible on account of these models being unable to mimic the yield stress and/or shear thinning behavior, which is also reflected by the relatively large model biases in \autoref{fig:modelbias}. For the models with a small bias, \emph{i.e.}, the Herschel-Bulkley model and the biviscous power law model, the results deviate marginally from the case where the model error is omitted. This is expected, as the inferred model bias is very small for these models (\autoref{fig:modelbias}). The variance of the plausibility estimator is, however, observed to be reduced compared to the case in which model bias was omitted, indicating that the inference of noise makes the plausibility calculations more reliable.

\section{Constitutive model selection in a squeeze flow}
\label{sec:squeezeflow}
In this section, we consider Bayesian model selection for a more complex fluid flow, \emph{viz.} a squeeze flow. In \autoref{sec:squeeze_data}, we introduce the experimental setup and discuss data acquisition procedures. \autoref{sec:squeeze_model} introduces the squeeze flow model for the admissible rheological laws presented in \autoref{sec:rheologymmodels}. We then detail the probabilistic setting -- modulo the specification of the rheological priors -- in \autoref{sec:squeeze_probabilistic}. In \autoref{sec:squeeze_results}, we consider two distinct model selection approaches. In the first approach, referred to as \emph{rheo-informed inference}, we define prior information for the rheological parameters based on the inference discussed in \autoref{sec:results_rheometry}. In the second approach, referred to as \emph{expert-informed inference}, we omit rheological measurements and instead rely on broader priors informed by expert opinion.

\subsection{Experimental data acquisition}\label{sec:squeeze_data}
\begin{figure}
     \centering
     \subfloat[Setup schematic]{\includegraphics[width=0.46\textwidth]{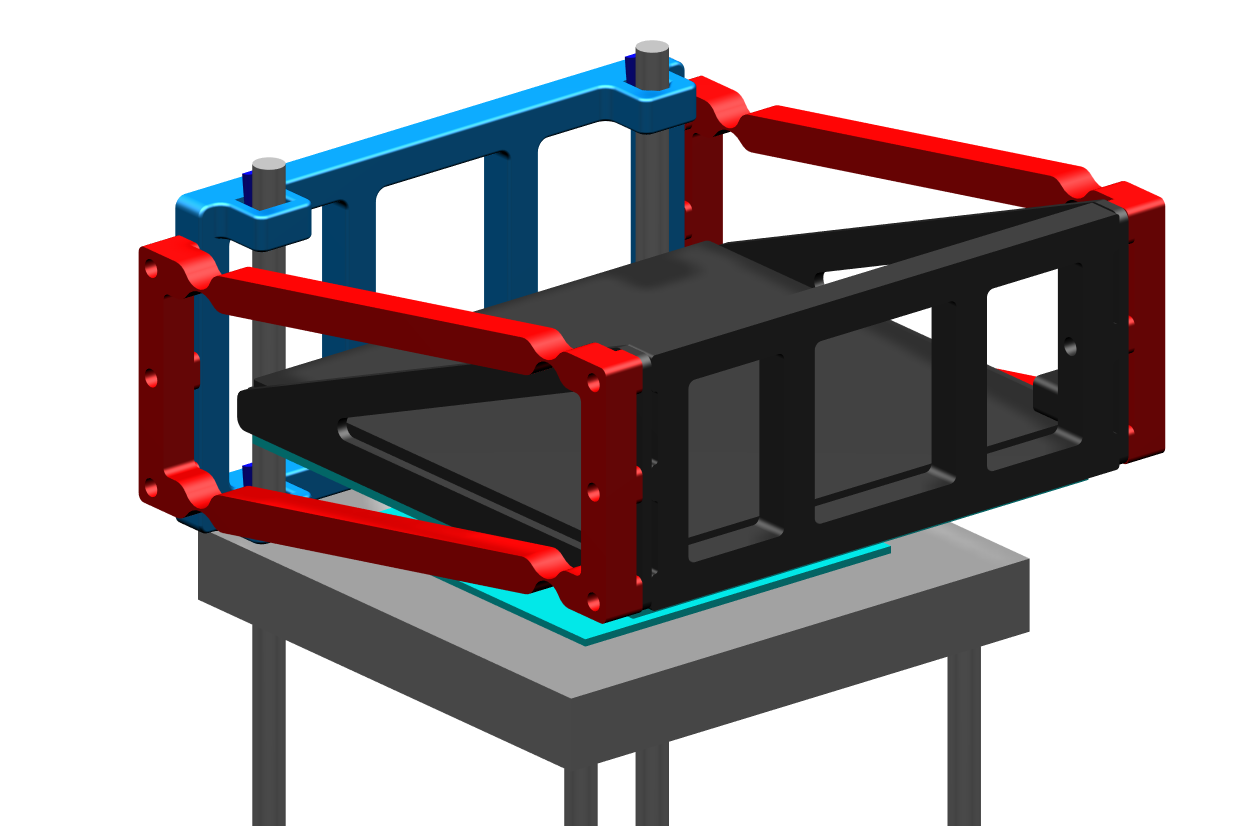}
     \label{fig:setup_schem_a}}
     \subfloat[Camera image]{\includegraphics[width=0.3\textwidth]{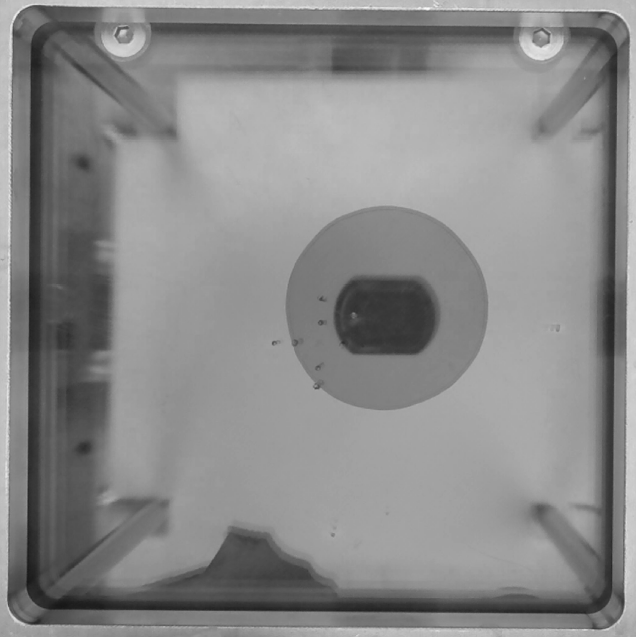}
     \label{fig:setup_schem_b}}
    \caption{Squeeze flow experiment.}\label{fig:setup_schem}
\end{figure}

We use the squeeze flow setup described in \citet{rinkens2023uncertainty}; see \autoref{fig:setup_schem_a}. The system consists of two parallel plates that compress a fluid placed between them. A camera positioned below the setup captures the fluid's motion, as illustrated in \autoref{fig:setup_schem_b}. Parallel motion of the plates is maintained by a set of leaf springs. The top plate is mounted to a support structure, which is intentionally colored black to facilitate post-processing of the recorded images. The own weight of the top part of the setup is approximately equal to \SI{350}{\gram}. A counterweight is applied to reduce the force exerted on the fluid. A net weight of \SI{150}{\gram} (\emph{i.e.}, a \SI{200}{\gram} counterweight) was empirically found to be suitable for studying the flow regime where yield stress effects are significant. That is, the exerted force is low enough to preserve yielding behavior, yet high enough to limit experimental uncertainties associated with, among others, the stiffness of the leaf springs.

We consider the same yield stress fluid as used in the rheological measurements in \autoref{sec:rheoexperiment}, \emph{i.e.}, Carbopol 980 at 0.5 wt\%. Since deposition of this fluid using a syringe proved impractical, we use a removable cylindrical container to accurately control the volume of the material. Once the fluid is in place, we remove the cylinder and lower the top plate until it reaches the release mechanism. The top plate is lowered slowly, and the sample is allowed to rest for five minutes before initiating the experiment to minimize pre-loading effects.

\begin{figure}
     \centering
     \subfloat[Linear scale]{\includegraphics[width=0.5\linewidth]{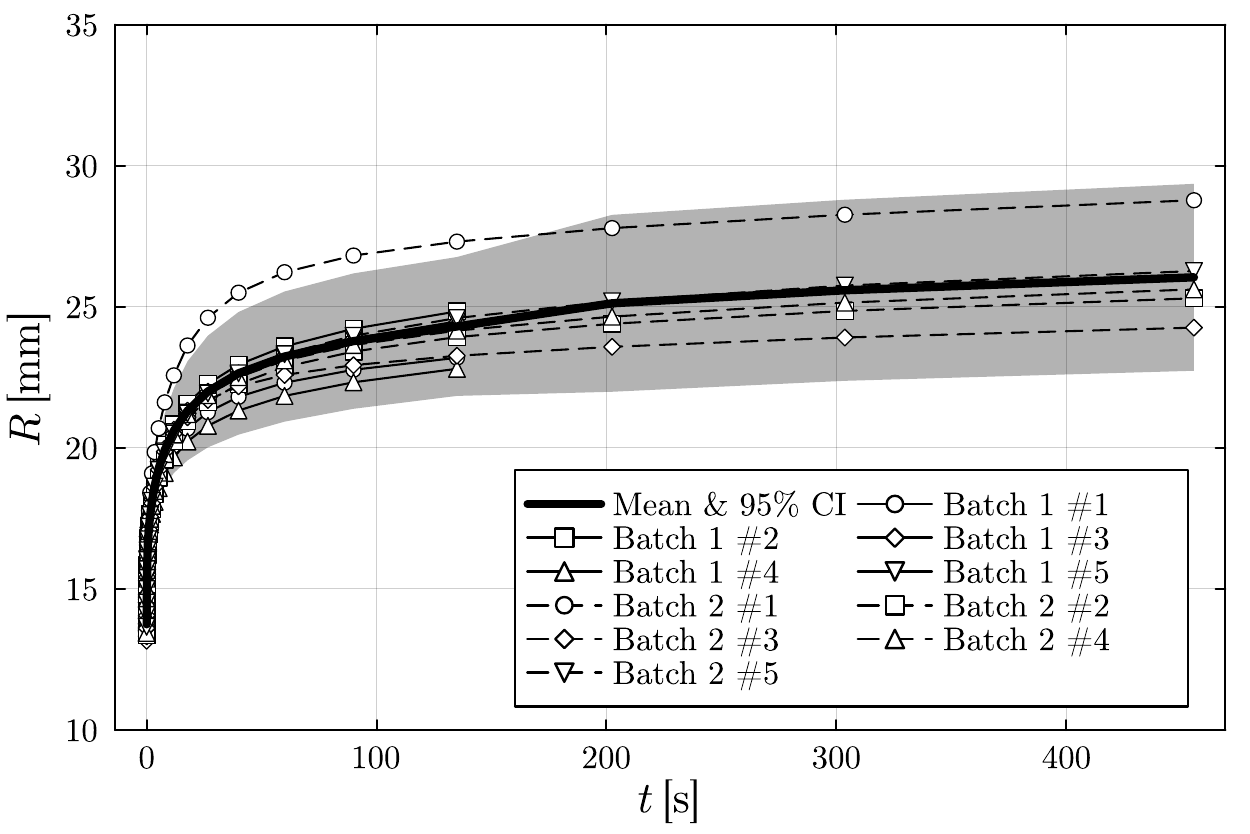}}
     \subfloat[Logarithmic scale]{\includegraphics[width=0.5\linewidth]{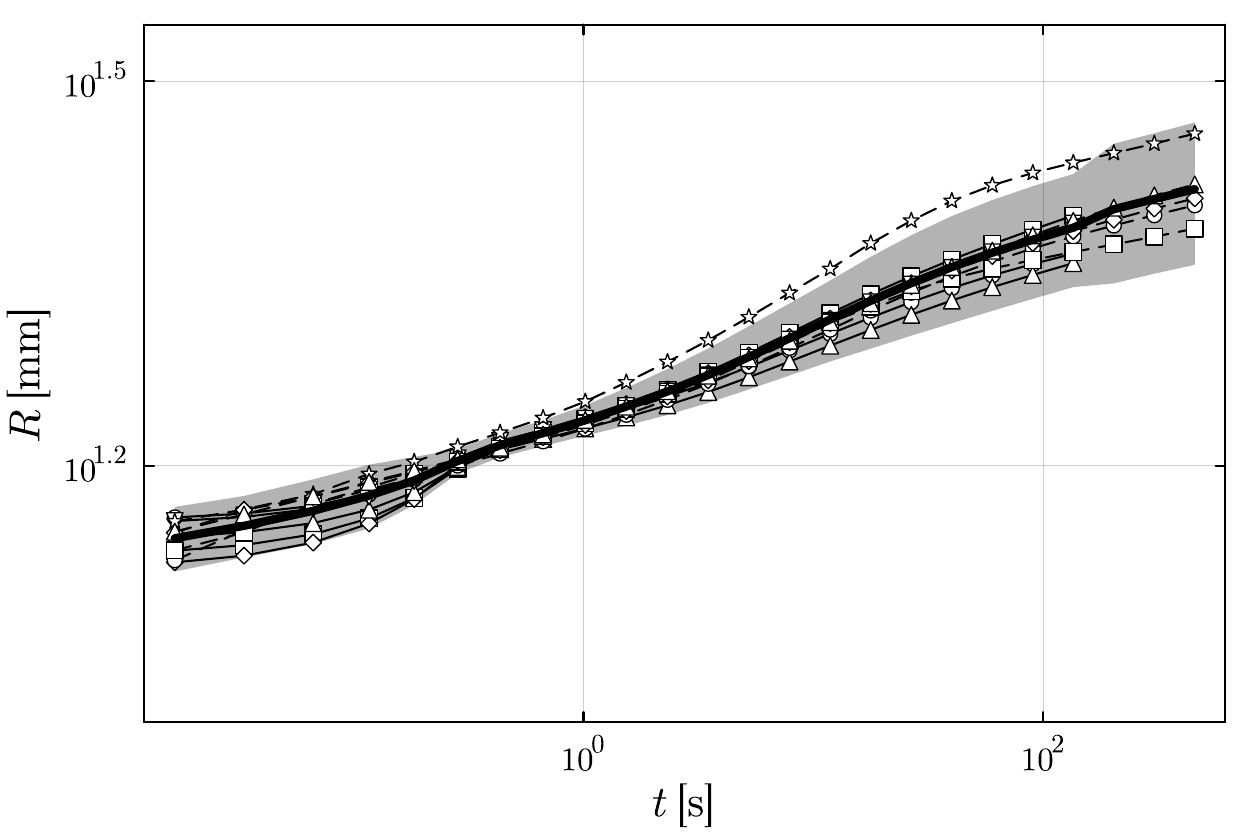}}
    \caption{Squeeze flow data for Carbopol 980 for two material batches.}
     \label{fig:squeezeflow_data}
\end{figure}

We start the experiment by releasing the top plate, allowing it to compress the fluid. We record the fluid’s motion using the bottom-mounted camera. From the video, we extract both the initial and subsequent frames for post-processing. The radius, $R(t)$, of the fluid layer is detected from these frames. We determine the radius at  exponentially distributed points in time, \emph{i.e.}, $t_{j+1}=t_j + \Delta t_j$ with $\Delta t_j=1.5^j\Delta t_0$ and $\Delta t_0 = \SI{17.7}{\milli\second}$, and collect these in the observation vector $\boldsymbol{y}_i$. We conducted $n=10$ experiments distributed over two batches, which are visualized in \autoref{fig:squeezeflow_data}. Note that the experiment duration varied between the batches.

\subsection{The squeeze flow model}
\label{sec:squeeze_model}
\begin{figure}
    \centering
    \includegraphics[width=0.5\linewidth]{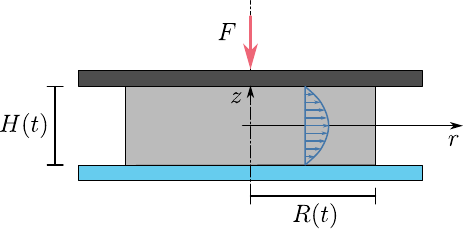}
    \caption{Squeeze flow model schematic.}
    \label{fig:squeezedomain}
\end{figure}

We consider the evolving axisymmetric domain $[0, R(t)] \times [-H(t)/2, H(t)/2] \ni (r, z)$ (\autoref{fig:squeezedomain}). The volume of the deposited fluid, $V$, is assumed constant on account of the fluid's incompressibility.  The initial height of the domain is set to $H_0$, with corresponding radius $R_0 = \sqrt{V/ (\pi H_0)}$. The height is assumed to be much smaller than the radius, such that lubrication assumptions hold \cite{Szeri2010FluidLubrication}. Specifically, the pressure is assumed constant over the height of the domain and all components of the stress and strain rate tensors other than the shear stress, $\tau$, and shear rate $\dot{\gamma} = \frac{\partial v}{\partial z}$ are assumed negligible. We confirm the validity of these assumptions in Appendix~\ref{app:sf_vandv} and refer to \citet{rinkens2023uncertainty} for details. At the contact between the fluid and the plates, we assume no slip conditions. At the fluid front, we assume capillary effects to be small and hence, the fluid pressure to be equal to the ambient pressure.

Under the above conditions, the time-dependent evolution of the domain and pressure field are governed by the initial boundary value problem:
\begin{subequations}
\begin{empheq}[left=\scalebox{1.1}{$\empheqlbrace$}~]{alignat=2}
  \multicolumn{4}{l}{
    \text{Find } $p(r,t)$ \text{ and } $H(t)$ \text{ for all } 
    $r \in [0,R(t)]$ \text{ and }
  } \nonumber \\
  \multicolumn{4}{l}{
    $t \in [0,T]$, \text{ with } $R(t) = \sqrt{V/ (\pi H(t))}$, \text{ such that:}
  } \nonumber \\[0.5em]
  Q\left( \frac{\partial p}{\partial r} \right) &= - \frac{r}{2} \frac{\partial H}{\partial t}
  &\qquad & r \in (0, R(t)),~t \in [0,T], \label{eq:ibvpa} \\
  \int_0^{R(t)} p  r \,dr &= \frac{F}{2 \pi}
  & & t \in [0,T], \label{eq:ibvpb} \\
  p &= 0
  & & r =R(t),~t \in [0, T], \label{eq:ibvpc}\\
  H &= H_0
  & & t = 0. \label{eq:ibvpd}
\end{empheq}
\label{eq:ibvp}%
\end{subequations}
The functional relation between the flux
\begin{align}
  Q(r) &= \int_{-H/2}^{H/2} v(r, z) \, dz 
  \label{eq:fluxrelation}
\end{align}    
and the pressure gradient, $\frac{\partial p}{\partial r}$, follows from the momentum balance:
\begin{subequations}
\label{eq:flowprofile}
\begin{empheq}[left=\scalebox{1.1}{$\empheqlbrace$}~]{alignat=2}
\multicolumn{4}{l}{
    \text{Given } $\frac{\partial p}{\partial r}$, \text{ find } $v(z)$ \text{ such that:} 
  } \hspace{0.3\linewidth} \nonumber \\[0.5em]
\tau \left(  \frac{\partial v}{\partial z} \right) &= \frac{\partial p}{\partial r} z &\qquad   & z \in (-H/2,H/2),\\
 v &=  0  &\qquad &z= \pm H/2.
\end{empheq}%
\end{subequations}
Upon substitution of one of the rheological models introduced in \autoref{sec:rheologymmodels}, this momentum balance can be integrated to find the velocity profile $v(z)$ as a function of the pressure gradient $\frac{\partial p}{\partial r}$, which, upon substitution in \cref{eq:fluxrelation} gives an expression for the flux.  

For the squeeze flow model \eqref{eq:ibvp} to be mathematically well-posed, the functional relationship between the flux and the pressure gradient must be invertible. In other words, for every admissible flux, there must exist a unique corresponding pressure gradient. This condition is violated by models such as the Bingham and Herschel-Bulkley models, which assume a rigid behavior (\emph{i.e.}, zero flow) below a critical yield stress. In these models, a finite range of pressure gradients corresponds to zero flux, breaking the invertibility requirement. Physically, this leads to what is known as the \emph{squeeze flow paradox}~\cite{wilson1993squeezing,estelle2007squeeze}, which asserts that squeeze flow cannot occur under such models. At the center of the squeezing plates, the shear rates remain below the yield threshold, implying that the material behaves as a solid and cannot be compressed, contradicting the observed behavior in practical squeeze flow scenarios. For this reason, we exclude the Bingham and Herschel-Bulkley models from our analyses.

The flow profile problem \eqref{eq:flowprofile} can be solved analytically for all considered models, \emph{i.e.}, the Newtonian model, the biviscous model, and the biviscous power law model. The main complication in solving the squeeze flow model \eqref{eq:ibvp} is that, with the exception of the Newtonian case, the relation between the flux and the pressure gradient is nonlinear, as elaborated in Appendix~\ref{app:sf_fluxrelation}. Our numerical procedure to solve this nonlinear problem is detailed in Appendix~\ref{app:sf_procedure}.

\begin{remark}[Shear rates in the squeeze flow problem]
For the squeeze flow problem, negative shear rates, $\dot{\gamma}$, occur. Consequently, the simplified mathematical descriptions of \autoref{sec:rheologymmodels} must be amended to account for these negative values. We refer the reader to the references in \autoref{sec:rheologymmodels} for details.
\end{remark}

\subsection{Probabilistic modeling}\label{sec:squeeze_probabilistic}
In our Bayesian framework, the parameters associated with the squeeze flow experiment are modeled probabilistically. Below, we specify the prior information for the experimental parameters, excluding those related to the rheological models. We also define the likelihood function, which accounts for both experimental noise and model bias. These priors and the likelihood formulation are used in both the \emph{rheo-informed inference} (\autoref{sec:rheoinformed}) and the \emph{expert-informed inference} (\autoref{sec:expertinformed}) approaches. Since the prior information for the fluid parameters differs between these two approaches, their specification is discussed in the respective sections.

\subsubsection{Prior information for the setup}
The prior information for the experimental parameters is obtained through additional independent measurements. The applied force, $F$, is determined via the counterweights. Counterweight is removed from the balanced system in order to apply a force. The removed counterweight then directly determines the force. A log-normal prior distribution is assigned to $F$ to ensure physically realistic values. The fluid volume, $V$, is determined by weighing a sample and dividing by the known density, and a log-normal distribution based on ten measurements is used as its prior. The initial radius, $R_0$, is extracted from the initial frame of the squeeze flow experiment, and a log-normal distribution based on ten measurements is also used here. The prior distributions are summarized in \autoref{tab:priorparam_sf}.

{ 

\sisetup{
  detect-all,
  round-mode=figures,
  round-precision=3,
  scientific-notation=true,
  retain-explicit-plus = true
}

\begin{table*}
\centering
\caption{Squeeze flow experiment posteriors for the rheo-informed and expert-informed analyses across models.}\label{tab:priorparam_sf}
\resizebox{\textwidth}{!}{%
\begin{tabular}{
  l@{\hspace{3.em}}  
  l@{\hspace{1.em}}  
  S@{\hspace{1.em}}  
  S@{\hspace{1.em}}  
  l@{\hspace{1.em}}  
  S@{\hspace{1.em}}  
  S@{\hspace{1.em}}  
  l@{\hspace{1.em}}  
}
\toprule
\textbf{Parameter} & \textbf{Prior} & \multicolumn{3}{c}{\textbf{Rheo-informed posterior}} & \multicolumn{3}{c}{\textbf{Expert-informed posterior}} \\
\cmidrule(lr){3-5} \cmidrule(lr){6-8}
 & & \textit{Mean} & \textit{Std.~dev.} & \textit{Coef.~of~var.} & \textit{Mean} & \textit{Std.~dev.} & \textit{Coef.~of~var.} \\
\midrule
\multicolumn{8}{l}{\textit{Newtonian}} \\
\midrule
$V$ [\si{\meter^3}] & $\text{log}\mathcal{N}(7.57 \times 10^{-7}, 2.78 \times 10^{-9})$ & 7.5272e-7 & 3.5772e-9 & \phantom{0}0.5\% & 7.5741e-7 & 2.5998e-9 & \phantom{0}0.3\% \\
$R_0$ [\si{\meter}]                   & $\text{log}\mathcal{N}(1.37 \times 10^{-2}, 4.25 \times 10^{-5})$ & 0.013702  & 4.4085e-5 & \phantom{0}0.3\% & 0.0137111 & 4.0758e-5 & \phantom{0}0.3\% \\
$F$ [\si{\newton}]                    & $\text{log}\mathcal{N}(1.50, 0.15)$ & 0.379292  & 0.066782  &           17.6\% & 1.5465    & 0.174106  & 11.3\% \\
\midrule
\multicolumn{8}{l}{\textit{Biviscous}} \\
\midrule
$V$ [\si{\meter^3}] & $\text{log}\mathcal{N}(7.57 \times 10^{-7}, 2.78 \times 10^{-9})$ & 7.5813e-7 & 2.8160e-9 & \phantom{0}0.4\% & 7.5726e-7 & 2.7837e-9 & \phantom{0}0.4\% \\
$R_0$ [\si{\meter}]                   & $\text{log}\mathcal{N}(1.37 \times 10^{-2}, 4.25 \times 10^{-5})$ & 0.013766  & 3.9405e-5 & \phantom{0}0.3\% & 0.013709  & 3.4966e-5 & \phantom{0}0.3\% \\
$F$ [\si{\newton}]                    & $\text{log}\mathcal{N}(1.50, 0.15)$ & 3.0227    & 0.19488   & \phantom{0}6.4\% & 1.5251    & 0.14534   & \phantom{0}9.5\% \\
\midrule
\multicolumn{8}{l}{\textit{Biviscous power law}} \\
\midrule
$V$ [\si{\meter^3}] & $\text{log}\mathcal{N}(7.57 \times 10^{-7}, 2.78 \times 10^{-9})$ & 7.5707e-7 & 3.4299e-9 & \phantom{0}0.5\% & 7.5708e-7 & 2.6408e-9 & \phantom{0}0.3\% \\
$R_0$ [\si{\meter}]                   & $\text{log}\mathcal{N}(1.37 \times 10^{-2}, 4.25 \times 10^{-5})$ & 0.0137287 & 5.0980e-5 & \phantom{0}0.4\% & 0.013691  & 3.8848e-5 & \phantom{0}0.3\% \\
$F$ [\si{\newton}]                    & $\text{log}\mathcal{N}(1.50, 0.15)$ & 3.6269    &  0.198505 & \phantom{0}5.5\% & 1.48044   & 0.173446  &           11.7\% \\
\bottomrule
\end{tabular}
}
\end{table*}
} 

Since the Bingham model and the Herschel-Bulkley model cannot be used in the squeeze flow (\autoref{sec:squeeze_model}), we assign a zero prior probability to these model classes. We assign a uniform discrete prior to the other model classes, \emph{i.e.}, Newtonian, biviscous, and biviscous power law, reflecting that we have no prior information regarding their suitability.

\subsubsection{The likelihood function}
The likelihood function for the squeeze flow experiment is defined identically to that for the rheological measurements as discussed in \autoref{sec:rheolikelihood}. The covariance matrix of the likelihood function \eqref{eq:likelihoodformcombined} again comprises an experimental noise component, $\boldsymbol{\Sigma}_{\boldsymbol{y}}$, which is based on the squeeze flow experiments (\autoref{sec:squeeze_data}), and a model bias component with standard deviation $\sigma_{\rm bias}$ and correlation function
\begin{equation}
    \rho(t, t') = \exp{\left( -\frac{\left| t - t' \right|}{l_{\rm bias}} \right)}.
    \label{eq:correlationfunctionsqueeze}
\end{equation}
We set the correlation time to $l_{\rm bias}=\SI{1}{\second}$ and model the bias probabilistically using an exponential distribution with prior standard deviation of \SI{2e-2}{\meter}.

\subsection{Results}\label{sec:squeeze_results}
With the Bayesian setting completely defined, we now consider the two inference cases: \emph{rheo informed} (\autoref{sec:rheoinformed}) and \emph{expert informed} (\autoref{sec:expertinformed}). For both cases we include the model bias as a hyperparameter. All results are based on MCMC simulations with two walkers per parameter and \num{1000} steps per walker. The Gelman-Rubin diagnostic converged up to a tolerance of 1.1 for all chains.

{ 

\sisetup{
  detect-all,
  round-mode=figures,
  round-precision=3,
  scientific-notation=true,
  retain-explicit-plus = true
}

\begin{table*}
\centering
\caption{Rheological posteriors for the rheo-informed and expert-informed squeeze flow analyses.}\label{tab:posterior_sf}
\resizebox{\textwidth}{!}{%
\begin{tabular}{
  l@{\hspace{3.em}}  
  S@{\hspace{1.em}}  
  S@{\hspace{1.em}}  
  l@{\hspace{1.em}}  
  S@{\hspace{1.em}}  
  S@{\hspace{1.em}}  
  l@{\hspace{1.em}}  
}
\toprule
\textbf{Parameter} & \multicolumn{3}{c}{\textbf{Rheo-informed posterior}} & \multicolumn{3}{c}{\textbf{Expert-informed posterior}} \\
\cmidrule(lr){2-4} \cmidrule(lr){5-7}
 & \textit{Mean} & \textit{Std.~dev.} & \textit{Coef.~of~var.} & \textit{Mean} & \textit{Std.~dev.} & \textit{Coef.~of~var.} \\
\midrule
\multicolumn{7}{l}{\textit{Newtonian}} \\
\midrule
$\eta$ [\si{Pa \cdot s}] & 6.66218   & 0.58843     & \phantom{0}8.8\% & 44.273    & 5.6464    &           12.8\% \\
$\sigma_{\rm bias}$ [\si{\meter}]    & 0.0011190 & 0.000205860 &           18.3\% & 0.0011639 & 9.3678e-5 & \phantom{0}8.0\% \\
\midrule
\multicolumn{7}{l}{\textit{Biviscous}} \\
\midrule
$\log_{10}{\eta_0}$ [\si{Pa \cdot s}] & 2.8874    & 0.0079715  & \phantom{0}0.3\% & 1.9596     & 0.073348  & \phantom{0}3.7\% \\
$\eta_1$ [\si{Pa \cdot s}]            & 1.6558    & 0.08236    & \phantom{0}5.0\% & 6.5066     & 1.0211    &           15.7\% \\
$\tau_y$ [\si{Pa}]                    & 125.31    & 2.5171     & \phantom{0}2.0\% & 50.758     & 8.1186    &           16.0\% \\
$\sigma_{\rm bias}$ [\si{\meter}]     & 0.0013294 & 0.00013992 &           10.5\% & 0.00028423 & 4.8452e-5 &           17.0\% \\
\midrule
\multicolumn{7}{l}{\textit{Biviscous power law}} \\
\midrule
$\log_{10}{\eta_0}$ [\si{Pa \cdot s}]  & 2.95195    & 0.0120911   & \phantom{0}0.4\% & 2.53095    & 0.42317   &           16.7\% \\
$K$ [\si{Pa \cdot s^n}]                  & 50.749     & 6.2079      &           12.2\% & 43.712     & 7.52368   &           17.2\% \\
$\tau_y$ [\si{Pa}]                     & 75.108     & 3.5443      & \phantom{0}4.7\% & 8.8034     & 5.78750   &           65.7\% \\
$n$ [-]                                & 0.59160    & 0.026111    & \phantom{0}4.4\% & 0.44669    & 0.036377  & \phantom{0}8.1\% \\
$\sigma_{\rm bias}$ [\si{\meter}]      & 0.00028428 & 6.524916e-5 &           22.9\% & 0.00020306 & 4.2233e-5 &           20.7\% \\
\bottomrule
\end{tabular}
}
\end{table*}
} 

\subsubsection{Rheo-informed inference}
\label{sec:rheoinformed}
For the rheo-informed model selection we use the posterior distributions following from the rheological measurements as prior information. Specifically, we consider the posteriors obtained when including model bias (\autoref{sec:includedmodelbias}). This prior selection is summarized in \autoref{tab:prior_posterior_rheo}.

\begin{figure*}
    \centering
    \subfloat[Newtonian]{\includegraphics[width=0.45\linewidth]{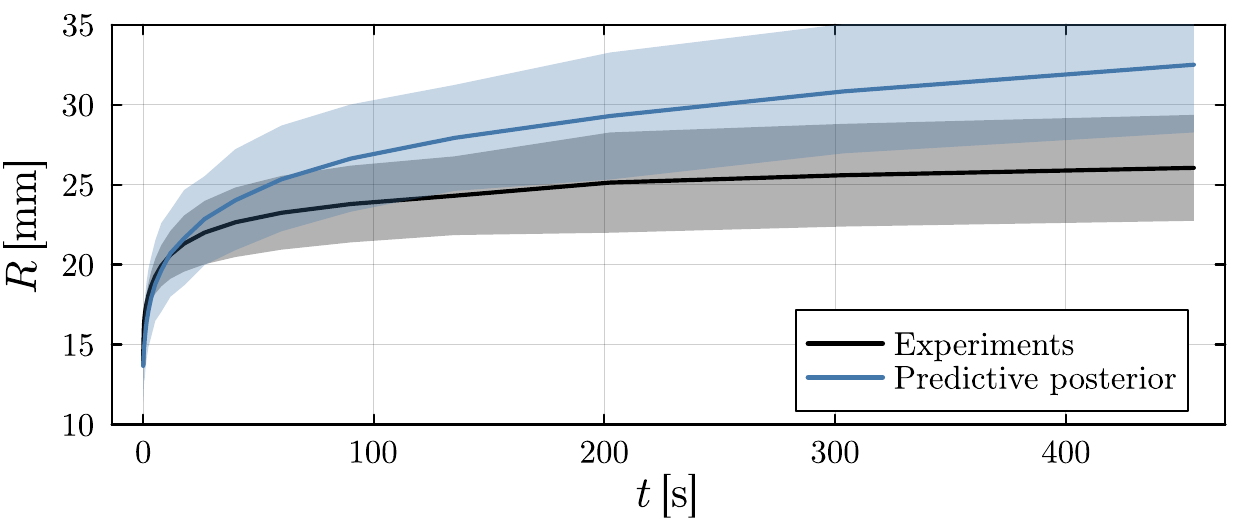}
    \includegraphics[width=0.45\linewidth]{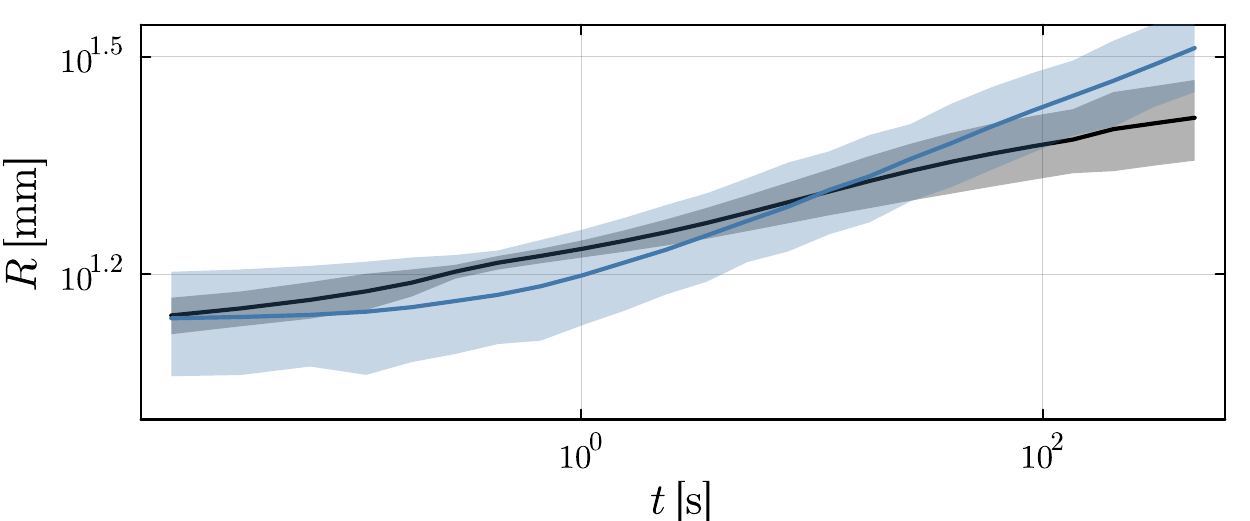}
    \label{fig:ppd_N_rheo}}\\[1em]
    \subfloat[Biviscous]{\includegraphics[width=0.45\linewidth]{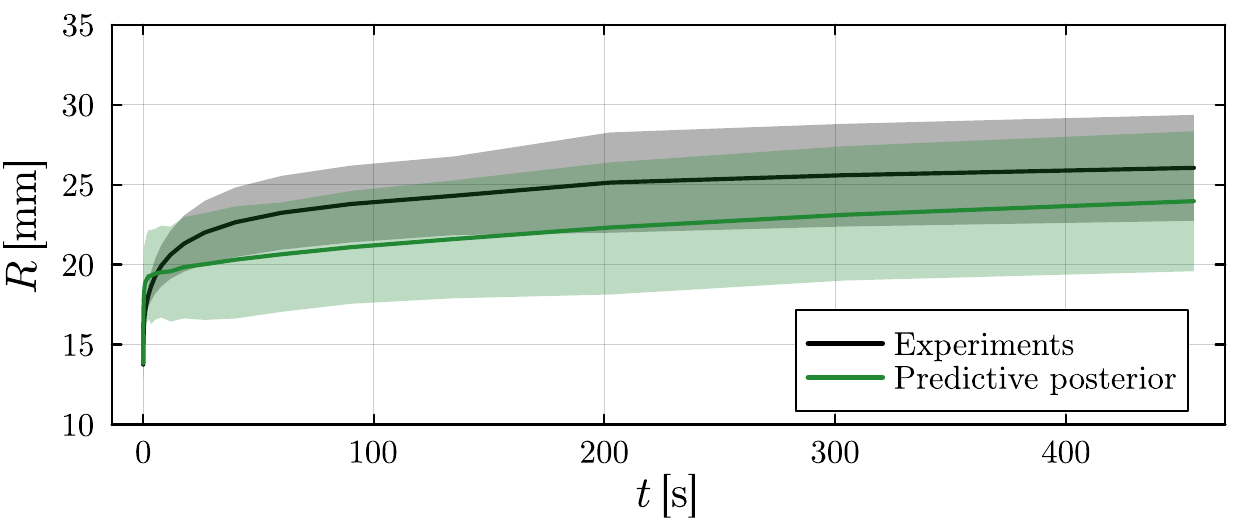}
    \includegraphics[width=0.45\linewidth]{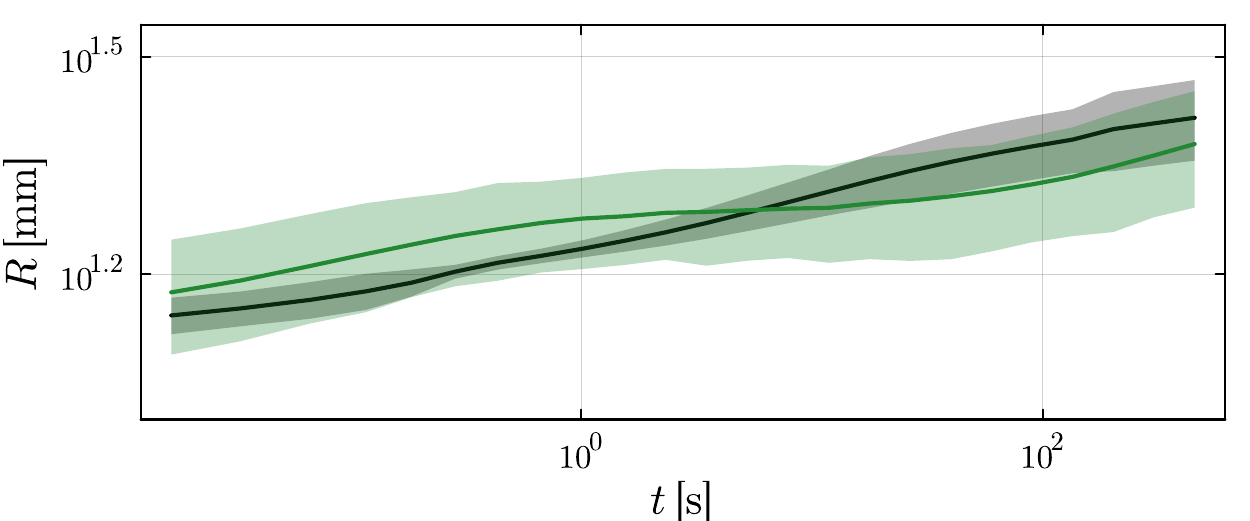}}\\[1em]
    \subfloat[Biviscous power law]{\includegraphics[width=0.45\linewidth]{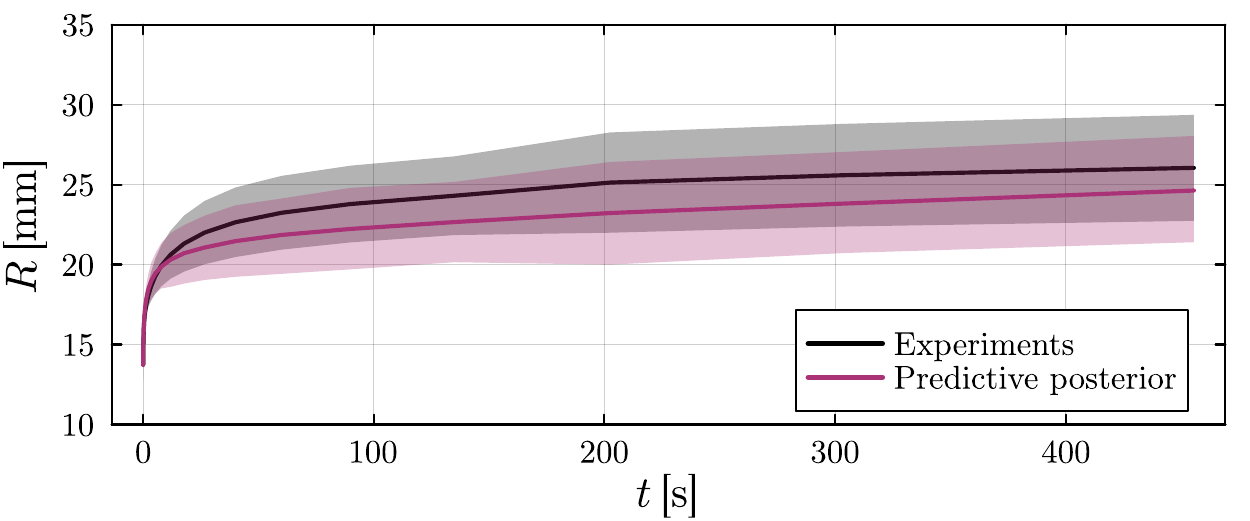}
    \includegraphics[width=0.45\linewidth]{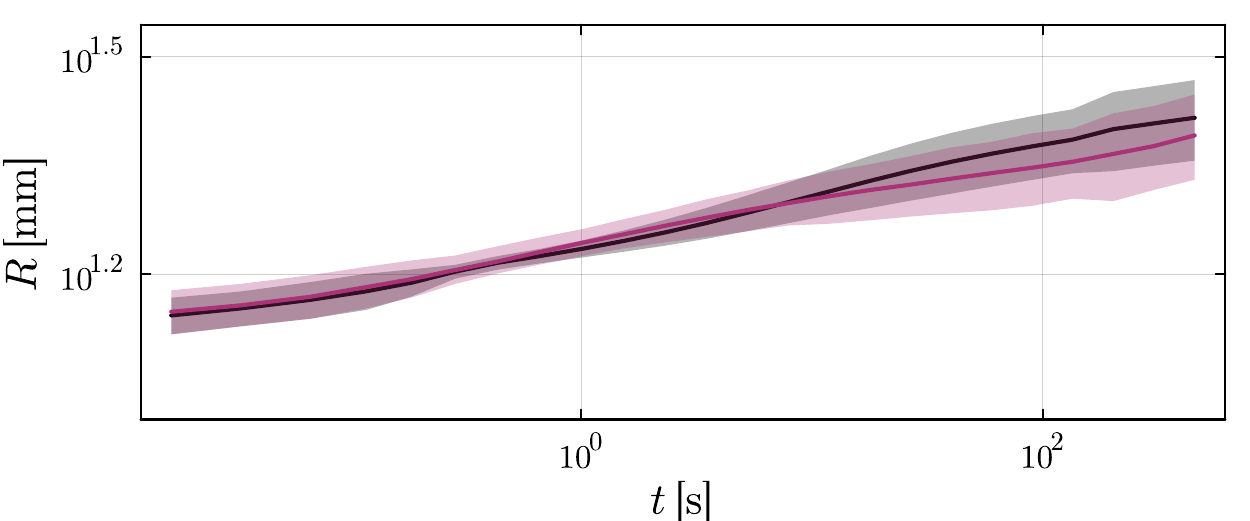}\label{fig:ppd_BVPL_rheo}}
    \caption{Rheo-informed predictive posterior distribution per model. The left column shows all models on a linear scale, whereas the right columns shows the same results on a logarithmic scale.}
    \label{fig:sf_ppdcombined_info}
\end{figure*}

\begin{figure*}
    \centering
    \subfloat[Newtonian]{\includegraphics[width=0.33\linewidth]{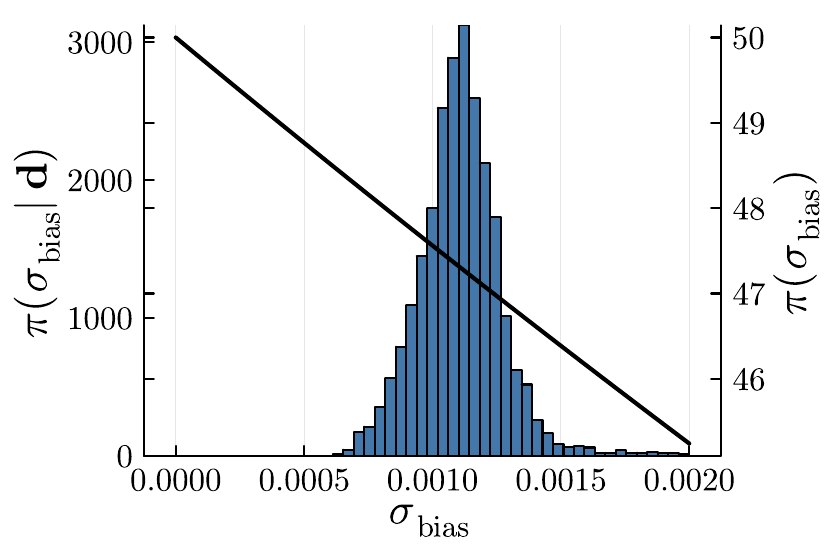}}
    \subfloat[Biviscous]{\includegraphics[width=0.33\linewidth]{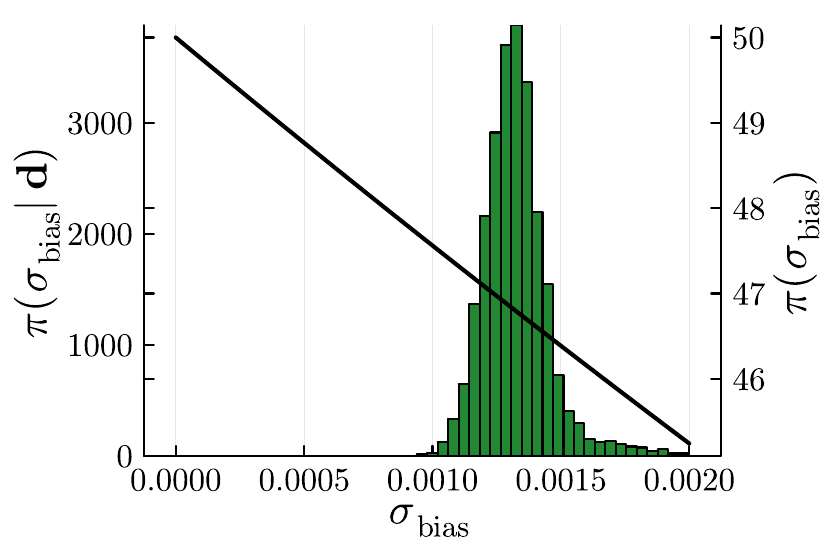}}
    \subfloat[Biviscous power law]{\includegraphics[width=0.33\linewidth]{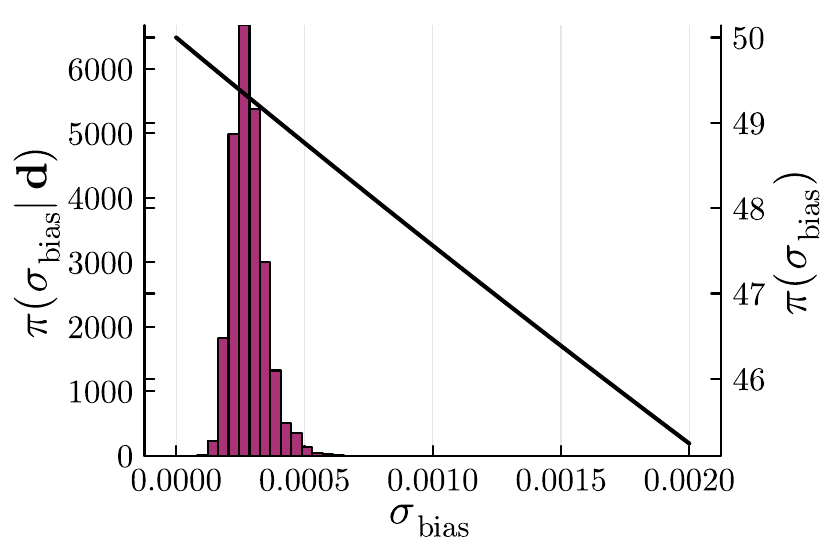}}
    \caption{Posterior (histogram) versus prior (black line) of the inferred model bias for all rheo-informed models.}
    \label{fig:sf_rheomodelbias}
\end{figure*}

\autoref{fig:sf_ppdcombined_info} shows the predictive posterior distributions for the Newtonian model, the biviscous model, and the biviscous power law model. For the first two models, throughout most of the duration of the experiment, the observed 95\% credibility intervals are substantially broadened compared to the corresponding confidence intervals of the experiments, indicating that the model bias is the dominant source of uncertainty for these models. This is confirmed by the model bias distributions shown in \autoref{fig:sf_rheomodelbias}. For the biviscous power law model, the model bias is observed to be substantially lower than that of the other two models, which is also reflected by the improved match with the experiments in \autoref{fig:ppd_BVPL_rheo}. When computing the model plausibilities, the probability of the biviscous power law model is equal to one, whereas the other models have zero probability. However, also for the biviscous power law model the mismatch with the experimental data remains significant.

\begin{figure}
    \centering
    \subfloat[Biviscous]{\includegraphics[width=0.5\linewidth]{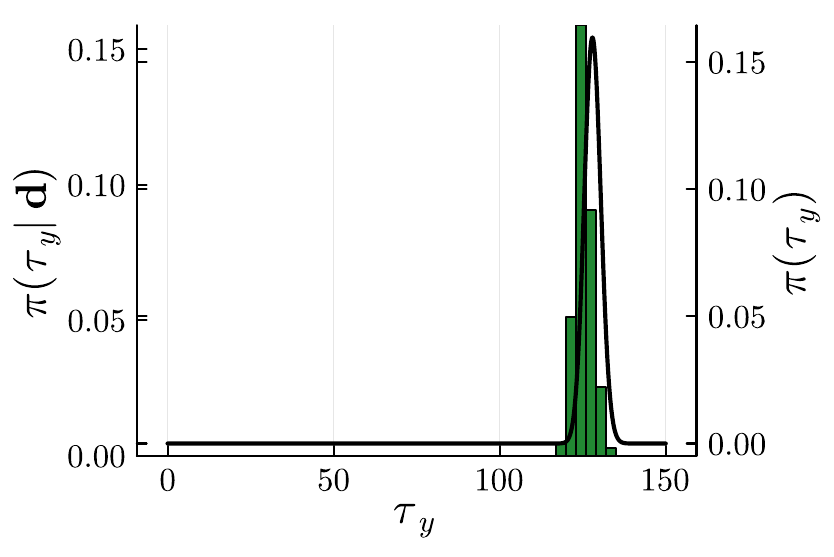}}
    \subfloat[Biviscous power law]{\includegraphics[width=0.5\linewidth]{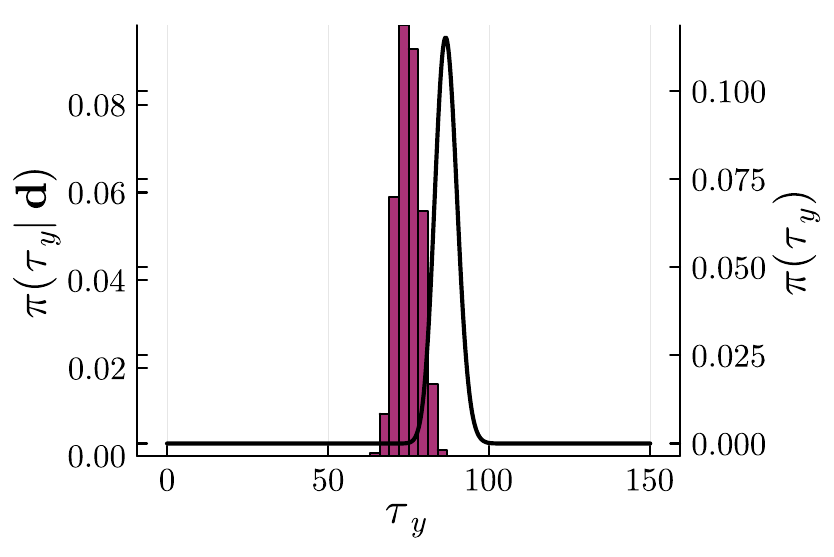}}
    \caption{Posterior (histogram) versus prior (black line) of the inferred yield stress for the rheo-informed models.}
    \label{fig:informedyieldstress}
\end{figure}

The main reason for the significant mismatch between the calibrated models and the experiments is that the prior yield stresses were determined from rheological steady shear measurements. These values (e.g., a mean of \SI{128}{\pascal} with a standard deviation of \SI{2.5}{\pascal} for the yield stress in the biviscous model) are not representative of the squeeze flow considered here, which is non-viscometric and involves non-uniform and mixed deformation modes as well as transient effects. Inspection of the posterior distribution of the yield stress in \autoref{fig:informedyieldstress} reveals that the squeeze flow inference tends to lower the yield stress, but that its rheo-informed priors prevent it from being lowered enough as to explain the squeeze flow data. This means that the yield stress prior is strongly biased, resulting in unrealistically high calibrated force values (\autoref{tab:priorparam_sf}) and inaccurate model predictions.  

\subsubsection{Expert-informed inference}
\label{sec:expertinformed}
For the expert-informed inference we omit the information obtained from the rheological measurements. The priors for the rheological model parameters are the same as those used for the rheological inference in \autoref{sec:results_rheometry}, as summarized in \autoref{tab:prior_posterior_rheo}.

\begin{figure*}
    \centering
    \subfloat[Newtonian]{\includegraphics[width=0.45\linewidth]{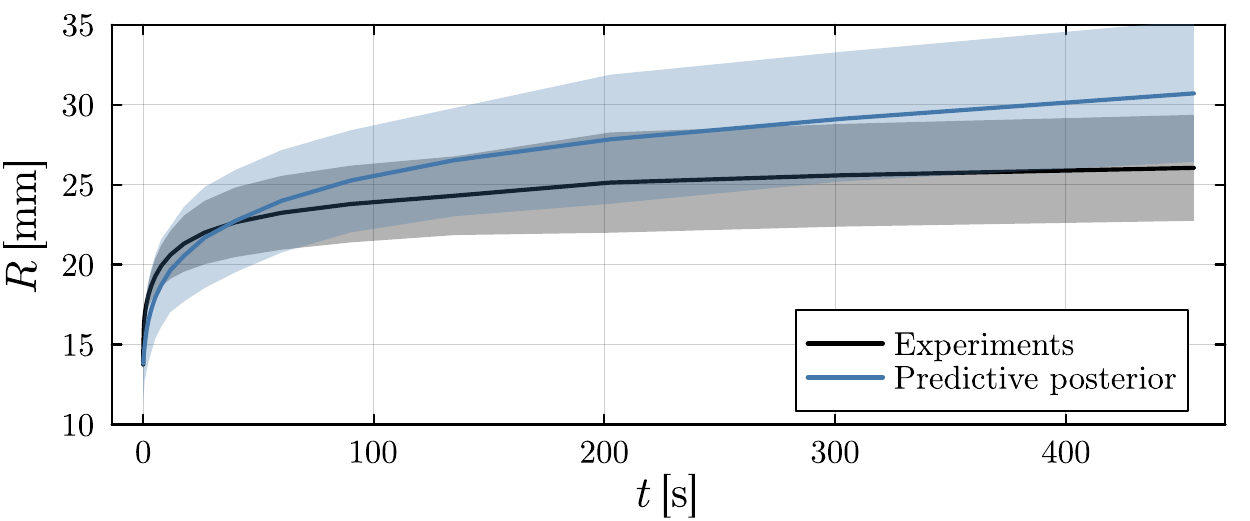}
    \includegraphics[width=0.45\linewidth]{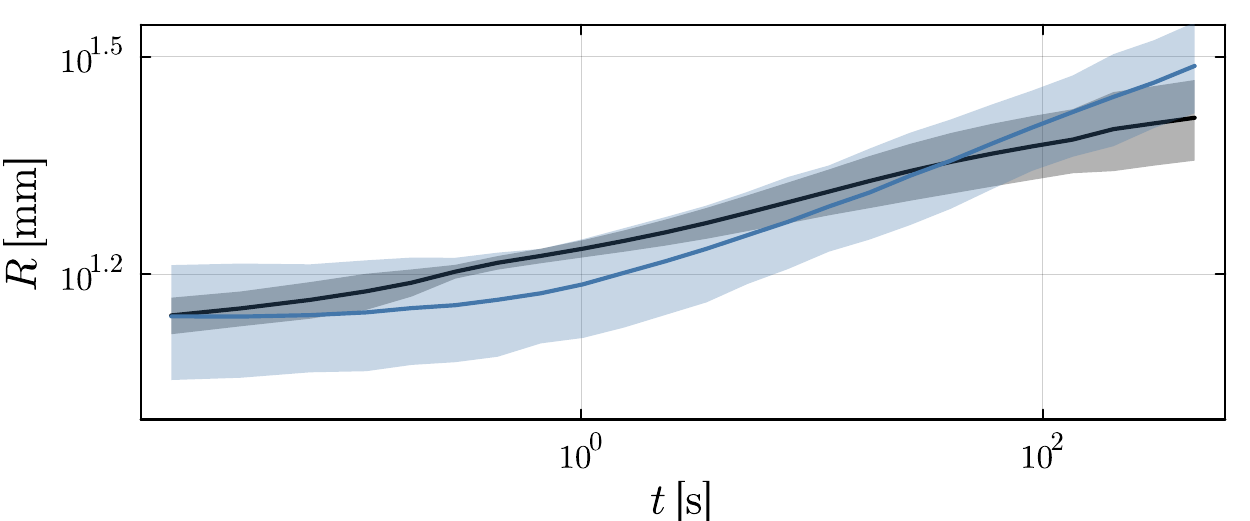}
    \label{fig:ppd_N_expert}}\\[1em]
    \subfloat[Biviscous]{\includegraphics[width=0.45\linewidth]{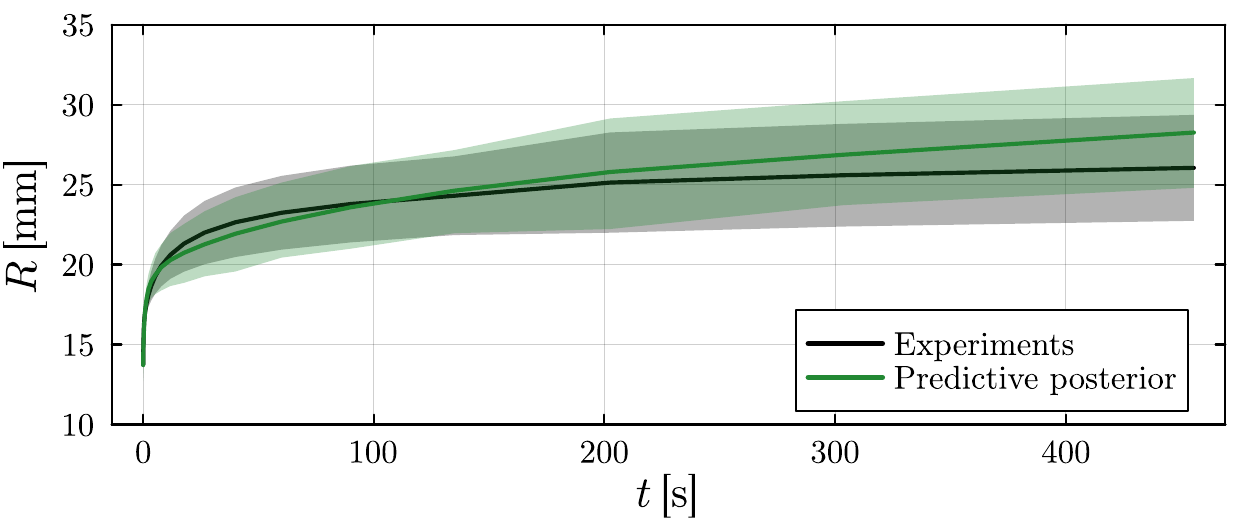}
    \includegraphics[width=0.45\linewidth]{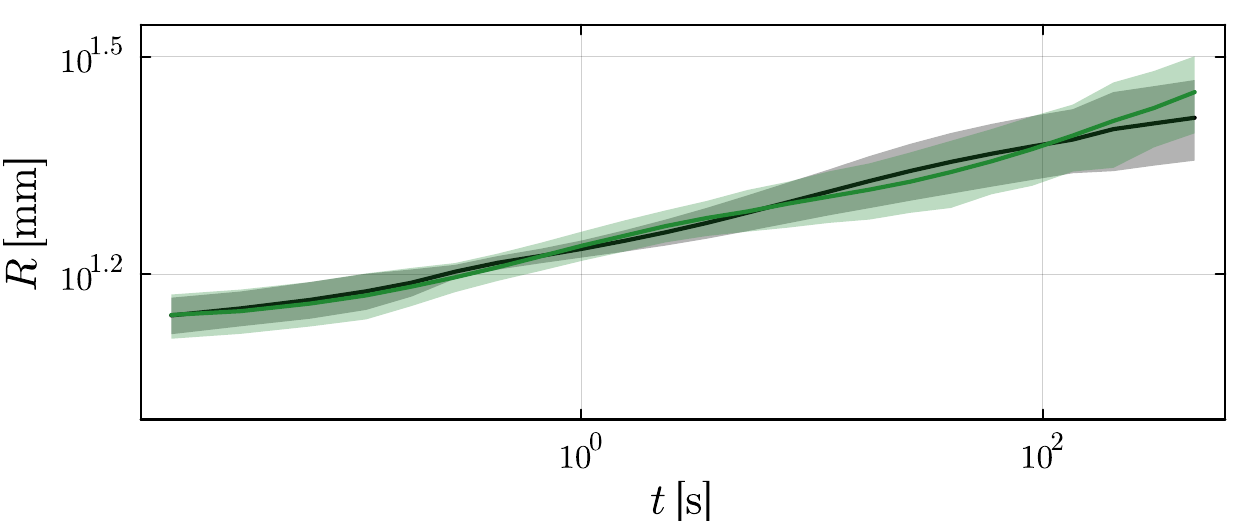}}\\[1em]    
    \subfloat[Biviscous power law]{\includegraphics[width=0.45\linewidth]{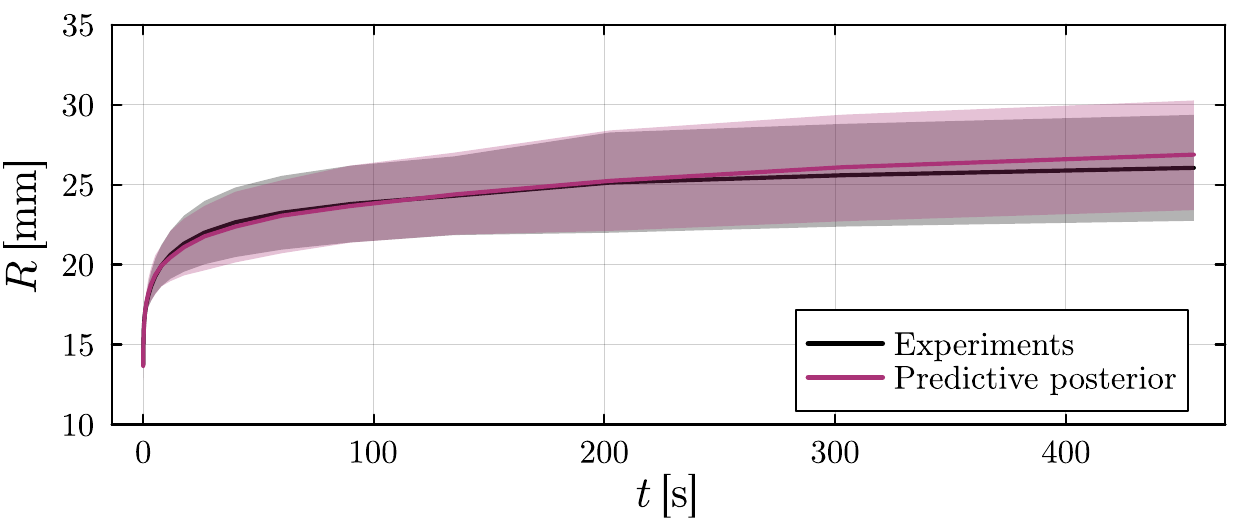}
    \includegraphics[width=0.45\linewidth]{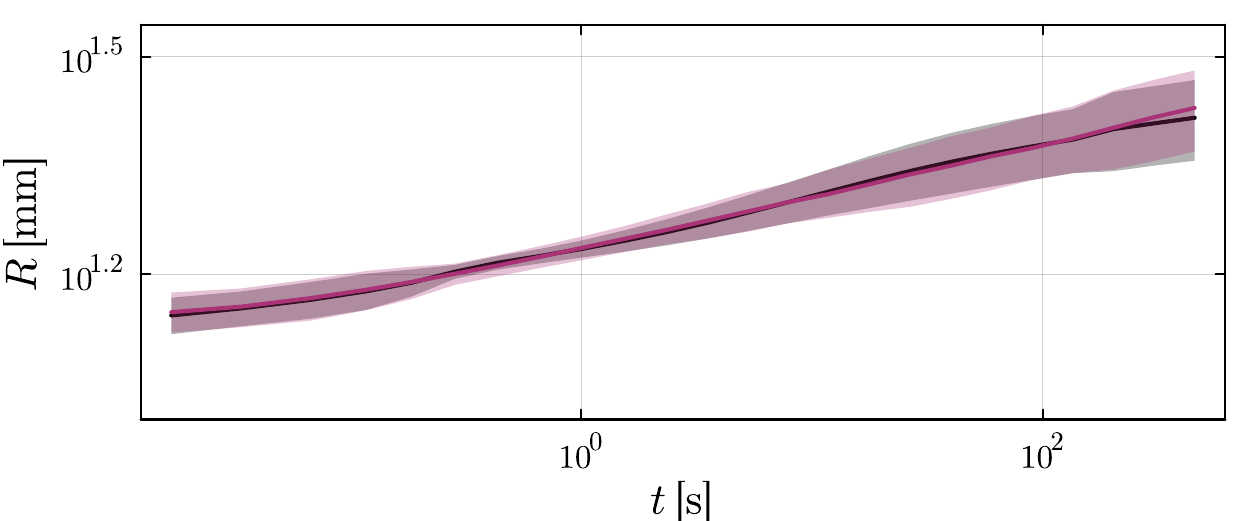}\label{fig:ppd_BVPL_expert}}
    \caption{Expert-informed predictive posterior distribution per model. The left column shows all models on a linear scale, whereas the right columns shows the same results on a logarithmic scale.}
    \label{fig:sf_ppdcombined_uninfo}
\end{figure*}

\begin{figure*}
    \centering
    \subfloat[Newtonian]{\includegraphics[width=0.33\linewidth]{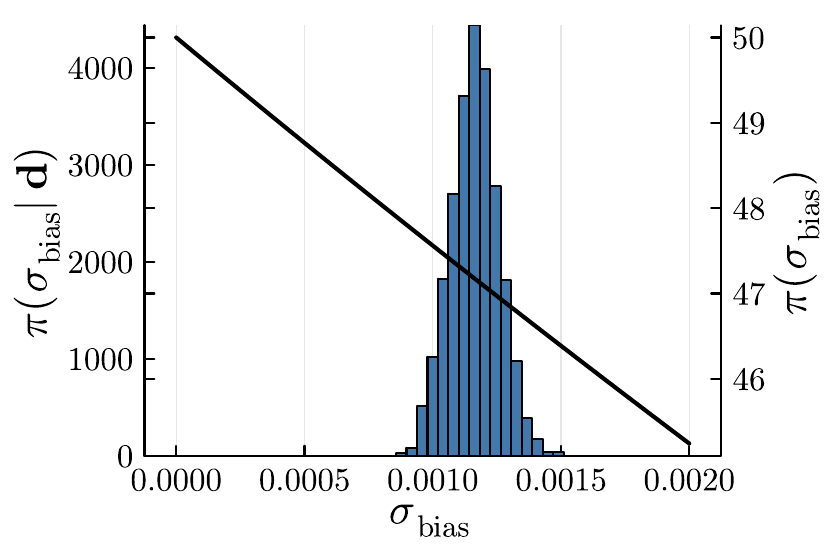}}
    \subfloat[Biviscous]{\includegraphics[width=0.33\linewidth]{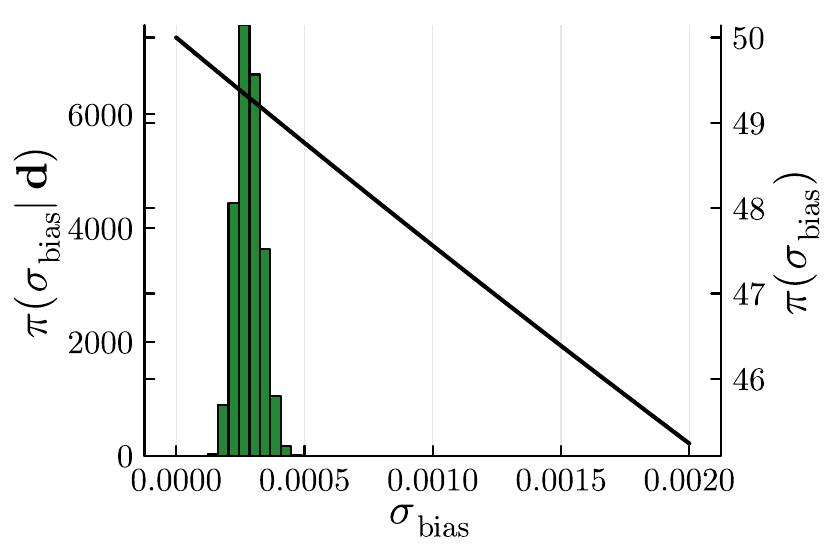}}
    \subfloat[Biviscous power law]{\includegraphics[width=0.33\linewidth]{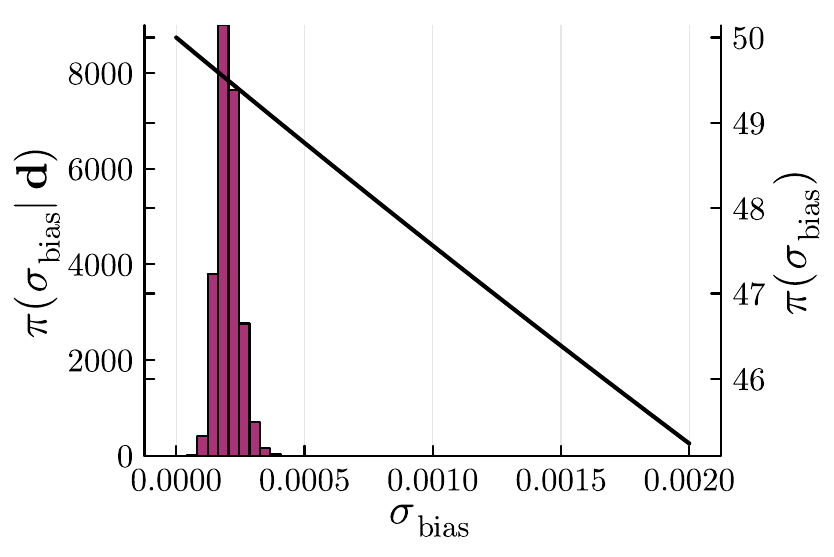}}
    \caption{Posterior (histogram) versus prior (black line) of the inferred model bias for all expert-informed models.}
    \label{fig:sf_expertmodelbias}
\end{figure*}

In \autoref{fig:sf_ppdcombined_uninfo} we show the predictive posterior distributions for the expert-informed scenario considered here. A clear distinction can now be made between the three considered models based on the shown credibility intervals. Although the biviscous model, which omits thinning effects, is able to capture the experimental data well in the early stages of the experiment, it deviates from the data at later times. This contrasts the behavior of the biviscous power law model, which does include thinning effects. This observed difference is confirmed by the distribution of the model bias in \autoref{fig:sf_expertmodelbias}, which reveals that the inferred model bias of the latter model is lower than that of the former. For the biviscous power law model it is observed that the model bias with expert-informed priors is substantially smaller than that with rheo-informed priors. When evaluating the plausibilities, for all three rheological models the expert-informed cases are found to have probability one and the rheo-informed cases have probability zero. This conveys that the impact of misinforming a complex fluid flow analysis with non-representative rheological information can compromise modeling accuracy and uncertainty.

\begin{figure}
    \centering
    \subfloat[Biviscous]{\includegraphics[width=0.5\linewidth]{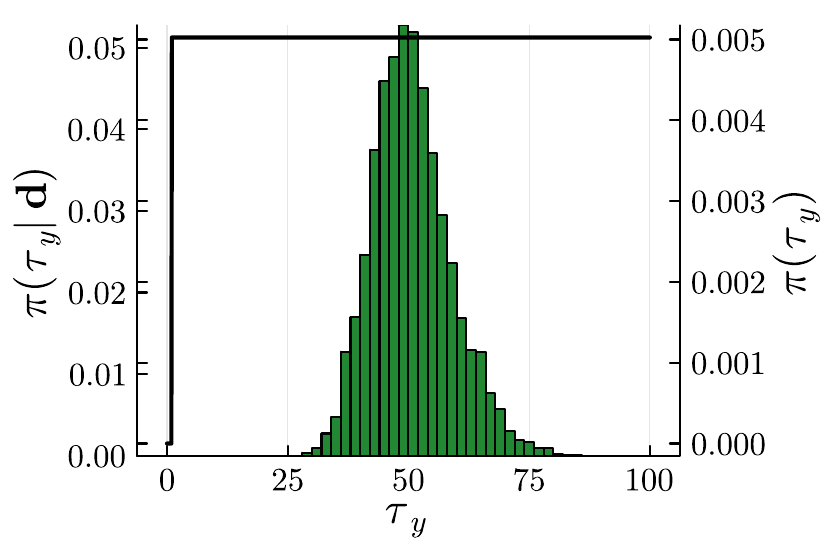}}
    \subfloat[Biviscous power law]{\includegraphics[width=0.5\linewidth]{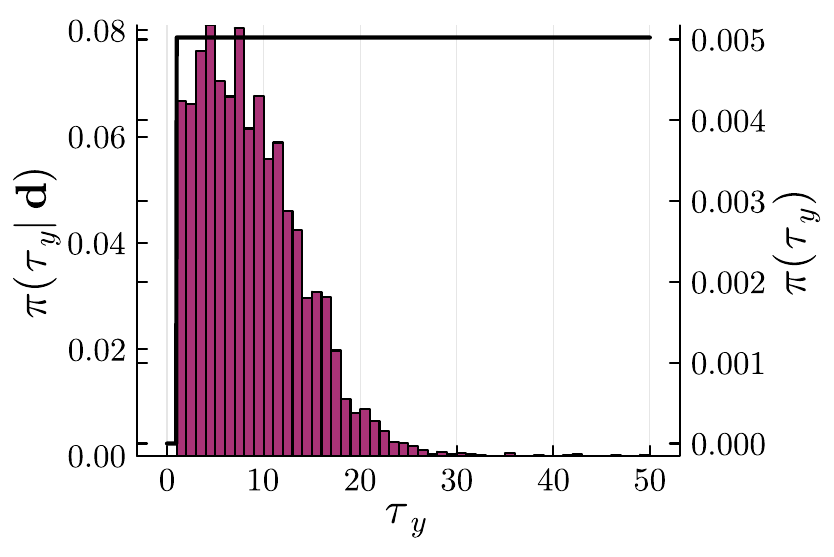}}
    \caption{Posterior (histogram) versus prior (black line) of the inferred yield stress for the expert-informed models.}
    \label{fig:expertyieldstress}
\end{figure}

\autoref{fig:expertyieldstress} shows the calibrated yield stress for the biviscous model and the biviscous power law model. Compared to the rheo-informed case in \autoref{fig:informedyieldstress}, for the expert-informed case considered here substantially lower yield stresses are found. This is a consequence of the broad prior considered for all rheological parameters, including the yield stress. The calibrated force values (\autoref{tab:priorparam_sf}) are realistic for the expert-informed case -- this in contrast to the rheo-informed case -- as now the data can be explained through calibration of the rheological parameters. The uncertainty of the inferred rheological parameters, in the particular the yield stress, increases substantially as a result.

\begin{figure*}
    \centering
    \subfloat[Rheo-informed case]{\includegraphics[width=0.5\linewidth]{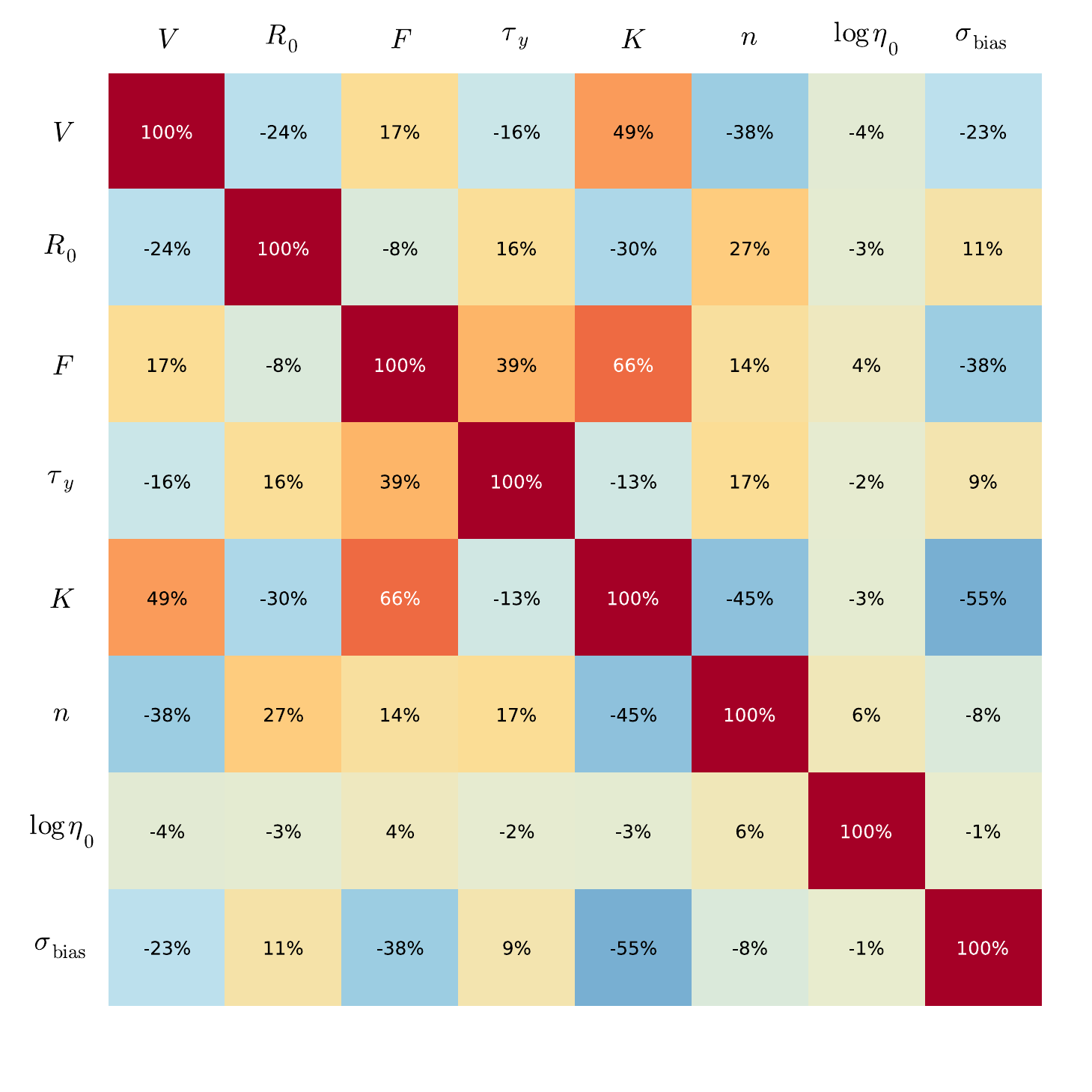}\label{fig:correlationsrheo}}
    \subfloat[Expert-informed case]{\includegraphics[width=0.5\linewidth]{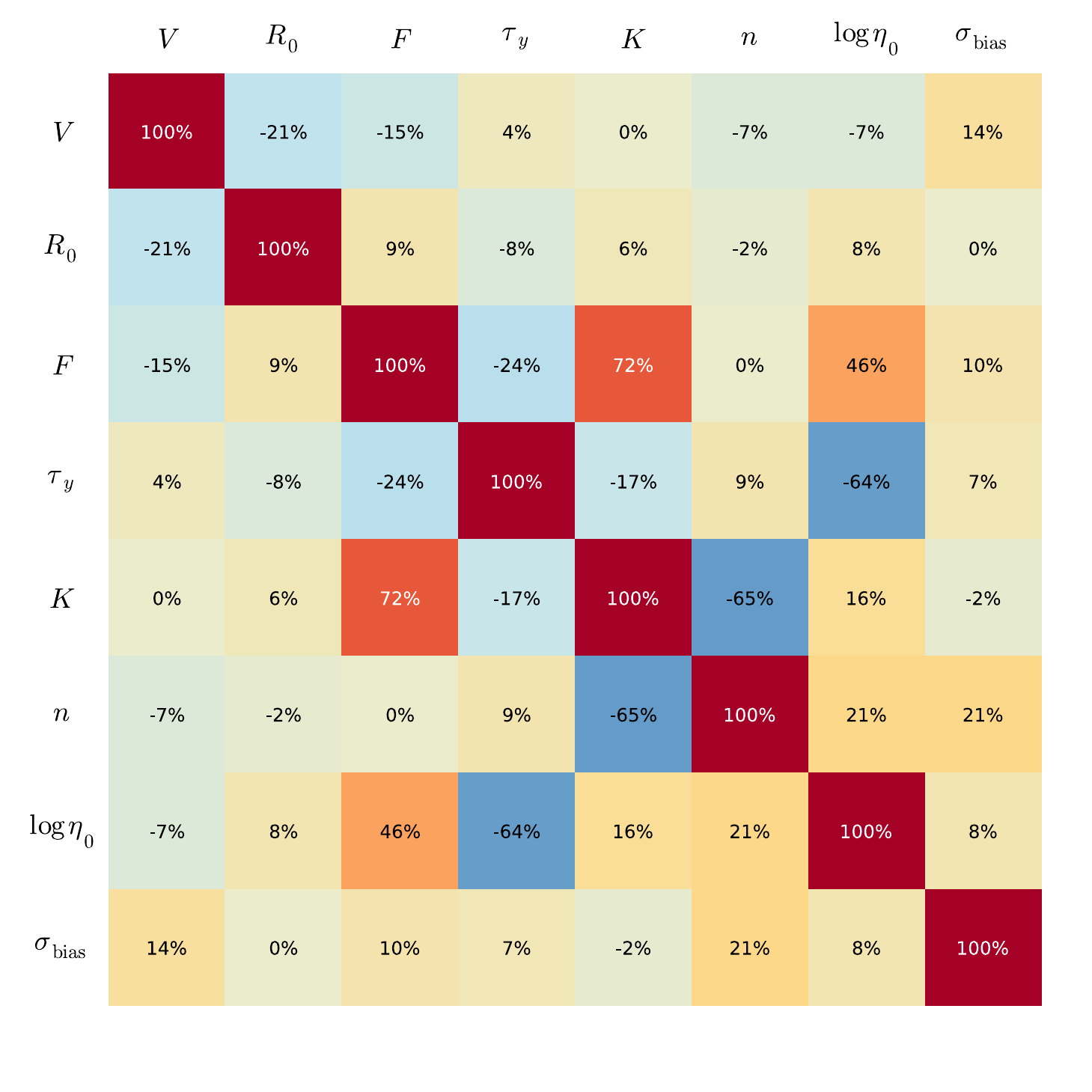}\label{fig:correlationsexpert}}
    \caption{Correlation matrices for the biviscous power law model.}
    \label{fig:correlations}
\end{figure*}

It is important to note that the observed reduction in inferred yield stress is a result of the calibration over the full parameter space. The posterior distribution over this space reveals strong correlations between various sets of parameters; see \autoref{fig:correlations}. These correlations complicate the interpretation of the calibration results, as changes in parameters are not independent. Moreover, the correlation structure of the posterior is strongly influenced by the priors, and hence differs substantially between the rheo-informed case and the expert-informed case. For example, a strong negative correlation is observed between the yield stress and the pre-yield viscosity for the biviscous power law model in the expert-informed case (\autoref{fig:correlationsexpert}). This implies that if the pre-yield viscosity was to be lowered, the yield stress would increase. This implies that when priors are introduced that favor a lower pre-yield viscosity, a higher yield stress would be observed. This steering of the solution is clearly visible when comparing the correlation structure to the rheo-informed case (\autoref{fig:correlationsrheo}), from which it is, e.g., observed that the strong correlation of the yield stress with the pre-yield viscosity as observed for the expert-informed case is suppressed by the highly informed priors for the rheological parameters. This stipulates the importance of careful prior selection in Bayesian analyses.

\section{Conclusions and recommendations}
\label{sec:conclusion}

We applied Bayesian model selection to compare various generalized Newtonian constitutive models for yield stress fluids in both rheological and more complex fluid flow settings. Our study encompasses rheological measurements of Carbopol 980 using a rotational rheometer. Squeeze flow experiments on the same fluid were considered as an example of a complex flow case.

In the rheological setting, we examined the effect of including model bias in the inference procedure. In the squeeze flow setting, we explored two distinct approaches: \emph{rheo-informed inference} and \emph{expert-informed inference}. In the rheo-informed approach we used the posterior distributions obtained from Bayesian inference on the rheological measurements as prior information for the squeeze flow analysis. In the expert-informed approach, we adopted uninformative priors. The first approach emulates the setting in which detailed lab experiments are available, whereas the second approach resembles a situation in which predictions of complex flows are to be made in the absence of a detailed characterization of the fluid.

In the rheological setting, we found that including or excluding the model bias had limited impact on the relative plausibility rankings of the models. However, excluding the model bias led to a misleading impression of predictive confidence, in particular when extrapolating beyond the observed data. Models that failed to capture key rheological features of the considered yield stress fluid, such as the Newtonian, Bingham, and biviscous models, exhibited substantial modeling errors. These errors translated directly into lower plausibility scores. The models that fit the data well, \emph{i.e.}, the Herschel–Bulkley and the biviscous power law models, could not be distinguished based on modeling error alone. In these cases, observational noise dominated the uncertainty, and Bayesian model selection favored models that avoided unnecessary complexity \emph{cf.}~Occam's razor.

In relation to the introduction of unnecessary complexity, we found the plausibility of the biviscous power law model to be sensitive to the choice of prior for the pre-yield viscosity. When steering the prior toward relatively large values, the plausibilities of the Herschel-Bulkley model and biviscous power law model coincide, since the biviscous power law model then converges toward the Herschel-Bulkley model. When steering the prior toward lower values, the Herschel-Bulkley model becomes more probable, as the prior then encompasses pre-yield viscosities that are inconsistent with the experimental data. These observations highlight the interplay between prior specification and model ranking, and emphasize the importance of carefully chosen priors in Bayesian frameworks.

For the squeeze flow problem, the \emph{rheo-informed approach} resulted in a notable discrepancy between model predictions and experimental observations, even for models that incorporate the essential physical mechanisms. In contrast, the \emph{expert-informed approach} yielded a significantly better fit to the squeeze flow data. The root cause of the rheo-informed approach's inability to explain the data lies in its priors, which bias the models toward yield stress values that are substantially higher than those inferred by the expert-informed approach. This discrepancy is attributed to the more complex stress states present in squeeze flow compared to those encountered in standard rheological measurements. The squeeze flow data is best explained by the expert-informed biviscous power law model. The bias associated with this model was found to be substantially lower than that of any other model, which directly translates into a very high model plausibility.

These findings underscore the well-known challenges of modeling yield stress fluids in complex flow scenarios: fluid characterization based solely on simple rheological tests can lead to misleading predictions. While the yield stress may appear to be accurately inferred from rheological measurements, its direct translation to complex flow settings introduces a bias that can result in significant deviations from observed behavior. A key strength of the Bayesian model selection framework is its ability to quantify this model bias and thereby illuminate the limitations of rheological inference in complex flow environments. In cases where a detailed rheological characterization is not available, the Bayesian framework allows experts to explicitly account for uncertainties, thereby facilitating decision-making processes in inherently uncertain environments.

Our Bayesian framework relies on Markov Chain Monte Carlo (MCMC) samplers, which require (at least) thousands of model evaluations to produce accurate results. Consequently, the framework currently depends on fast-to-evaluate models, imposing practical limitations on the complexity of the simulations that can be incorporated. To address this challenge, several complementary developments are needed. First, high-performance computing techniques -- such as parallelization and GPU acceleration -- can significantly reduce computational time. Second, the development of reduced-order modeling (ROM) approaches is essential to enable the integration of more advanced simulations, such as finite-element-based models, without compromising computational feasibility. Third, more efficient methods for evaluating Bayes' rule could further accelerate the inference process. These techniques collectively support the extension of the framework to industrially relevant applications. Theoretical advancements in assessing convergence and sampling quality remain critical to ensure robustness and reliability in practical deployments.

From the perspective of constitutive modeling of yield-stress fluids, a wide range of additional effects can be considered, including elastoviscoplasticity, thixotropy, aging, spatial inhomogeneities, and temperature dependence. Other effects, such as complex boundary conditions (e.g., wall slip) and geometrical effects, may also be incorporated. Bayesian uncertainty quantification, through its built-in Occam’s razor effect, provides a natural means to assess whether such model extensions, which generally introduce additional complexity, are justified by the available data. Finally, we note that inter- and intra-laboratory variability in material preparation and experimental setups, as discussed at length during the VPF workshop, can lead to significant uncertainties. Bayesian uncertainty quantification provides a structured and rigorous framework to address these challenges.

\section*{Acknowledgments}
The research contributions of CV and NJ are partially sponsored by the DAMOCLES project within the EMDAIR program of the Eindhoven Artificial Intelligence Systems Institute (EAISI).

\section*{Declaration of generative AI and AI-assisted technologies in the manuscript preparation process}
During the preparation of this work the authors used Microsoft Copilot for textual editing. After using this service, the authors reviewed and edited the content as needed and take full responsibility for the content of the published article.


\appendix


\section{Model evidence variance estimator}
\label{sec:APPvariance}

To assess the quality of the model evidence estimator \eqref{eq:importancesamplingestimator}, we here study its approximation behavior for different sample sizes using ten independently generated Markov chains for each sample size. Our analysis is based on the calibration of the Herschel-Bulkley model on the rough plate rheological measurement data, with model bias included.

\begin{figure}
    \centering
    \includegraphics[width=0.5\linewidth]{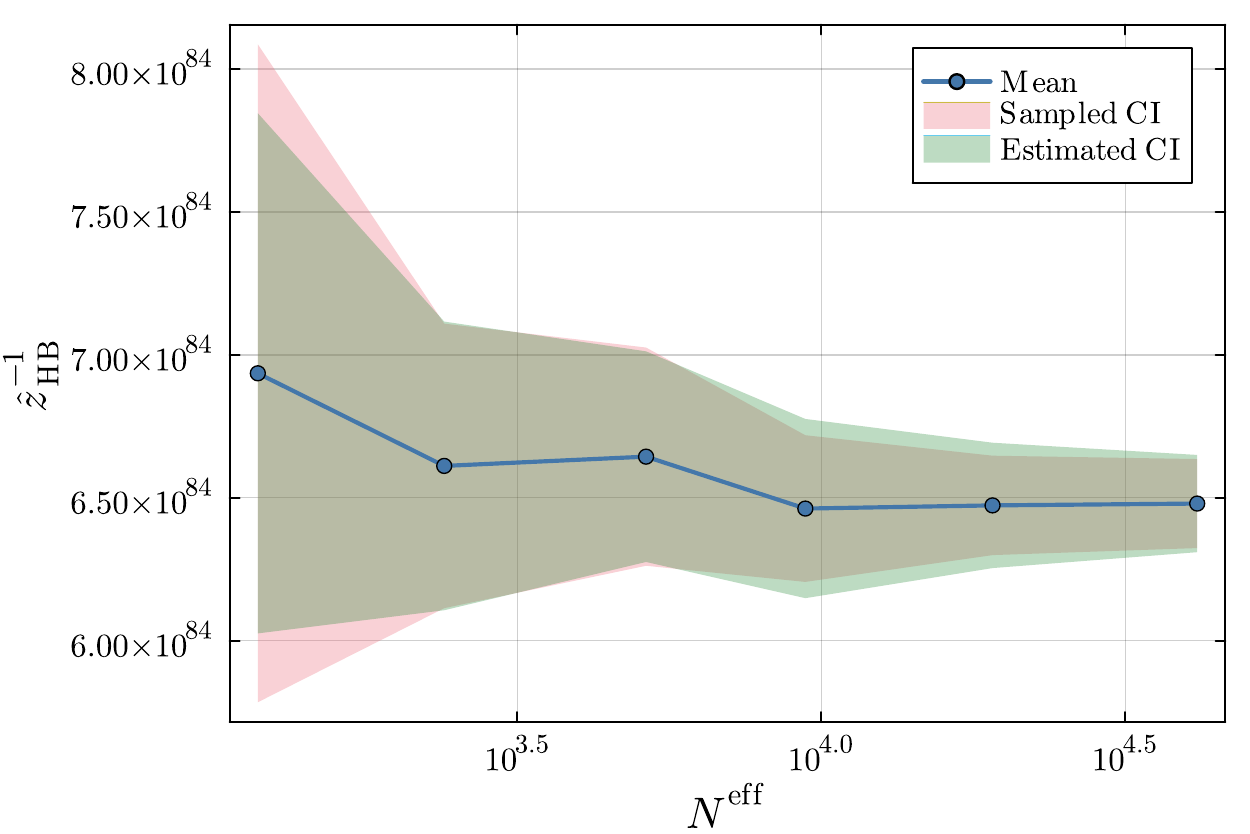}
    \caption{Dependence of the inverse evidence estimator on the (effective) sample size. The displayed 95\% confidence intervals (CI) are based on the variance of ten independent chains (Sampled CI) and on the variance estimate  $\text{Var}\left[ \hat{z}_{\rm HB}^{-1} \right]$ (Estimated CI).}
    \label{fig:evidenceestimatorcomparison}
\end{figure}

In \autoref{fig:evidenceestimatorcomparison} we display the estimated inverse evidence, $\hat{z}^{-1}_{\mathrm{HB}}$, for different sample sizes. The displayed mean value is the average over the ten independent chains. The sampled 95\% confidence interval is based on the standard deviation over the chains, whereas the estimated confidence interval corresponds to the evidence variance estimate \eqref{eq:varianceestimator}, averaged over the independent chains. \autoref{fig:evidenceestimatorcomparison} conveys that the estimator converges upon increasing the sample size. In addition, the evidence variance estimate is observed to be in good agreement with that computed based on the ten independent chains.

\begin{figure}
    \centering
    \includegraphics[width=0.5\linewidth]{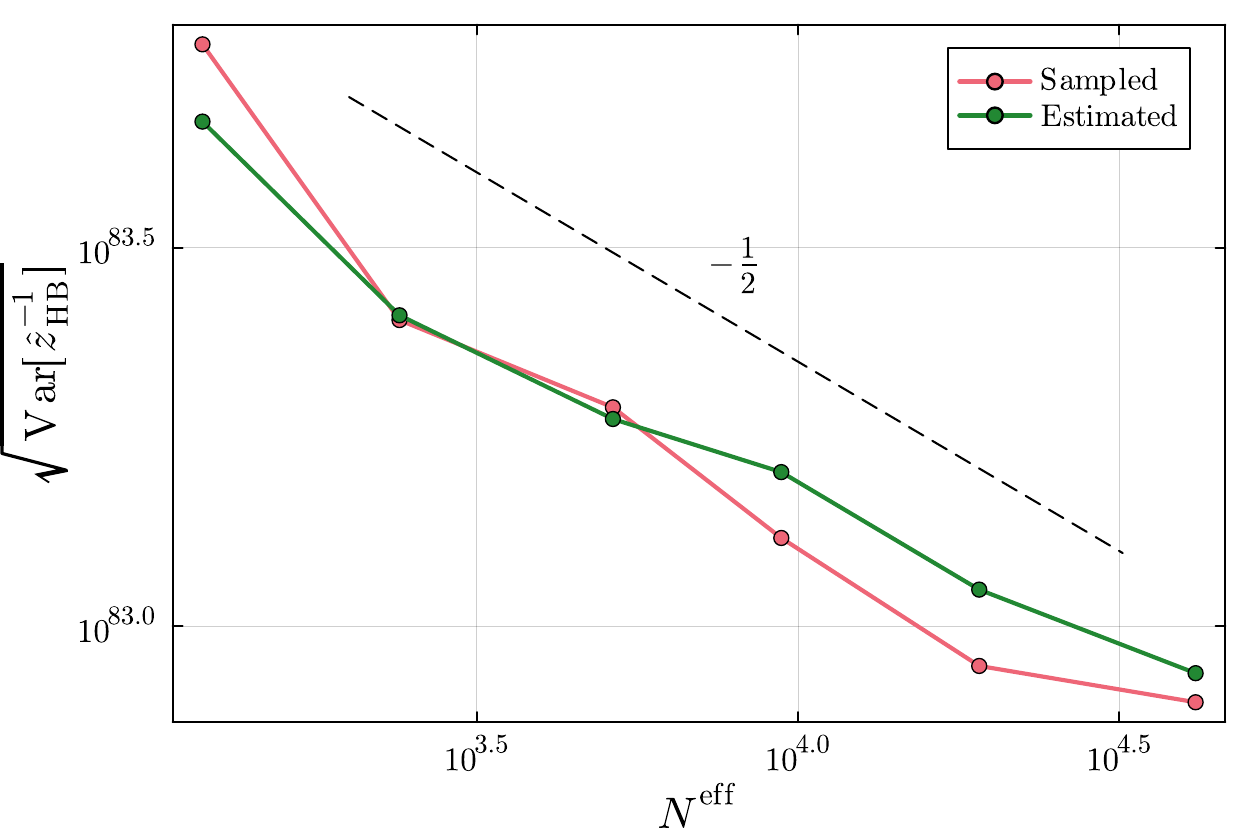}
    \caption{Convergence of the variance of the evidence estimator, comparing the variance obtained from ten independent chains (Sampled) with that evaluated using the estimate $\text{Var}\left[ \hat{z}_{\rm HB}^{-1} \right]$ (Estimated).}
    \label{fig:evidenceestimatorrate}
\end{figure}

In \autoref{fig:evidenceestimatorrate} the increase in accuracy with increasing sample size is studied in detail by comparing the standard deviation of the evidence estimator computed based on the ten independent chains with that computed using \cref{eq:varianceestimator}. A good resemblance of these variance estimates is observed, with both converging with the square root of the sample size, in line with theoretical expectations. In summary, the estimator \eqref{eq:varianceestimator}, which can be evaluated based on a single chain, is observed to provide a good estimate of the variance of the evidence estimator \eqref{eq:importancesamplingestimator}.

\section{The squeeze flow solver}
\label{sec:solver}
This appendix discusses the developed squeeze flow solver. In Appendix~\ref{app:sf_fluxrelation} we first detail the flux relations for the various rheological models. In Appendix~\ref{app:sf_procedure} we then detail the numerical solution procedure. Finally, in Appendix~\ref{app:sf_vandv} we verify and validate our numerical squeeze flow model.

\subsection{Velocity profile and flux}\label{app:sf_fluxrelation}
For each of the rheological models in \autoref{sec:rheologymmodels}, the squeeze flow velocity profile can be found using \cref{eq:flowprofile}. We here derive the velocity profile for the biviscous power law model, as all other rheological models are special cases thereof.

Integration of \cref{eq:flowprofile} yields the velocity profile $v(z)$ for $z \in [-H/2,H/2]$ as
\begin{equation}
   v( z ) = \begin{cases}
    v_{u}(z) &  | z | < w_y \quad\text{(unyielded)}, \\
    v_{y}(z) &  | z |  \geq w_y \quad\text{(yielded)},
    \end{cases}
    \label{eq:velocityprofile}
\end{equation}
where
\begin{equation}
 w_y = \begin{cases}
     \frac{H}{2} & \tau_y \geq |p'| \frac{H}{2},\\
     \frac{\tau_y}{ \left| p' \right|}  & \text{otherwise},
 \end{cases}
 \label{eq:yieldsurface}
\end{equation}
is the boundary between the unyielded region at the center and the yielded region near the plates. The velocity contributions in \cref{eq:velocityprofile} read
\begin{subequations}
\begin{align}
    v_{u}(z) &= \frac{n}{(n+1)\sqrt[n]{K} p' } \left( ( |p'|w_y  -  \tau_0  )^{\frac{n+1}{n} } -  (  |p'| \tfrac{H}{2} -  \tau_0 )^{ \frac{n+1}{n} } \right)  \nonumber \\    
    &\phantom{=} + \frac{p'}{2 \eta_0}  \left(  z^2 -w_y^2\right),\\
    v_{y}(z) &= \frac{n}{(n+1)\sqrt[n]{K} p' } \left( ( |p'| z  -  \tau_0  )^{\frac{n+1}{n} } -  ( |p'| \tfrac{H}{2} -  \tau_0 )^{ \frac{n+1}{n} } \right) , 
\end{align}
\label{eq:BPLvelocity}
\end{subequations}
where $p' = \frac{\partial p}{\partial r}$.

Note that for a positive yield stress and finite pressure gradient, there will always be a small unyielded zone at the center. As the pressure gradient becomes smaller in magnitude, the width of the unyielded zone increases \emph{cf.} \cref{eq:yieldsurface}, until there is no more yielding when $w_y = \frac{H}{2}$. Also note that the yield boundary definition \eqref{eq:yieldsurface} considers the boundary position in the top half of the profile, but that there is a symmetric yield boundary at $z=-w_y$ in the bottom half of the profile.

Upon substitution of \cref{eq:BPLvelocity} in \cref{eq:fluxrelation}, the flux relation for the biviscous power law model is obtained in analytical form as
\begin{align}
    Q &= Q_u + Q_y,
\end{align}
with
\begin{subequations}
\begin{align}
    Q_u &= \frac{2 n w_y \left( ( |p'|w_y  -  \tau_0  )^{\frac{n+1}{n} } -  (  |p'| \tfrac{H}{2} -  \tau_0 )^{ \frac{n+1}{n} } \right)}{(n+1)\sqrt[n]{K} p' } - \frac{2 p' w_y^3}{3 \eta_0},\\
    Q_y &= \frac{2 n^2 \left( ( |p'| \tfrac{H}{2}  -  \tau_0  )^{\frac{2n+1}{n} } - ( |p'| w_y  -  \tau_0  )^{\frac{2n+1}{n} } \right)}{(n+1)(2n+1)\sqrt[n]{K} p' | p' |}  \nonumber \\
     &\phantom{=} - \frac{n ( H - 2 w_y)  ( |p'| \tfrac{H}{2} -  \tau_0 )^{ \frac{n+1}{n} }}{(n+1)\sqrt[n]{K} p' }.
\end{align}    
\label{eq:BPLflux}%
\end{subequations}
The flux for the Herschel-Bulkley model is obtained by setting $\eta_0 \rightarrow \infty$ and replacing $\tau_0$ with $\tau_y$. The biviscous model is obtained by setting $n$ to one. The Bingham model is obtained by combining these modifications. And finally, the Newtonian model is obtained by setting $w_y=\tfrac{H}{2}$ and substituting $\eta_0$ by $\eta$. We note that the Newtonian model can also be obtained through other simplifications of \cref{eq:BPLflux}.

\subsection{Nonlinear solution procedure}\label{app:sf_procedure}
To solve the nonlinear problem \eqref{eq:ibvp}, we discretize the domain in $N_r$ segments of size $\Delta R(t_i) = R(t_i) / N_r$, where $t_i$ are discrete time instances following a geometric series of size $N_t$, with the initial time step size set to $\Delta t_0 = t_1 - t_0$.

For every time instance we compute the height rate by solving \cref{eq:ibvpa,eq:ibvpb,eq:ibvpc} and explicitly update the height by
\begin{equation}
    H ( t_{i+1} ) =  H ( t_{i} ) + \Delta t \dot{H}(t_{i}),
\end{equation}
which we initialize using \cref{eq:ibvpd}. The corresponding radius follows as $R(t_i) = \sqrt{V/ (\pi H(t_i))}$.

To determine the height rate, $\dot{H}$, at any moment in time, we solve the discretized form of \cref{eq:ibvpb}, \emph{i.e.}, 
\begin{equation}
    \sum_{j=1}^{N_r} \bar{p}_j(\dot{H}) \bar{r}_j \Delta R = \frac{F}{2 \pi},
    \label{eq:discretemassbalance}
\end{equation}
where $\bar{p}_j = \tfrac{1}{2}(p_{j-1} + p_{j})$ and $\bar{r}_j = \tfrac{1}{2}(r_{j-1} + r_{j})$ with $j=1,\ldots, N_r$ are the midpoint pressures and radii. Using the boundary condition \eqref{eq:ibvpc}, the discrete pressures are found as
\begin{equation}
    p_j = \Delta p (H) - \Delta R \sum_{k={j+1}}^{N_r} \bar{p}'_k,
\end{equation}
where the midpoint pressure gradients follow from \cref{eq:ibvpa} as
\begin{equation}
    \bar{p}_k' = Q^{-1}\left( - \frac{\bar{r}_k \dot{H}}{2}  \right).
    \label{eq:fluxinverse}
\end{equation}

We use Newton-Raphson iterations to solve the nonlinear problem \eqref{eq:discretemassbalance}, as well as for the inversion of the flux relation \eqref{eq:fluxinverse}. To improve the robustness of this nested Newton procedure, we augment the Newton algorithm with a bracket in which the solution is to be found and fall back to a bisection update if the Newton step proposes an update outside of this bracket \cite{Press2007Numerical}. In addition, we allow for adaptive time step refinement of the geometric series based on the number of Newton iterations of the outer Newton algorithm and on the predicted height rate. If the relative height change $| \dot{H} \Delta t_i / H(t_i) |$ is larger than a set maximum, the time step is limited to match this maximum. 

We have implemented our squeeze flow solver in a performant Julia \cite{Julia-2017} package.

\subsection{Validation and verification}\label{app:sf_vandv}
To optimize the mesh size and time step size for our squeeze flow solver, we study these discretization parameters in Appendix~\ref{app:mesh} and Appendix~\ref{app:timestep}, respectively. In Appendix~\ref{app:TFEM} we validate the result of our solver using a high-fidelity finite element solution.

For our verification and validation study we make use of the representative test scenario in \autoref{tab:test_scenario}. The relative error of the radius is measured at $t \in [0.125, 0.25, 0.5, \ldots, 32.0, 64.0]~[\si{\second}]$, which resembles the way in which we post-process our experimental results. The errors are defined relative to the reference setting in \autoref{tab:test_scenario}. Simulation times are normalized with respect to the reference case.

{ 
\sisetup{
  detect-all,
  round-mode=figures,
  round-precision=3,
  scientific-notation=true,
  retain-explicit-plus = true
}

\begin{table*}
\centering
\caption{Parameters of the validation and verification scenario.\label{tab:test_scenario}}
\begin{tabular}{
  r@{\hspace{3.em}}  
  l@{\hspace{3.em}}  
}
\toprule
\textbf{Parameter} & \textbf{Value} \\
\midrule
\multicolumn{2}{l}{\textit{Fluid}} \\ \midrule
$\eta_0$ & $\SI{100000}{\pascal\second}$ \\
$K$ & $\SI{50}{\pascal\second}^{n}$ \\
$\tau_y$ & $\SI{10}{\pascal\per\second}$ \\
$n$ & $\SI{0.5}{}$ \\ \midrule
\multicolumn{2}{l}{\textit{Experiment}} \\ \midrule
$R_0$ & $\SI{0.01}{\meter}$\\
$V$ & $\SI{1e-6}{\meter^3}$\\
$F$ & $\SI{1}{\newton}$\\
$T$ & $\SI{100}{\second}$\\
\bottomrule
\end{tabular}
\hspace{1cm}
\begin{tabular}{
  r@{\hspace{3.em}}  
  l@{\hspace{3.em}}  
}
\toprule
\textbf{Parameter} & \textbf{Value} \\
\midrule
\multicolumn{2}{l}{\textit{Reference case}} \\ \midrule
$N_t$ & $8192$ \\
$N_r$ & $256$\\
$\Delta t_0$ & $\SI{0.0001220703125}{\second}$\\ \midrule
\multicolumn{2}{l}{\textit{Optimized case}} \\ \midrule
$N_t$ & $256$ \\
$N_r$ & $8$\\
$\Delta t_0$ & $\SI{0.001953125}{\second}$\\
\bottomrule \\
& 
\end{tabular}
\end{table*}
}

\subsubsection{Mesh convergence}\label{app:mesh}
\begin{figure}
    \centering
    \includegraphics[width=0.5\linewidth]{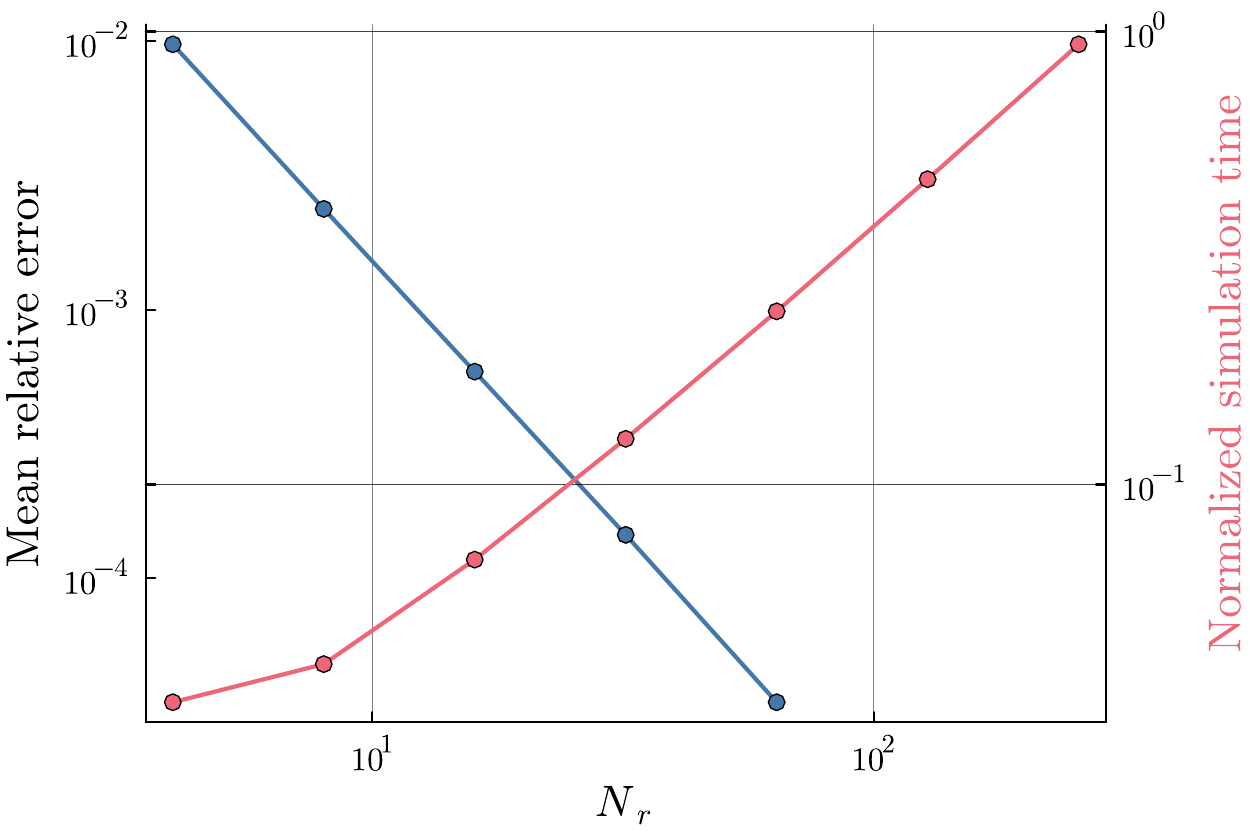}
    \caption{Mesh convergence results}
    \label{fig:Nr_convergence}
\end{figure}

\autoref{fig:Nr_convergence} shows the relative error in the radius averaged over all measurement times for various mesh sizes. The number of time steps and initial time step size are kept constant at their reference values. The error is observed to converge linearly with the mesh size, with an error of 0.2\% obtained for $N_r=8$ segments. The corresponding relative simulation times are also shown, revealing a linear scaling with the number of segments. At $N_r=8$ the simulation time is an order of magnitude smaller than that for the reference setting.

\subsubsection{Time-step size convergence}\label{app:timestep}
\begin{figure}
     \centering
     \includegraphics[width=0.5\linewidth]{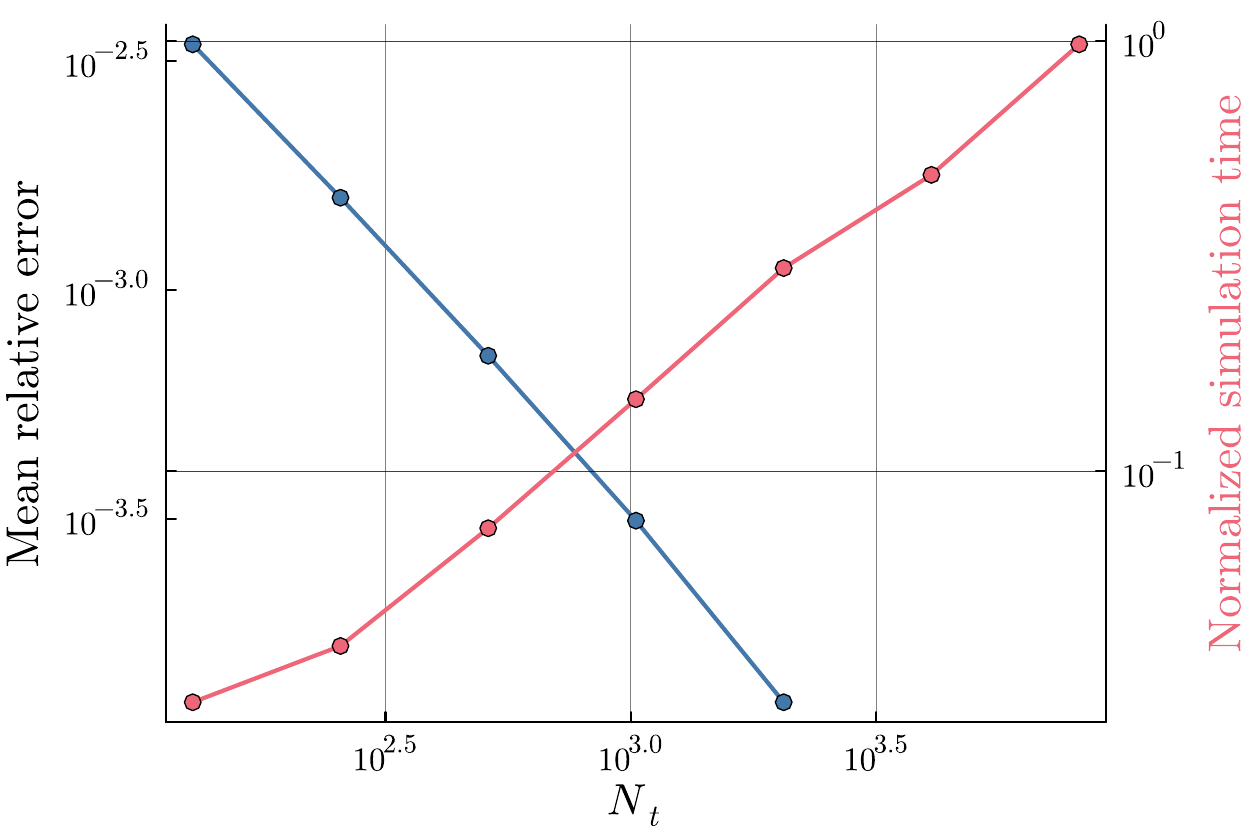}
     \caption{Time steps convergence results}
     \label{fig:Nt_convergence}
\end{figure}

\autoref{fig:Nt_convergence} shows the error versus the number of time steps, with the mesh size and initial time step size being kept at their reference values. The error is observed to decrease linearly with the time step size. At $N_t=256$ time steps, the relative error is 0.2\%. The corresponding simulation times are shown to scale linearly, with the simulation time for $N_t=256$ an order of magnitude smaller than that for the reference case.

\begin{figure}
    \centering
    \includegraphics[width=0.5\linewidth]{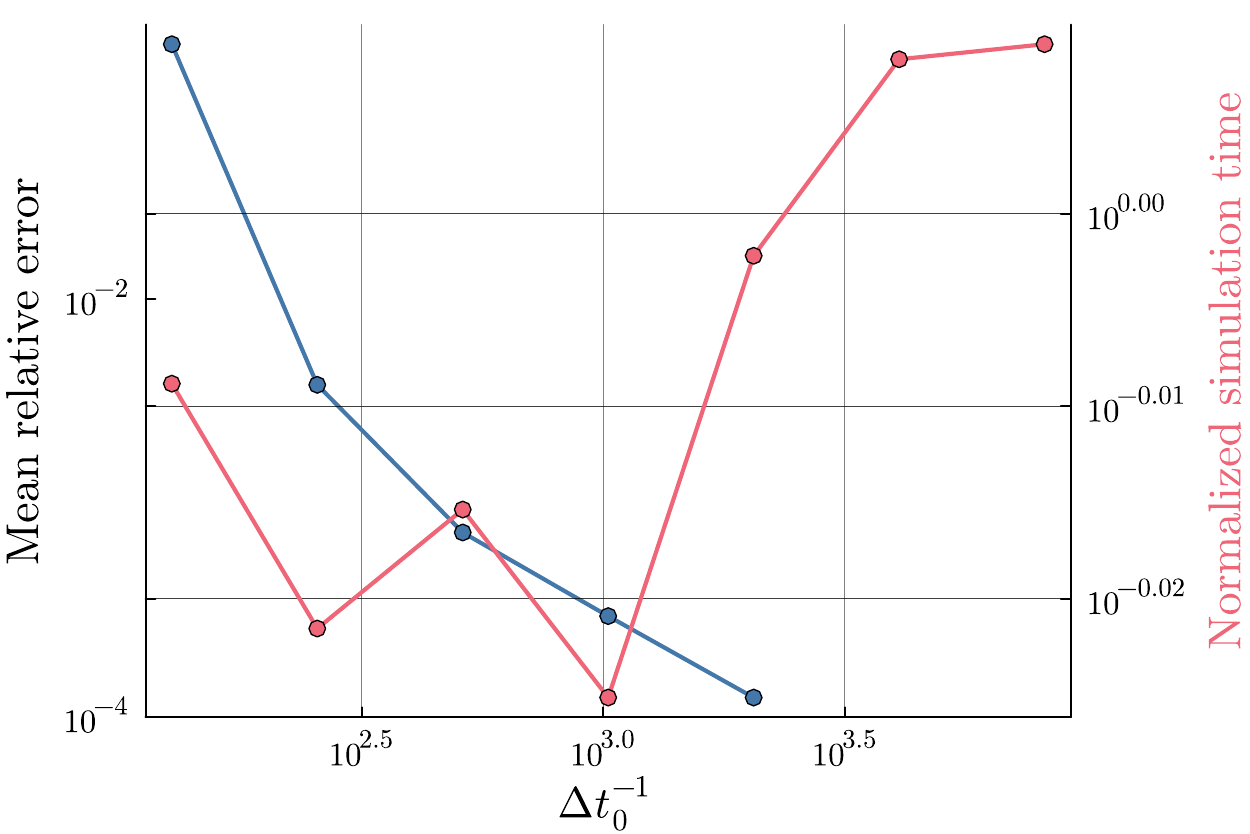}
    \caption{Initial time step size convergence results}
    \label{fig:Dt0_convergence}
\end{figure}

\autoref{fig:Dt0_convergence} shows the convergence behavior of the initial time step size. At $\Delta t_0 = \SI{1.95}{\milli\second}$ the average error is observed to be less than 0.1\% compared to the reference solution. The simulation time is observed to be hardly affected by the initial time step size, as it does not affect the number of time steps. 

\subsubsection{Comparison with a full-dimensional finite element solution}\label{app:TFEM}
\begin{figure}
     \centering
     \begin{subfigure}[b]{0.49\textwidth}
         \centering
         \includegraphics[width=\textwidth]{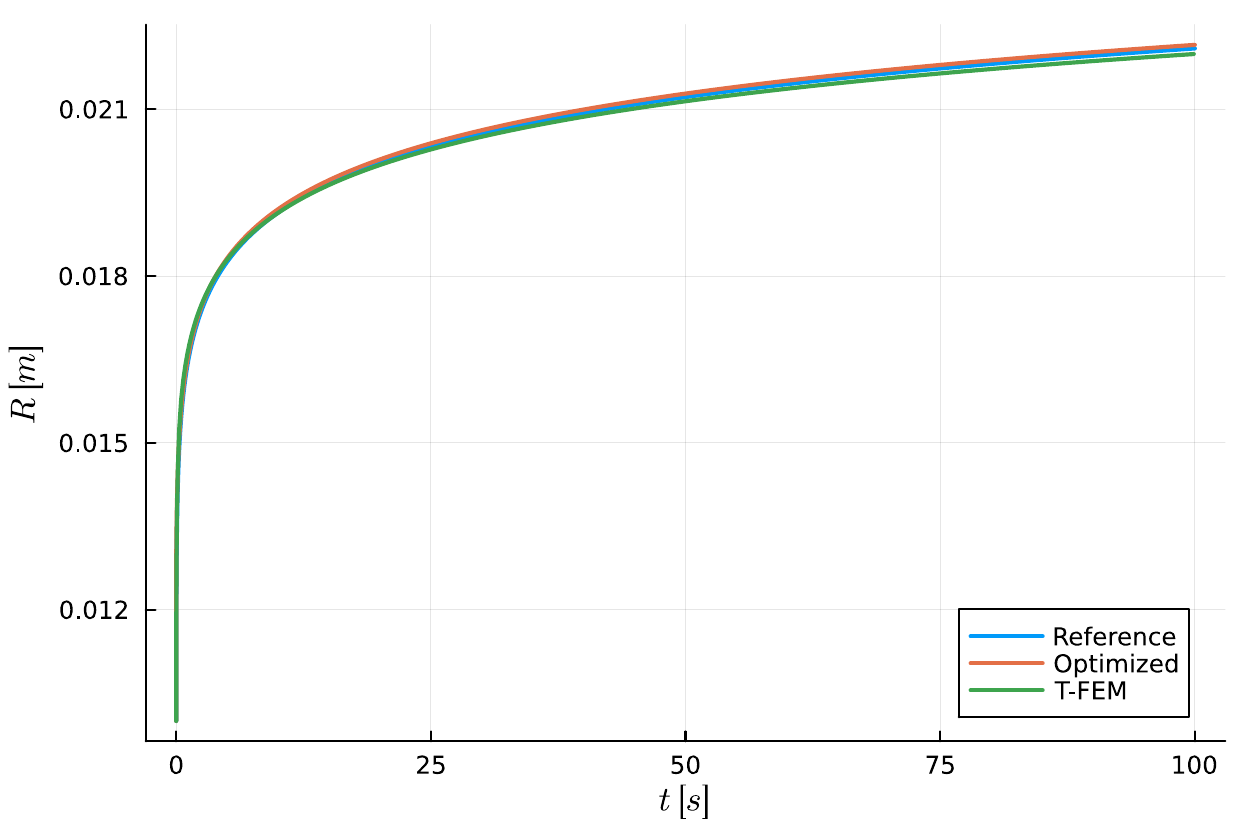}
         \caption{Radius evolution}
         \label{fig:R_vs_t}
     \end{subfigure}
     \hfill
     \begin{subfigure}[b]{0.49\textwidth}
         \centering
         \includegraphics[width=\textwidth]{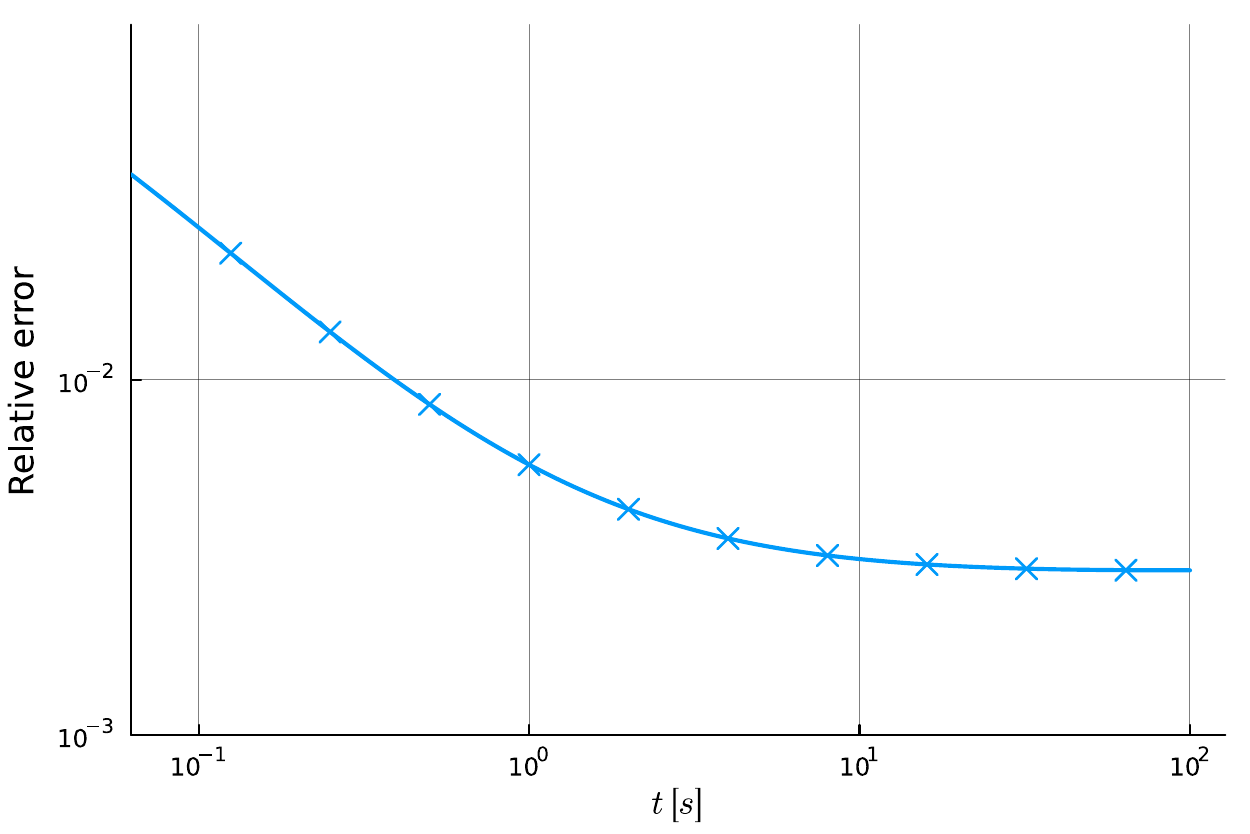}
         \caption{Optimized error}
         \label{fig:error_vs_t}
     \end{subfigure}
          \hfill
     \begin{subfigure}[b]{0.49\textwidth}
         \centering
         \includegraphics[width=\textwidth]{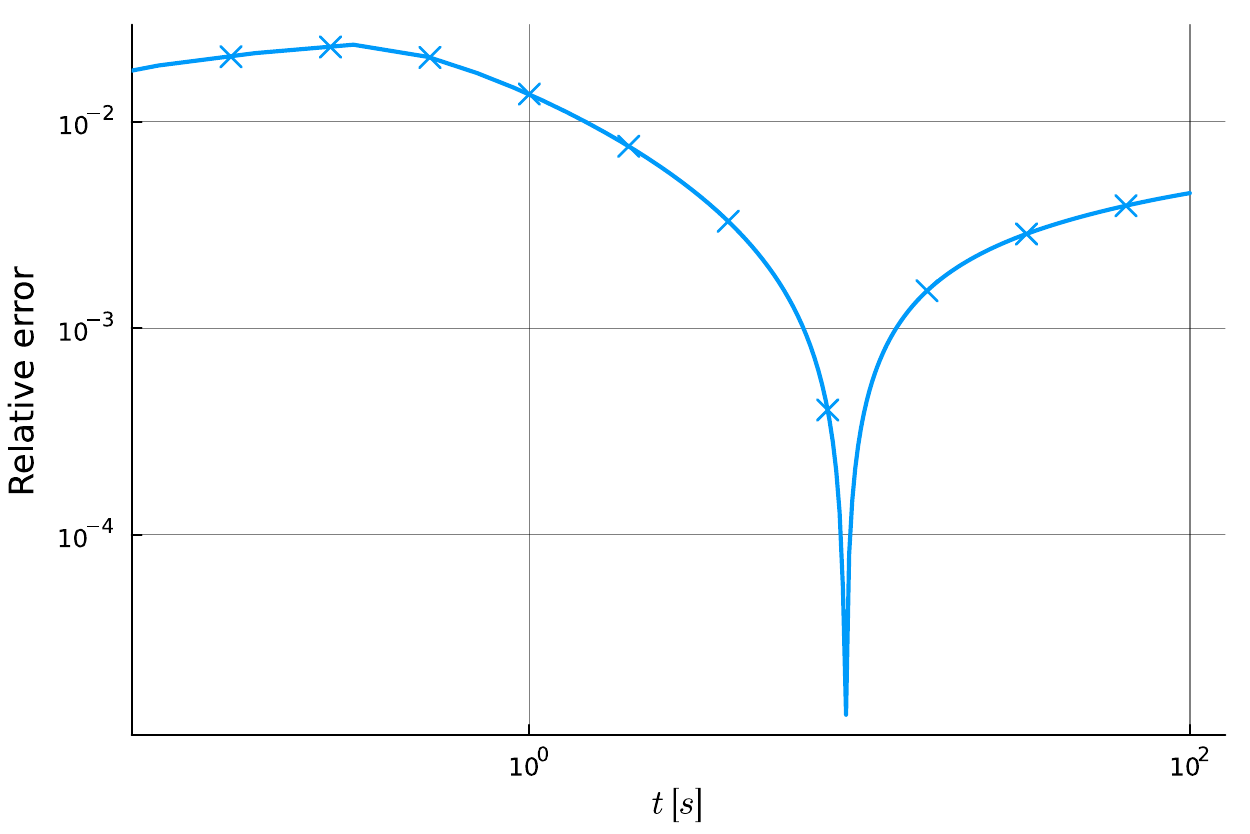}
         \caption{T-FEM error}
         \label{fig:TFEMerror_vs_t}
     \end{subfigure}
        \caption{Temporal error analysis}
        \label{fig:convergenceintime}
\end{figure}

Based on the convergence study presented above, we have defined the optimized settings as reported in \autoref{tab:test_scenario}. As observed from \autoref{fig:convergenceintime}, on average, the relative radius error is around 0.2\%, which is insignificant in comparison to the uncertainties we study. In fact, the difference between the reference case and the optimized case can hardly be observed when inspecting the radius evolution curve (\autoref{fig:R_vs_t}). The simulation time corresponding to the optimized case is 200 times smaller than that of the reference case, which translates to optimized single-core simulation times of well below a second on a modern personal computer.

To validate our squeeze flow model, we compare its results with a high-fidelity finite element (FEM) result based on the full-dimensional Stokes equations. In the FEM simulations, the balance of momentum and balance of mass are solved in a cylindrical coordinate system, and by using axisymmetry, the problem is solved on a 2D domain. The full rate-of-deformation tensor $\boldsymbol{D}$ and stress tensor $\boldsymbol{\tau}$ are retained, \emph{i.e.}, none of its components are assumed negligible. In contrast to the lubrication assumption, this implies that in the cylindrical coordinate system used here, the stress tensor can have non-zero components on its diagonal, in addition to shear components.

For numerical stability, we employ the regularization of the Herschel-Bulkley model by \citet{Papanastasiou87} to determine the stress as
\begin{equation}
       \boldsymbol{\tau} = 2 \eta(\dot{\gamma}_\text{e}) \boldsymbol{D} \quad \text{with} \quad
       \eta(\dot{\gamma}_\text{e})=K\dot{\gamma}_\text{e}^{n-1} + \frac{\tau_\text{y}(1-\exp(- \varrho\dot{\gamma}_\text{e}))}{\dot{\gamma}_\text{e}},
\end{equation}
where $\dot{\gamma}_\text{e}=\sqrt{2\boldsymbol{D}:\boldsymbol{D}}$ is the effective shear rate, and $\varrho$ is the regularization parameter. In the limit of $\varrho$ going to infinity, the Papanastasiou model approaches the Herschel-Bulkley model.

The non-linear system of equations is solved using an in-house finite element code. To deal with the non-linearity, initially, a number of Picard iterations are used followed by Newton-Raphson iterations. In order to make the simulations more efficient, we exploit the symmetry of the problem with respect to the mid-plane, \emph{i.e.}, only half of the domain is simulated. A force is prescribed on the top wall, and a constraint is implemented to ensure the entire wall moves with the same velocity. The velocity of the wall is then solved for as unknown, and is used to update the domain with a forward Euler time stepping scheme. The free surface of the fluid is kept straight, and it is displaced such that volume is conserved. We use iso-parametric, quadrilateral $Q_2/Q_1$ (Taylor-Hood) elements for the velocity/pressure. A mesh of 20 $\times$ 20 elements is employed, which is shown to give converged solutions.

\begin{figure}
    \centering
    \includegraphics[width=0.5\linewidth]{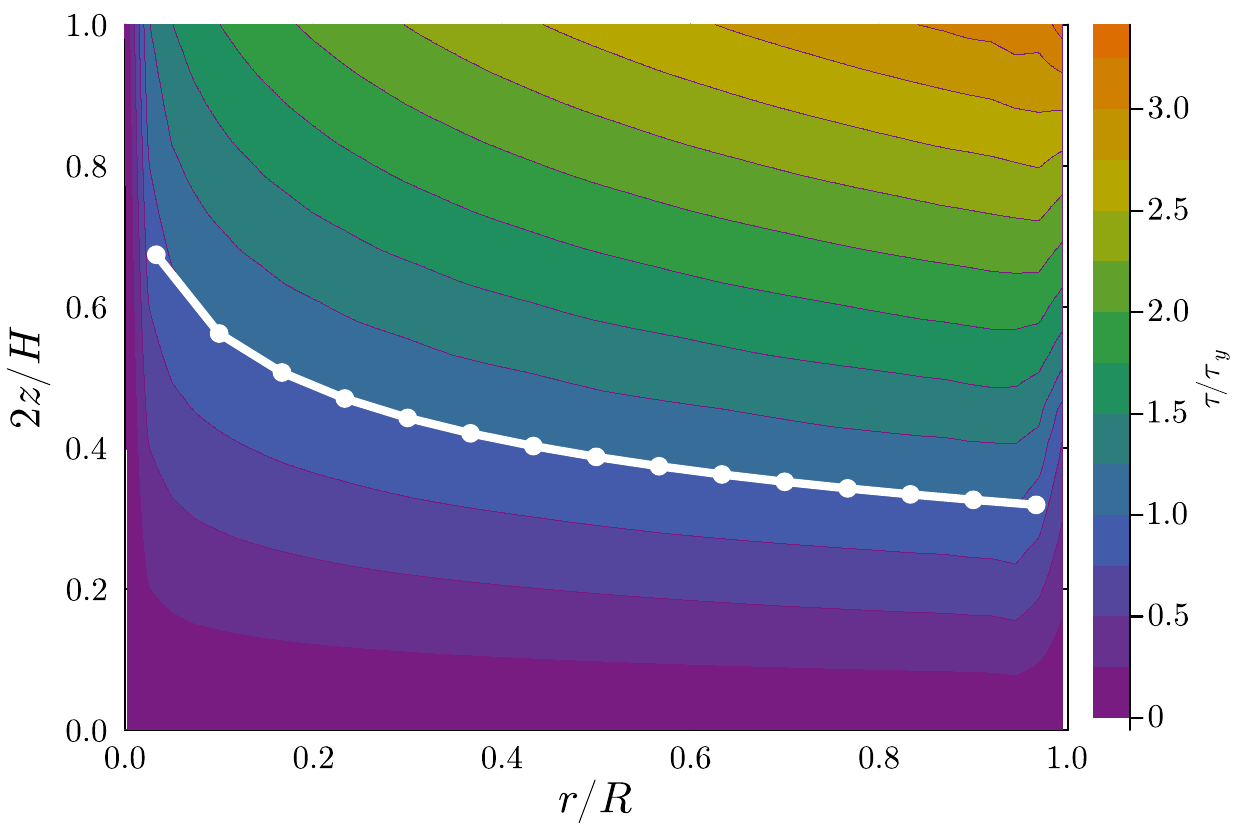}
    \caption{Yield boundary comparison between the high-fidelity FEM simulation (contour) and our squeeze flow model (white line).}
    \label{fig:taucontour}
\end{figure}

\begin{figure}
    \centering
    \includegraphics[width=0.5\linewidth]{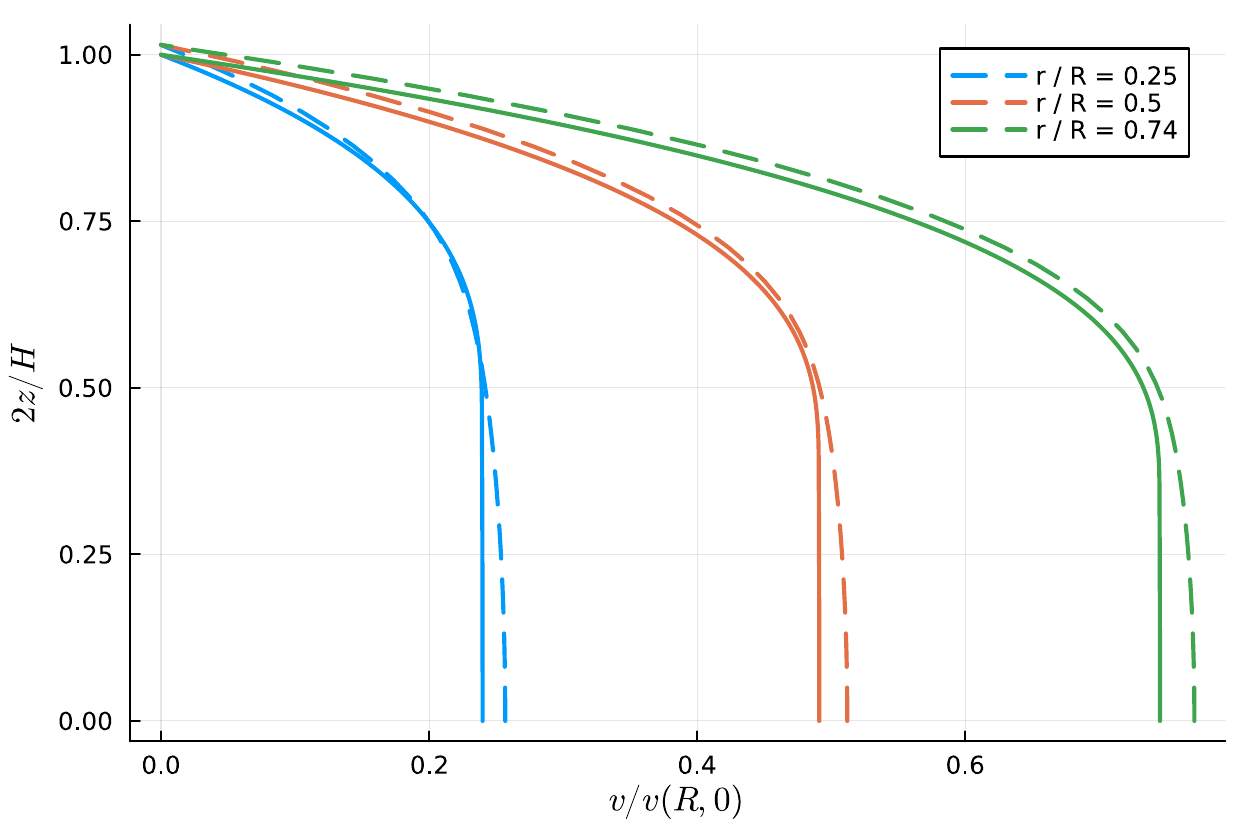}
    \caption{Velocity profile comparison between the high-fidelity FEM simulation (dashed lines) and our squeeze flow model (solid lines).}
    \label{fig:profilecomparison}
\end{figure}

The radius evolution prediction by the FEM simulation is shown in \autoref{fig:R_vs_t}, from which only minor differences are observed. The relative error in the radius over time is shown in \autoref{fig:TFEMerror_vs_t}, which shows that on average the error remains below 1\%. In \autoref{fig:taucontour} we compare the yield surface prediction of our optimized squeeze flow solver with that of the FEM simulation at $t=T$. The finite element result is shown by the contour of the shear stress, $\tau_{rz}$, normalized by the yield stress $\tau_y$. That is, the yield surface corresponds to the isocontour with value one. For our squeeze flow model, the height of the yield surface is given by \cref{eq:yieldsurface}. It is observed that our model predicts the yield surface with a high degree of accuracy throughout the domain, with the most notable difference emanating from the way in which boundary conditions are applied. When inspecting the velocity profiles in \autoref{fig:profilecomparison}, also minor differences between our optimized model and the full FEM simulation are observed.

\bibliographystyle{unsrtnat} 
\bibliography{references.bib}

\end{document}